\documentclass[a4paper, 11pt]{article}

\usepackage[utf8]{inputenc}   \usepackage[T1]{fontenc}
\usepackage{lmodern}          \usepackage{amsfonts,amscd,amssymb,amsmath,amsthm}
\usepackage{array,mathtools,mathrsfs,bm}
\usepackage{booktabs,caption,longtable,multicol,multirow}
\usepackage{subcaption,lscape,enumitem}
\usepackage[linesnumbered,boxruled,lined]{algorithm2e}
\usepackage[colorlinks,citecolor=purple,linkcolor=blue]{hyperref}
\usepackage[margin=0.6in]{geometry}
\usepackage[authoryear]{natbib}
\usepackage{arydshln,xfrac}
\usepackage{csquotes,etoc}
\allowdisplaybreaks
\usepackage{tikz,pgfplots}
\usepackage{adjustbox}
\definecolor{ao(english)}{rgb}{0.0, 0.5, 0.0}
\definecolor{cadmiumgreen}{rgb}{0.0, 0.42, 0.24}
\definecolor{darkpastelgreen}{rgb}{0.01, 0.75, 0.24}
\definecolor{rblue}{RGB}{156,166,192}
\usetikzlibrary{shadings}
\usetikzlibrary{arrows.meta}
\usetikzlibrary{positioning}
\usetikzlibrary{backgrounds}
\usepgfplotslibrary{patchplots}
\usepgfplotslibrary{fillbetween}
\pgfplotsset{layers/standard/.define layer set={background,axis background,axis grid,axis ticks,axis lines,axis tick labels,pre main,main,axis descriptions,axis foreground}{
            grid style={/pgfplots/on layer=axis grid},tick style={/pgfplots/on layer=axis ticks},axis line style={/pgfplots/on layer=axis lines},label style={/pgfplots/on layer=axis descriptions},legend style={/pgfplots/on layer=axis descriptions},title style={/pgfplots/on layer=axis descriptions},colorbar style={/pgfplots/on layer=axis descriptions},ticklabel style={/pgfplots/on layer=axis tick labels},axis background@ style={/pgfplots/on layer=axis background},3d box foreground style={/pgfplots/on layer=axis foreground},},
}

\usepackage{tcolorbox}
\usepackage{pgfplots}
\usepackage{cleveref}
\tcbuselibrary{listings,theorems,breakable,skins}

\newtheorem{theorem}{Theorem}[section]
\newtheorem{lemma}{Lemma}[section]
\newtheorem{corollary}{Corollary}[section]
\newtheorem{definition}{Definition}
\newtheorem{assumption}{Assumption}
\newtheorem{remark}{Remark}
\newtheorem{example}{Example}

\newtheorem{fact}{Fact}

\crefname{observation}{Observation}{Observations}
\crefname{theorem}{Theorem}{Theorems}
\crefname{lemma}{Lemma}{Lemmas}
\crefname{definition}{Definition}{Definitions}
\crefname{assumption}{Assumption}{Assumptions}
\crefname{corollary}{Corollary}{Corollaries}
\crefname{fact}{Fact}{Facts}
\crefname{remark}{Remark}{Remarks}
\crefname{example}{Example}{Examples}
\crefname{section}{Section}{Sections}
\crefname{subsection}{Section}{Sections}
\crefname{subsubsection}{Section}{Sections}
\crefname{equation}{Eq.}{Eqs.}

\setlist[itemize]{leftmargin=*,label=$\bullet$,topsep=2pt,itemsep=0pt}
\setlist[enumerate]{label=$(\arabic*)$, leftmargin=*,topsep=2pt,itemsep=0pt}

\usepackage{titlesec}
\titleformat{\subsubsection}[runin]{\normalfont\normalsize\itshape}{\thesubsubsection}{1em}{}

\RequirePackage{tgtermes}
\RequirePackage{newtxtext}
\usepackage[scaled=1.05,vvarbb,smallerops]{newtxmath}
\RequirePackage{bm}

\DeclareMathSymbol{\shortminus}{\mathbin}{AMSa}{"39}
\newcommand{\st}{\mathrm{s.t.}~}
\newcommand{\diag}{\mathrm{diag}}

\newcommand{\real}{\mathbb R}
\newcommand{\preal}{\mathbb R_{\scriptscriptstyle >0}}

\newcommand{\intp}{\mathrm{int}}

\newcommand{\conv}{\mathrm{conv}}

\newcommand{\trace}{\mathrm{tr}}

\newcommand{\ex}{\vmathbb{E}}

\newcommand{\qlq}{~\Rightarrow~}
\newcommand{\qaq}{~\quad\text{and}\quad~}

\newcommand{\softmax}[1]{\mathrm{softmax}(#1)}

\newcommand{\poly}{\mathrm{poly}}
\newcommand{\TV}{\mathrm{TV}}
\newcommand{\diff}{\mathrm{d}}

\newcommand{\1}{\mathbf{1}}
\newcommand{\zero}{\mathbf{0}}

\newcommand{\inner}[2]{\langle #1,#2\rangle}

\definecolor{rosemary}{RGB}{190,88,69}
\definecolor{stone}{RGB}{42,121,10}
\definecolor{deepblack}{RGB}{90,90,90}

\newcommand{\emphr}[1]{{\bf\color{rosemary}#1}}
\hypersetup{
    colorlinks,
    linkcolor={deepblack},
    citecolor={rosemary},
    urlcolor={rosemary}
}

\makeatletter
\def\@algocf@capt@plain{top}
\renewcommand{\algocf@makecaption@plain}[2]{\algocf@makecaption@ruled{#1}{#2}}
\makeatother
\definecolor{algobg}{RGB}{244,245,248}
\newenvironment{myalgo}[2][]{\begin{figure}[!htbp]\centering
        \begin{tcolorbox}[colback=algobg,colframe=algobg,boxrule=0pt,arc=1mm,left=4pt,right=4pt,top=3pt,bottom=3pt]\RestyleAlgo{plain}
            \begin{algorithm}[H]\caption{#2}\ifx\\#1\\\else\label{#1}\fi }{\end{algorithm}\end{tcolorbox}\end{figure}}
\makeatletter
\newif\ifeqalignlabel
\newenvironment{eqalign}[1]{\setlength{\abovedisplayskip}{5pt}\setlength{\belowdisplayskip}{3pt}\def\temp{#1}\def\tempempty{}\ifx\temp\tempempty
        \eqalignlabelfalse
        \begin{equation*}\aligned[b]\else
            \eqalignlabeltrue
            \equation\label{#1}\aligned[b]\fi
            }{\endaligned
            \ifeqalignlabel
            \endequation \else
        \end{equation*}\fi
    \ignorespacesafterend
}
\makeatother

\newenvironment{subalign}[1]
{\setlength{\abovedisplayskip}{5pt}\setlength{\belowdisplayskip}{3pt}\subequations\label{#1}\align}
{\endalign\endsubequations}
              
\definecolor{deepblue}{RGB}{0,70,170}

\makeatletter
\newcommand{\myparagraph}{\@startsection{paragraph}{4}{\z@}{0.6ex \@plus 0.2ex \@minus 0.1ex}{-0.5em}{\normalfont\normalsize\bfseries}}
\makeatother

\newenvironment{myproof}[1][Proof]{\par\noindent\textit{#1.}~\ignorespaces }{\hfill$\blacksquare$\par}

\let\oldequation\equation
\let\endoldequation\endequation
\renewenvironment{equation}
{\setlength{\abovedisplayskip}{5pt}\setlength{\belowdisplayskip}{3pt}\oldequation}
{\endoldequation}

\everydisplay{\setlength{\abovedisplayskip}{5pt}\setlength{\belowdisplayskip}{3pt}}

\makeatletter
\newcommand\NoIndentNext{\@afterindentfalse\@afterheading}
\makeatother
\newcommand{\NoIndentAfter}[1]{\AfterEndEnvironment{#1}{\NoIndentNext}}
\forcsvlist{\NoIndentAfter}{theorem,fact,myproof,lemma,proposition,corollary,definition,remark,proof,condition,assumption,example}
\allowdisplaybreaks

\makeatletter
\def\thm@space@setup{\thm@preskip=0.3\baselineskip
    \thm@postskip=\thm@preskip
}
\makeatother

\titlespacing*{\section}      {0pt}{1.6ex plus .2ex minus .2ex}{1.0ex plus .1ex}
\titlespacing*{\subsection}   {0pt}{1.4ex plus .2ex minus .2ex}{0.8ex plus .1ex}
\titlespacing*{\subsubsection}{0pt}{1.2ex plus .2ex minus .2ex}{1em}

\def\a{\mathbf{a}}

\def\b{\mathbf{b}}

\def\c{\mathbf{c}}

\def\d{\mathbf{d}}

\def\e{\mathbf{e}}

\def\f{\mathbf{f}}

\def\g{\mathbf{g}}

\def\h{\mathbf{h}}

\def\p{\mathbf{p}}

\def\q{\mathbf{q}}

\def\s{\mathbf{s}}

\def\t{\mathbf{t}}

\def\u{\mathbf{u}}

\def\v{\mathbf{v}}

\def\w{\mathbf{w}}

\def\x{\mathbf{x}}

\def\y{\mathbf{y}}

\def\z{\mathbf{z}}

\def\bA{\mathbf{A}}

\def\rB{\mathscr{B}}

\def\cF{\mathcal{F}}

\def\cG{\mathcal{G}}

\def\rH{\mathscr{H'}}
\def\cI{\mathcal{I}}

\def\vbI{\vmathbb{I}}

\def\bI{\mathbf{I}}

\def\cJ{\mathcal{J}}

\def\bJ{\mathbf{J}}

\def\cK{\mathcal{K}}

\def\cL{\mathcal{L}}

\def\bM{\mathbf{M}}

\def\bbN{\mathbb{N}}

\def\bN{\mathbf{N}}

\def\cO{\mathcal{O}}

\def\cP{\mathcal{P}}

\def\vbP{\vmathbb{P}}

\def\cQ{\mathcal{Q}}

\def\bQ{\mathbf{Q}}

\def\cR{\mathcal{R}}

\def\cS{\mathcal{S}}

\def\bU{\mathbf{U}}
\def\rU{\mathscr{U'}}
\def\cV{\mathcal{V}}

\def\cW{\mathcal{W}}

\def\rW{\mathscr{W'}}
\def\cX{\mathcal{X}}

\def\rX{\mathscr{X'}}

\def\bbZ{\mathbb{Z}}

\def\bgamma{\boldsymbol{\gamma}}

\def\bzeta{\boldsymbol{\zeta}}
\def\bTheta{\boldsymbol{\Theta}}
\def\btheta{\boldsymbol{\theta}}

\def\bmu{\boldsymbol{\mu}}

\def\bnu{\boldsymbol{\nu}}
\def\bXi{\boldsymbol{\Xi}}
\def\bxi{\boldsymbol{\xi}}

\def\bomega{\boldsymbol{\omega}}

\makeatother

\newcommand{\cg}{\texttt{CG}}

\newcommand{\fw}{\texttt{FW}}

\newcommand{\adcg}{\texttt{CG(AD)}}
\newcommand{\adfw}{\texttt{FW(AD)}}

\newcommand{\classces}{\rH^{\mathrm{CES}}}
\newcommand{\classlin}{\rH^{\mathrm{Lin}}}
\newcommand{\classleon}{\rH^{\mathrm{Leon}}}
\newcommand{\classcobbs}{\rH^{\mathrm{Cobbs}}}

\title{Rationalizing collective revealed preferences with an application in fair resource allocation}

\usepackage[auth-sc]{authblk}

\author[1]{Chuwen Zhang\footnote{Corresponding author. Email: \url{chuwen.zhang@chicagobooth.edu}}}
\author[2]{Zhiyun Guo}
\author[3]{Zizhuo Wang}
\author[2,4]{Yinyu Ye}

\affil[1]{Booth School of Business, University of Chicago, Chicago, IL 60637, USA}
\affil[2]{Antai College of Economics and Management, Shanghai Jiao Tong University, Shanghai 200240, China}
\affil[3]{School of Data Science, School of Management and
Economics, The Chinese University of Hong Kong, Shenzhen 518172, China.}
\affil[4]{Department of Management Science and Engineering, Stanford University, Stanford, CA 94305, USA}

\begin{document}

\maketitle

\begin{abstract}
    This paper presents a revealed preference approach for rationalizing collective consumption behavior.
We introduce the Constructive Rationalization Method (CRM), which approximates the real market via a surrogate market of artificial consumers, called androids, with easy-to-compute demand functions. CRM uses observed aggregate demand and adds artificial consumers on the fly, while redistributing wealth under an empirical risk minimization principle.
Unlike classical revealed preference approaches, CRM provides guarantees on the generalization risk for learning the aggregate demand function, while respecting the privacy of the underlying consumers in the real market. As an application, CRM can be used to provide reliable predictions for collective consumption behavior. Specifically, we show how to apply CRM to approximate allocations that are proportionally fair without requiring the knowledge of individual utilities. \end{abstract}

\etocdepthtag.toc{mainmatter}
\section{Introduction}

Demand analysis lies at the heart of decision-making problems in operations research.
In revenue management, we are interested in deciding the prices of products to offer customers with a view to maximizing expected revenues \citep{talluriRevenueManagementGeneral2004}.
In resource allocation, we design mechanisms that allocate scarce goods so that what each agent receives matches their preferences \citep{varianEquityEnvyEfficiency1974,kellyChargingRateControl1997}.
These downstream decisions rest on learning the demand and its sensitivity to control variables such as prices.

Broadly speaking, demand is only a revelation of economic primitives like individuals' preferences or utilities.
For example, consider the canonical problem with a set of goods $[n] := \{1, \ldots, n\}$, and a population of agents $[m] := \{1, \ldots, m\}$ as consumers.
If we assume the rationality of an agent $i$, then its demand $\x_i$ should be consistent with the behavior that optimizes its preference relations $\succeq_i$ over the goods.  The aggregate demand $\d$ summarizes the demands of all agents.

While individual-level preference elicitation is appealing, it is sometimes impractical and ethically sensitive in large-scale economic systems. The number of individuals may be enormous, and collecting or analyzing individual preference data can raise concerns related to privacy and regulation \citep{euRegulationEU20162016}. In many applications,
the available data consist only of a finite collection of aggregate demand observations, with the identities of the contributing individuals anonymized or unobserved.
This forms the key question addressed in this paper: \textbf{How to efficiently support decision-making processes based only on collective and observable  consumption behavior?}

\subsection{Motivation and our approach}

We focus on the demand of consumers with Walrasian competitive budgets.
Assume that the goods are divisible, and each of them is associated with a price $p_j \in \real_+$ and a limited amount of supply.
For each agent $i$, its utility $u_i \in \rU$ is a function that encodes the preferences $\succeq_i$ and depends on the current market price $\p$ and its wealth function $w_i \in \rW$, which as well depends on $\p$.
The demand $\x_i \in \rX$ is induced by the Utility Maximization Problem\footnote{Strictly speaking, the existence of a utility function requires regularity conditions on the agent's preference relation $\succeq$.},
\begin{equation}\label{eq.def.ump}
    \tag{UMP}
    \x_i(\p, w_i(\p)) = \arg\max_{\x \in \real_+^n} ~u_i(\x), ~ \st \inner{\p}{\x} \le w_i(\p).
\end{equation}
Thus, the aggregate demand function of the market is
\begin{eqalign}{eq.def.aggregate.demand}
    \d(\p) = \smallsum_{i=1}^m \x_i(\p, w_i(\p)).
\end{eqalign}

One natural approach to analyzing $\d$ is to combine the inferences for each agent by rationalizing individual demand. This task has been a major topic in revealed preference theory \citep{samuelsonConsumptionTheoryTerms1948,chambersRevealedPreferenceTheory2016}.
Specifically, consider the case where for each agent $i$, $K$ observed price-demand pairs $\bXi_i = \{(\p_k, \x_{ik})\}_{k\in[K]}$ are provided. We hope to find such a $u_i: \real_+^n \mapsto \real_+$ that the observed demand $\x_{ik}$ maximizes the corresponding \ref{eq.def.ump} for all $k \in [K]$. This turns out to be possible only if the data $\bXi_i$ satisfies the Generalized Axiom of Revealed Preference (GARP), i.e., there are no preference cycles. A test can be performed by the Afriat-Varian inequalities \citep{afriatConstructionUtilityFunctions1967,varianNonparametricApproachDemand1982}:
\begin{eqalign}{eq.afriat.garp.pairs}
    u_{ik'}\le u_{ik} + \lambda_{ik} \inner{\p_k}{\x_{ik'}-\x_{ik}}, &\quad \forall k,k' \in [K], \\
    (u_{ik}, \lambda_{ik}) \in \real \times \real_+, &\quad\forall k \in [K].
\end{eqalign}
The infeasibility of \eqref{eq.afriat.garp.pairs} denies the existence of a utility function; otherwise, any feasible solution implies that GARP holds, and the data is rationalized by the following concave function:
\begin{eqalign}{eq.afriat.utility}
    u_i(\x)=\min_{k \in [K]}\big\{u_{ik} + \lambda_{ik} \inner{\p_k}{\x-\x_{ik}}\big\},
\end{eqalign}
which is Piecewise-Linear Concave (PLC) and monotone. To query the demand of a PLC agent like this, one needs to solve a linear programming problem. It is possible to use parametric models, but not surprisingly, preferences recovered therein are usually misspecified \citep{varianGoodnessoffitOptimizingModels1990,halevyParametricRecoverabilityPreferences2018}.

Leaving aside the scalability and privacy concerns in modern economic systems, the value of individual rationality seems quite limited.
For example, we are often interested in finding a Proportionally Fair Allocation (PFA) of goods to each of the $m$ agents.
Specifically, an allocation $(\x_1, ..., \x_m) \in \cX$ is said to be  proportionally fair if for all other feasible allocations $(\y_1, ..., \y_m) \in \cX$, the following condition holds,
{\small
        \begin{equation}\label{eq.def.prop.fair}
            \tag{PFA}
            \smallsum_{i\in[m]} w_i\frac{u_i(\y_i)-u_i(\x_i)}{u_i(\x_i)} \leq 0,~~
            \cX = \left\{(\x_1, ..., \x_m) \in \real_+^{n \times m} \mid \smallsum_{i\in[m]} \x_i \le \1\right\},
        \end{equation}}where the weights $w_i$ can be interpreted as the (constant) wealth of the agent $i$ \citep{varianEquityEnvyEfficiency1974,kellyChargingRateControl1997,bertsimasPriceFairness2011}.
\ref{eq.def.prop.fair} is a normative notion desirable for many applications, such as pricing and allocation in communication networks, spectrum management, online advertising, kidney transplantation, and electricity markets; see, e.g., \citet{srikantMathematicsInternetCongestion2004,luoDynamicSpectrumManagement2008,bertsimasFairnessEfficiencyFlexibility2013,azizanruhiOpportunitiesPriceManipulation2018,bateniFairResourceAllocation2022}.
Computationally, a \ref{eq.def.prop.fair} can be obtained by maximizing the weighted Nash Social Welfare function \citep{eisenbergConsensusSubjectiveProbabilities1959} over the set of feasible allocations $\cX$:
\begin{equation}\label{eq.def.welfare}
    \tag{NSW}
    \smallprod_{i\in[m]} u_i(\x_i)^{w_i},
\end{equation}
which is a convex optimization problem. A dilemma arises if the Afriat–Varian rationalization \eqref{eq.afriat.utility} is used for this purpose. On one hand, achieving an accurate utility representation requires a sufficiently large number of observations. On the other hand, incorporating more observations increases the size of the resulting NSW maximization. Moreover, Afriat–Varian rationalization only interpolates the observed data -- the required sample size may be unbounded.
An alternative route is to first find a Walrasian Equilibrium (WE) price $\p$, at which the aggregate demand $\d$ clears the supply of the market:
\begin{equation}\label{eq.def.we}
    \tag{WE}
    \real_+^n \ni \p \perp \big(\1 - \d(\p) \big) \in \real_+^n,
    \qquad \text{where } \d \text{ is defined in Eq.}~\eqref{eq.def.aggregate.demand}.
\end{equation}
Then, the allocation is obtained by assigning each agent its demand at the equilibrium price. Unlike maximizing \ref{eq.def.welfare} as a centralized planner with full knowledge of $\{(u_i,w_i)\}_{i\in[m]}$, a WE can in principle be reached through a decentralized price-adjustment process that relies only on demand information. However, computing a WE is generally more difficult. For example, with constant wealth and PLC utilities here, finding an approximate \ref{eq.def.we} is PPAD-hard \citep{chenSpendingNotEasier2009}. To keep this route tractable, we often restrict $\{u_i\}_{i\in[m]}$ to simple parametric utilities \citep{gaoOnlineMarketEquilibrium2021,bateniFairResourceAllocation2022,jalotaStochasticOnlineFisher2025}.

In view of the above obstacles, we consider utilizing the aggregate demand $\d$ holistically. Given the access to historical prices and the corresponding aggregate demands $\{(\p_k, \d(\p_k))\}_{k=1}^K$, we present an approach that constructively rationalizes this collective information.
Without identifying the \emph{real} agents in the market, we build a surrogate market to imitate the collective behavior by iteratively adding \emph{easy} artificial consumers, called \emph{androids}, and redistributing the wealth among them to produce a proper approximation.
While the real market might possess complex demand functions and nonlinear wealth, each android has an easy-to-compute demand function and linear wealth.
To ensure the quality of the approximation, androids are added on the fly until no further improvement is possible.

\subsection{Contributions}
We implement the above idea and make the following contributions.
We propose the Constructive Rationalization Method (CRM, \Cref{alg.cg}) with a surrogate market to approximate the aggregate demand $\d$ of the real market under an empirical risk minimization (ERM) principle.
Conceptually, because we use only collective information (even without knowledge of $\rU,\rX,$ and $\rW$), CRM respects agents' privacy while bypassing the drawbacks of individual rationality.

In comparison to the existing studies on rationality \citep{afriatConstructionUtilityFunctions1967,varianNonparametricApproachDemand1982} and learning of individual preferences \citep{beigmanLearningRevealedPreference2006,balcanLearningEconomicParameters2014,chambersRecoveringPreferencesFinite2021}, the method here is a revealed preference approach for collective behavior, without requiring prior information on individual primitives.
The main strength of CRM lies in its guarantees for generalization risk that does not deteriorate as the size of the surrogate market grows (\Cref{thm.gen_bound}).
Accordingly, performance guarantees for sample complexity (the required number of samples) and convergence rate (the required number of androids) are provided. We conduct numerical experiments to validate the prediction performance of CRM. Then, we demonstrate how CRM can be used to approximate \ref{eq.def.prop.fair} when full knowledge of agents' information is not available and when the number of agents is large.

Technically, CRM allows flexibility in designating the universe of androids and in the scheme for allocating wealth based on convex optimization. The overall principle is that the android should be tractable for computing and learning (e.g., bounded Rademacher complexity).
While CRM can accommodate non-homothetic androids and nonlinear wealth functions, we show that homothetic androids with linear wealth already suffice to approximate aggregate demand arbitrarily well (\Cref{thm.misspecification.linear}). We further establish a tight minimax-type lower bound on the approximation achievable by restricted classes of simple androids, including those with linear utilities (\Cref{thm.counterexample.informal}).

The main computational overhead of CRM lies at the ERM stage, where one seeks a new android that most improves the current surrogate market. We show that, for several prevalent utility classes, including linear, Leontief, and CES utilities, finding a globally optimal android is generally NP-hard and, in some cases, strongly NP-hard. Fortunately, the quality of the selected android is largely decoupled from the convergence rate; in practice, a reasonably good android suffices to improve the approximation.
Interestingly enough, in some classes there exists an FPTAS for a good-enough android or even a globally optimal android (cf. \Cref{tab.sep.complexity})

More broadly, our work is related to a long-standing theme in economics: what can be inferred from collective consumption behavior?
Classical results show that aggregate demand inherits remarkably little structure from individual rationality \citep{sonnenscheinWalrasIdentityContinuity1973,mantelCharacterizationAggregateExcess1974,debreuExcessDemandFunctions1974}, while the decomposition of collective behavior into individual preferences becomes difficult when wealth effects are unknown \citep{browningEfficientIntraHouseholdAllocations1998,chiapporiAggregationMarketDemand1999,cherchyeRevealedPreferenceApproach2011}. The constructive rationalization approach in this paper contributes an algorithmic tool for studying this fundamental topic.
 \section{Related literature}\label{sec.literature}
We review the most relevant research related to the current study.

\subsection{Rationalizability and learnability of individual preferences}
Afriat-Varian rationalization has been extensively studied in economics; see, e.g., \cite{chiapporiRevealedPreferencesDifferentiable1987,blundellNonparametricEngelCurves2003,blundellBestNonparametricBounds2008,blundellSharpSARPNonparametric2015} and the monograph \citet[Chapter 3-5]{chambersRevealedPreferenceTheory2016}.
It is also possible to quantify the extent of ``irrationality'' of the preferences by the inconsistency index \citep{varianGoodnessoffitOptimizingModels1990,echeniqueMoneyPumpMeasure2011}, but most of these indices cannot be computed very efficiently \citep{smeuldersGoodnessofFitMeasuresRevealed2014,cherchyeAreConsumersApproximately2025}.
To remedy nonsmoothness in Afriat-Varian rationalization, \citet{chiapporiRevealedPreferencesDifferentiable1987} show that if the same bundle cannot be induced by different prices, then the data can be rationalized by a smooth, infinitely differentiable, strictly concave utility function.
\citet{blundellNonparametricEngelCurves2003,blundellBestNonparametricBounds2008,blundellSharpSARPNonparametric2015} extends the classical revealed preference approach when the budget is also chosen by the agent, accommodating the effect of wealth changes.

The main drawback of the above methods is that they cannot provide guarantees for out-of-sample prediction.
It is also unlikely to be possible universally, since \citet{beigmanLearningRevealedPreference2006} showed that general demand is not always learnable under the Probably Approximately Correct (PAC) framework\footnote{In the same paper, learnable cases could be derived; unfortunately, the PLC utilities cannot be covered. PLC utilities can admit set-valued demand, clearly violating the condition in \citet{beigmanLearningRevealedPreference2006}.}.
Learnability is generally a feature that only holds for ``easy'' utilities.
For example, \citet{zadimoghaddamEfficientlyLearningRevealed2012} established polynomial bounds on the sample complexity if the utilities are restricted to be linear and separable concave with Lipschitz gradients.
\citet{balcanLearningEconomicParameters2014} provided sample complexity guarantees for a number of important classes, including linear, Leontief, and CES utility functions.
A recent line of economic literature has focused on the statistical learnability of general preferences and choices under uncertainty.
For example, \citet{basuFalsifiabilityLearnabilityDecision2020,chambersRecoveringPreferencesFinite2021} discussed learnable models with finite Vapnik-Chervonenkis (VC) dimensions, in which the choice in the Walrasian setting is one specific application.

\subsection{Rationalization and learning of collective consumption}
The problem studied in this paper is related to a question asked by
\citet{sonnenscheinUtilityHypothesisMarket1973,sonnenscheinWalrasIdentityContinuity1973}:
{What restrictions are necessary for a function to be the market excess demand function?}
This question was answered by
\citet{sonnenscheinWalrasIdentityContinuity1973,mantelCharacterizationAggregateExcess1974,debreuExcessDemandFunctions1974};
also see \citet{geanakoplosDisaggregationExcessDemand1980,shaferChapter14Market1982}.
The Sonnenschein-Mantel-Debreu (SMD) theorem shows that market excess demand can be very general: essentially, any continuous function satisfying homogeneity of degree zero and Walras' law can be the excess demand function of an Arrow-Debreu economy.
This generality has two closely related propositions.
First, there is little structure in the equilibrium set as well.
Indeed, any compact subset of the interior of the unit sphere can arise as the equilibrium set of a well-behaved exchange economy
\citep{mas-colellRecoverabilityConsumersPreferences1977}.
Second, SMD-type results imply that the convergence of price-adjustment processes (the formal definition is given at \eqref{eq.tatonnement} shortly below) cannot be guaranteed universally because the excess demand function can almost be arbitrary, while its convergence requires certain regularity conditions \citep{arrowStabilityCompetitiveEquilibrium1958}.

Characterization of market demand becomes more difficult when wealth functions are fixed \citep{shaferChapter14Market1982}.
In this setting, the decomposition of aggregate excess demand into a set of individual demand functions is only possible locally \citep{chiapporiAggregationMarketDemand1999} around a given price, but not globally
\citep{mcfaddenCharacterizationCommunityExcess1974,shaferChapter14Market1982}.
At the same time, because fixed wealth functions limit the arbitrariness allowed by SMD, the literature has obtained more positive results concerning the stability of the price-adjustment processes.

Several related decomposition problems have also been studied.
One line of work generalizes the SMD-type results to the incomplete markets, which is hard due to the Grassmannian from limited market span
\citep{radnerExistenceEquilibriumPlans1972,brownComputingEquilibriaWhen1996,chiapporiDisaggregationExcessDemand1999}.
Another line of literature studies collective consumption lying between purely individual and many-individual market behavior, such as households and firms, with observations of private and public consumption and Lindahl prices
\citep{browningEfficientIntraHouseholdAllocations1998,cherchyeNonparametricTestsCollectively2008,cherchyeRevealedPreferenceApproach2011}.
For these models, mixed-integer optimization approaches have been developed in a spirit analogous to Afriat-Varian rationalization.

Finally, our paper is also related to the literature on learning market primitives.
Since learning even a single agent's preferences is already difficult, the literature on learning collective consumption is quite limited.
To our knowledge, the closest work is \citet{beiLearningMarketParameters2016}, which studies an active-learning setting in which the learner can freely set prices and is therefore not a statistical learning problem.

\subsection{Fair resource allocation and market equilibrium}

Proportionally fair resource allocation (\ref{eq.def.prop.fair}) and Walrasian market equilibrium (\ref{eq.def.we}) are closely related concepts in economics and computer science \citep{mas-colellMicroeconomicTheory1995,nisanAlgorithmicGameTheory2007}.
While finding a fair allocation is generally solvable by convex optimization methods, finding a \ref{eq.def.we} is a fixed-point problem, for which not many cases can be computed efficiently.
The two concepts coincide when, for example, the wealth is \emph{fixed} and each agent's utility is homothetic.
In this case, a \ref{eq.def.prop.fair} can be found in polynomial time by maximizing the \ref{eq.def.welfare} \citep{eisenbergConsensusSubjectiveProbabilities1959}, and the dual variable to the resource constraint in $\cX$ is an equilibrium price \citep{devanurMarketEquilibriumPrimaldualtype2002,yePathArrowDebreu2008}.
Again, these methods require complete knowledge on the utilities and wealth, which are often difficult to obtain. For the same reason, they do not scale well with the number of agents.

A compelling rationale for pursuing \ref{eq.def.we} is that it allows us to harness the ``invisible hand'' without requiring complete information about individual agents. One may employ a mechanism that adjusts prices so that the demand matches the supply. The individual demand realized at the resulting equilibrium prices can then serve as a proxy for \ref{eq.def.prop.fair}.
Remarkably, even when utilities are not homothetic, every allocation induced by \ref{eq.def.we} achieves at least a 0.69 fraction of the maximum Nash welfare \citep{gargApproximatingCompetitiveEquilibrium2026}. Thus, this approach is not merely a theoretical aspiration but can be justified in a broad range of practical settings. As early as the nineteenth century, \citet{walrasElementsDEconomiePolitique1874} introduced the first such mechanism, the t\^atonnement process, which works as follows:
\begin{eqalign}{eq.tatonnement}
    \p \leftarrow \p + \cG \circ \left(\smallsum_{i\in[m]} \x_i(\p, w_i(\p)) - \1\right),
\end{eqalign}
where $\cG: \real^n \mapsto \real^n$ is some sign-preserving mapping.
The moral of this mechanism is very simple: it adjusts the price of good $j$ upward when demand exceeds supply and downward otherwise.
The convergence of the t\^atonnement and its variants to \ref{eq.def.we} was shown by \citet{arrowStabilityCompetitiveEquilibrium1958,smaleConvergentProcessPrice1976}. In fact, the usage of the t\^atonnement is not restricted to homothetic agents and fixed budgets. Worst-case complexity results have been established for linear wealth \citep{jainPolynomialTimeAlgorithm2007,yePathArrowDebreu2008,gargStronglyPolynomialAlgorithm2023}, and for utilities satisfying Weak Gross Substitutability (WGS) \citep{codenottiComputationMarketEquilibria2007} and bounded elasticity \citep{goktasTatonnementHomotheticFisher2023}, and for markets admitting a convex potential function \citep{jainEisenbergGaleMarkets2010,cheungTatonnementGrossSubstitutes2020}, and so on.

While price-adjustment processes do not require full preference revelation, they typically rely on prior structural assumptions on $(\rX,\rW,\rU)$ in order to guarantee convergence. For example, a mechanism may assume that agents' utilities satisfy WGS while remaining agnostic about specific parameterizations.
In light of SMD, however, such assumptions substantially restrict the class of aggregate demand functions that can arise in a market. Consequently, convergence guarantees derived under these assumptions may not extend to markets encountered in practice.

\section{Preliminaries}

In this section, we formally define the problem setting and discuss a few elements for later discussion.
\myparagraph{Notations.}
We use the following notations. The notation $\real,\bbZ,\bbN$ denotes the set of real, integer, and natural numbers; $\real_+^n$ and $\preal^n$ denote the sets of $n$-dimensional vectors with all nonnegative elements, and positive elements, respectively. For a subset $\cS \in \real^n$, we use $\intp(\cS)$ to denote the (relative) interior of $\cS$.
For a natural number $n\in \bbN$, we let $[n] = \{1, ..., n\}$.
We use $[a; b]$ (resp., $[a, b]$) to denote vertical (resp., horizontal) concatenation of arrays or numbers. For a subset $B \subseteq [n]$, we use $\x_{B}$ (resp., $\bA_{BB}$) to denote the subvector (resp., submatrix) of $\x$ (resp., $\bA$) obtained by keeping only the elements (resp., rows and columns) indexed by $B$.
We sometimes omit the brackets when there is no confusion. The symbol $\Delta_n$ means the $n$-dimensional simplex: $\Delta_n = \{\x:\inner{\x}{\1}=1,\x\in\real_+^n\}$. We use $\|\cdot\|$ to denote a proper semi-norm, and $\|\cdot\|_*$ to denote the dual norm; by default, they indicate the induced $\ell_2$ norm.

\subsection{Problem setting}

\myparagraph{Real market.}
The real market has $j=1,...,n$ divisible goods and $i=1,...,m$ agents.   Each good $j$ is associated with one unit of supply and a price $p_j$; the price lies in a closed convex cone, $\p \in \cP \subseteq \real_+^n.$
Each agent $i$ has a wealth function $w_i \in \rW: \cP \mapsto \real_+$, and a utility function $u_i \in \rU: \real_+^n \mapsto \real$.
Each $u_i$ is locally non-satiated and concave, and the wealth $w_i$ is bounded.
The total wealth in the market is $W(\p) = \sum_{i=1}^m w_i(\p)$.
As mentioned, the demand is $\x_i \in \rX$ for each agent $i$, and the aggregate demand function is given by:
\begin{eqalign}{}
    \d(\p) = \sum_{i=1}^m \x_i(\p, w_i(\p)), \qaq  \x_i(\p, w_i(\p)) = \arg \max_{\x \in \real_+^n} u_i(\x) ~~\st~ \inner{\p}{\x} \le w_i(\p).
\end{eqalign}
A deterministic tie-breaking rule is applied whenever the demand is set-valued.
Also, introduce the market excess demand function:
\begin{eqalign}{eq.def.z}
    \z(\p) = \d(\p) - \1.
\end{eqalign}
Now we define the inputs of our problem.

\myparagraph{Revealed preferences.}
We assume no a priori knowledge about the underlying agents; instead, the input of our problem is the sequence of price-demand pairs.
Let the distribution of price be $\tau$.
Given i.i.d. samples $\p_k\sim \tau, k\in[K]$,
the revealed preferences are presented by $\bXi = \{(\p_k, \d_k)\}_{k=1}^K$ with $\d_k \equiv \d(\p_k)$.
For simplicity, we assume $\p_k \in \Delta_n$ and $W(\p_k) = 1$ in the real market, which will be relaxed later.

\myparagraph{Surrogate market.}
To model the collective behavior of the real market, we introduce a surrogate market with $T$ artificial consumers, whose wealth and preferences are determined by our construction.
In the rest of this paper, we preserve the index $i$ and the term ``agent'' for a consumer in the real market, while $t$ and ``android'' for an artificial consumer.

\subsection{Expenditure shares}\label{sec.basic.ces}

For either an android or a real agent (suppressing the subscript for notational convenience), its demand function $\x(\p, w(\p))$ solves its own \ref{eq.def.ump}.
Instead of using the demand function explicitly, we define the expenditure share mapping as follows:
\begin{eqalign}{eq.def.gamma}
    \bgamma(\p, w(\p)) = \frac{\diag(\p)\x(\p, w(\p))}{w(\p)} \in \Delta_n,
\end{eqalign}
indicating the fraction of total wealth spent on each good. One reason to do so is that the demand functions can be unbounded and non-Lipschitz near the boundary (e.g., zero prices) under Euclidean geometry \citep{cheungDynamicsDistributedUpdating2018,zhangSecondorderTatonnementDecentralized2025}.
In the surrogate market, we limit an android's expenditure share from a prescribed universe $\rH$, the union of finitely many component classes:
\begin{eqalign}{eq.def.rh}
    \rH = \bigcup_{f \in [\bar f]} \rH^f, \quad \rH^f \subseteq \{ \bgamma: \Delta_n \times \real_+ \to \Delta_n \}, ~ \forall f \in [\bar f].
\end{eqalign}
For example, we may consider expenditure shares below.
\begin{example}[Additive homothetic classes]\label{example.additive.homothetic}
    Consider the utility function parameterized by $\c \in \real_+^n$ and $r \in [-\infty, 1]$: $u(\x) = \inner{\c}{\x^r}^{1/r}$.
    Depending on $r$, we have the following component classes: the linear class $\classlin$ for $r=1$, the Leontief case $\classleon$ for $r=-\infty$, and the CES class $\classces$ for $r \in (-\infty, 1)$, including the Cobb-Douglas class $\classcobbs$ if $r = 0$.
    By letting $\sigma = \tfrac{r}{1-r}$, these cases can be compactly represented as:
    \begin{eqalign}{}
        \classces &:= \left\{ \bgamma: \Delta_n \to \Delta_n \mid (\y, \sigma) \in \real^n \times (-1, \infty),~ \bgamma(\p) = \softmax{\y - \sigma \log(\p)} \right\}, \\
        \classcobbs &:= \left\{ \bgamma: \Delta_n \to \Delta_n \mid \y \in \real^n,~ \bgamma(\p) \equiv \bgamma = \softmax{\y} \right\}, \\
        \classleon &:= \left\{ \bgamma : \Delta_n \to \Delta_n ~\Big|~ \y \in \real^n, ~ \bgamma(\p) = \softmax{\y + \log(\p)} \right\}, \\
        \classlin &:= \left\{ \bgamma : \Delta_n \to \Delta_n ~\Big|~ \y \in \real^n, ~ \bgamma(\p) = \e_{j^\ast(\y, \p)}, ~ j^\ast(\y, \p) \in \arg\max_{j \in [n]} \{y_j - \log p_j\} \right\}.
    \end{eqalign}
\end{example}
We leave the derivation of expenditure shares (and for a few other classes) to \Cref{sec.proof.rh}.
In fact, we can simply write $\bgamma(\p)\equiv\bgamma(\p, w)$ independent of wealth $w(\p)$ if the preferences are homothetic (\Cref{lem.share.indep.iff.homothetic}); otherwise, the expenditure share remains to be a joint function of price and wealth.
In the real market, while the expenditure share $\bgamma_i$ of a real agent $i$ is not accessible, we can still acquire the market expenditure share:
\begin{eqalign}{eq.def.market.share}
    \g: \Delta_n \mapsto \Delta_n, \quad \g(\p) = \frac{\diag(\p)\d(\p)}{\sum_{i\in [m]} w_i(\p)} = \diag(\p)\d(\p) \in \Delta_n.
\end{eqalign}
The fact that $\g(\p) \in \Delta_n$ is apparent since the budget of each agent is exhausted by non-satiation, viz, $\inner{\1}{\g(\p)} = W(\p) = 1$. Thereby, the revealed preferences can be represented as,
\begin{eqalign}{}
    \bXi = \{\p_k, \g_k\}_{k=1}^K, \quad \g_k = \diag(\p_k)\d_k, ~\forall k\in[K].
\end{eqalign} \section{Constructive rationalization by the surrogate market}\label{sec.method}

With a collection of androids drawn from the universe $\rH$, we need a scheme to distribute the wealth among them.
To begin with, assume that an android's preferences are homothetic and the wealth function is exogenous to the price. In this case, the aggregate demand lies in the convex hull of the demand functions of the androids, so does the expenditure share. We propose extensions to accommodate non-homothetic androids and endogenous wealth functions.

\subsection{The wealth redistribution problem}\label{sec.method.constantw}
Let the wealth in the surrogate market be distributed according to a nonnegative Borel probability measure $\omega$ supported on $\rH$: $\omega \in \mathfrak M(\rH)$.
With homothetic androids, the expenditure share in the surrogate market can be written as a function $\h: \Delta_n \to \Delta_n$ in the convex hull of $\rH$:
$\h(\p) = \int_{\rH} \bgamma(\p) ~ \diff \omega.$
In short, $\h \in \conv(\rH)$. The goal is to minimize the mismatch between $\g(\cdot)$ for the real market and $\h(\cdot)$ for the surrogate market.
Given the real market shares $\g_k$, we write the {\bf W}ealth Redistribution in {\bf P}rimal Form (WP) as follows,
\begin{equation}\label{eq.cg.semi-infinite-primal}
    \tag{WP}
    \begin{aligned}[b]
        L^*
         & = \min_{\h \in \conv(\rH)}~ \frac{1}{K}\smallsum_{k\in [K]} \left\|~\g_k - \h(\p_k)\right\| = \min_{\omega \in \mathfrak M(\rH)}~ \frac{1}{K}\smallsum_{k\in [K]} \left\|~\g_k - \int_{\rH} \bgamma(\p_k) \diff \omega\right\|,
    \end{aligned}
\end{equation}
where $\|\cdot\|$ is any convex semi-norm on $\real^n$.
In view of the second equation, it is an infinite-dimensional convex optimization problem over the set of probability measures $\mathfrak M(\rH)$.
Here, the mismatch between the demands of two markets is not measured directly, but through a scaled local metric $\|\diag(\p_k)(\cdot)\|$, which is known to be the dual norm of the logarithmic barrier \citep{nesterovInteriorpointPolynomialAlgorithms1994}.

The dual problem, {\bf W}ealth Redistribution in {\bf D}ual Form, denoted WD, is the following:
\begin{equation}\label{eq.cg.semi-infinite-dual}
    \tag{WD}
    \begin{aligned}
        L^* = \max_{\bU, \mu}~ & \frac{1}{K}\left(\smallsum_{k\in [K]} \inner{\bU \e_k}{\g_k} - \mu\right), \quad\st~  (\bU, \mu) \in \cF(\rH),
    \end{aligned}
\end{equation}
where the dual feasible set, associated with the universe $\rH$, is given by,
{\small\begin{eqalign}{eq.dual.feasible.set}
            \cF(\rH) = \left\{(\bU, \mu) \in \real^{Kn}\times \real : \smallsum_{k\in [K]} \inner{\bU \e_k}{\bgamma(\p_k)} \le \mu, ~ \forall \bgamma \in \rH; ~ \|\bU \e_k\|_* \le 1, ~\forall k\in[K]\right\}.
        \end{eqalign}}Here, $\|\cdot\|_*$ denotes the dual norm, and $\e_k$ is the $k$-th canonical basis of $\real^K$. Any solution $(\bU, \mu)$ is invariant under scaling $(a \bU, a \mu), \forall a > 0$; the scale limits on $\bU$ only ensure boundedness. Indeed, $\cF$ is formed by the homogeneous linear inequalities, and thus $\cF(\rH)$ is a finite-dimensional convex set with a nonempty interior. Thus, strong duality holds, so the primal-dual pair shares the optimal value $L^*$ (cf. \Cref{lem.strong_duality}).

\subsection{The conceptual algorithm}

We are now ready to present the conceptual method (\Cref{alg.cg}).
With $T$ androids, the measure to distribute the wealth becomes a $T$-dimensional vector $\w \in \Delta_T$, and so $\h(\p) = \smallsum_{t=1}^T w_t \bgamma_t(\p, w_t).$
\begin{myalgo}[alg.cg]{{\bf CRM}, Constructive Rationalization Method (by the Surrogate Market)}
    \small
    \KwIn{Dataset $\bXi = \{\p_k, \g_k\}_{k=1}^K$; initial android collection $\{\bgamma_t\}_{t=1}^{T_0}$, $T_0 \ge 1$.}
    \For{$T = T_0, T_0+1, \ldots$}{
        Solve \eqref{eq.cg.dual} to obtain dual solutions $(\bU_T, \mu_T)$ and optimal wealth $\{w_t(\cdot)\}_{t=1}^T$\;
        Solve \eqref{eq.cg.pricing} to find $(\y_{T+1}, \sigma_{T+1})$, and corresponding distribution $\bgamma_{T+1}$\;
        \uIf{$\pi_{T+1} \le \mu_T$}{\textbf{break}\tcp*{no improving android}}
        Add android $(\y_{T+1}, \sigma_{T+1})$ to the collection.
    }
    \KwOut{Surrogate market $\{(\bgamma_t(\cdot), w_t(\cdot))\}_{t=1}^T$.}
\end{myalgo}

The algorithm starts with $T_0$ androids and iteratively adds new androids to the collection until no improving android exists.
We use the subscript $t \in [T]$ to denote related quantities of the $t$-th android.
Based on the discussion in \Cref{sec.basic.ces}, the $t$-th android will be denoted by its expenditure share $\bgamma_t(\p) \in \rH$.
At iteration $T$ of \Cref{alg.cg}, the finite realization of wealth redistribution problem reads:
\begin{subalign}{}
    \tag{WP$_T$}\label{eq.cg.master}   L_T =~& \min_{\w \in \Delta_T}~ \tfrac{1}{K}\smallsum_{k\in [K]} \left\|~\g_k - \h(\p_k)\right\|,\\
    \tag{WD$_T$}\label{eq.cg.dual}   =~&\max_{\|\bU \e_k\|_* \le 1, \bmu\in\real^n}  ~ \tfrac{1}{K}\left(\smallsum_{k\in [K]} \inner{\bU \e_k}{\g_k} - \mu\right)                \\
    \nonumber   &\quad \st~    \smallsum_{k\in [K]} \inner{\bU \e_k}{\bgamma_t(\p_k)} - \mu \le 0, \quad \forall t\in[T]. \end{subalign}
At each iteration $T$, one can think of some lump-sum payments being transferred among the androids.
Given optimal duals $(\bU_T, \mu_T)$, we find the most improving android,
\begin{equation}\label{eq.cg.pricing}
    \tag{SEP}
    \pi_{T+1} = \max_{\bgamma \in \rH} \smallsum_{k=1}^K \inner{\bU_T \e_k}{\bgamma(\p_k)}.
\end{equation}
For example, if one uses the \Cref{example.additive.homothetic}, this means to find the optimal parameters $(\y_{T+1}, \sigma_{T+1})$; if $\pi_{T+1} > \mu_T$, we add the corresponding android with spending share function $\bgamma_{T+1}$ to the market.
The above problem is equivalent to querying a separation oracle in $\rH$, thus denoted as SEP.
For the flow of our presentation, we leave details on solving SEP to \Cref{sec.solve.sep}.

For one thing, CRM can be seen as a boosting algorithm \citep{bartlettBoostingMarginNew1998,demirizLinearProgrammingBoosting2002} to solve the empirical risk minimization problem.
Specifically, for each android $t$, its expenditure share $\bgamma_t(\p)$ can be seen as a (weak) learner of $\g(\p)$. It follows that $\rH$ is the \emph{hypothesis class} of the base learners, the final predictor satisfies $\h \in \conv(\rH)$.  For each sample $k$, the loss is computed by:
\begin{eqalign}{eq.cg.boosting-loss}
    \ell(\h(\p_k), \g_k) := \|\h(\p_k) - \g_k\|.
\end{eqalign}
Thus, the empirical risk over $\bXi$ is given by the primal problem with $T$ base learners:
\begin{eqalign}{eq.cg.boosting-empirical-risk}
    L(\h) := \tfrac{1}{K}\smallsum_{k=1}^K \ell(\h(\p_k), \g_k) \underset{\eqref{eq.cg.master}}{=} L_T.
\end{eqalign}
The above viewpoint helps to derive a generalization bound at unseen prices.
The equivalence to a boosting algorithm also clears the apprehension that a large number of androids may deteriorate the generalization risk.

For another thing, CRM can be seen as a Cutting Plane Method (CPM) for the dual problem \eqref{eq.cg.dual} with potentially many homogeneous linear separation hyperplanes.
Thus, the number of androids $T$ needed to find an approximate solution follows explicitly from the rate of convergence of a CPM.
For example, directly solving \eqref{eq.cg.dual} and obtaining $(\bU, \mu)$ as an extreme point reduces to Kelly's method, which may require an exponential number of iterations to converge \citep{nesterovLecturesConvexOptimization2018}. To obtain better convergence rates, CRM should be implemented in correspondence to CPMs with complexity guarantees; see, e.g., \citet{lemarechalNewVariantsBundle1995,vaidyaNewAlgorithmMinimizing1996,diazOptimalConvergenceRates2023} and the references therein.
The following two remarks are helpful to understand the problem.
\begin{remark}\label{remark.fw}
    CRM is, in fact, a conceptual method with the flexibility to obtain a dual solution.
    It is also possible to implement CRM as a Frank-Wolfe method \citep{frankAlgorithmQuadraticProgramming1956,jaggiRevisitingFrankWolfe2013,lacoste-julienGlobalLinearConvergence2015} that works directly on the primal problem \eqref{eq.cg.semi-infinite-primal}.
    These types of methods are widely adopted in revenue management literature \citep{jagabathulaConditionalGradientApproach2020,huLearningMixedMultinomial2025a}.
\end{remark}
\begin{remark}[Non-homothetic androids]\label{sec.method.non-homothetic}
    Directly using non-homothetic androids makes \ref{eq.cg.master} nonconvex, because $\h = \smallsum_{t\in[T]} \bgamma_t(\p_k, w_t) w_t$ is nonlinear in $w_t$, and thus only weak duality holds.
    One simple strategy is to partition the androids into a homothetic block $H$ and a non-homothetic block $N$. Each non-homothetic android $t \in N$ is added to the master with a prefixed budget $w_0 \in (0, 1]$, and only $|N| \le \lfloor \tfrac{1}{w_0} \rfloor$ non-homothetic androids are allowed.
    Namely, \ref{eq.cg.master} becomes:
    \begin{eqalign}{}
        L^\ast = \min_{w_t \ge 0} ~ & \tfrac{1}{K}\smallsum_{k=1}^K \left\|\smallsum_{t \in H} w_t \,\bgamma_t(\p_k) + \smallsum_{t \in N} w_t \, \bgamma_t(\p_k, w_0) - \g_k\right\|,                                                                                                       \\
        \st ~
        & \smallsum_{t \in [T]} w_t = 1, \qquad w_t = w_0,~ \forall t \in N.
    \end{eqalign}
    The same strategy can be used to solve the separation problem as before.
\end{remark}

\subsection{Extensions to linear wealth}\label{sec.method.linearw}

So far, the wealth of the androids is \emph{exogenous} to the price. Allowing price-dependent wealth functions can improve the modeling power of the surrogate market.
Specifically, we consider a first-order wealth function $w_t(\p) = \inner{\p}{\b_t}$, with $\b_t \in \real_+^n$ being the endowment of android $t$. The predictor yields:
\begin{eqalign}{eq.cg.predictor.ad}
    \h(\p_k) = \smallsum_{t\in[T]} \inner{\p_k}{\b_t}\,\bgamma_t(\p_k) \in \Delta_n,
\end{eqalign}
The wealth redistribution becomes literally endowment redistribution:
{\small
\begin{eqalign}{}
    L_T^{\mathrm{AD}} =~ & \min_{\b_t \ge \zero}~ \tfrac{1}{K}\smallsum_{k\in [K]} \left\|~\g_k - \smallsum_{t\in[T]} \inner{\p_k}{\b_t}\,\bgamma_t(\p_k)\right\|,
    \quad \st ~ \smallsum_{t\in[T]} \b_t = \1.\\
    =~                    & \max_{\|\bU \e_k\|_* \le 1, \bmu\in\real^n} ~ \tfrac{1}{K}\left(\smallsum_{k\in [K]} \inner{\bU \e_k}{\g_k} - \inner{\bmu}{\1}\right),
    \quad \st~ \smallsum_{k\in [K]} \inner{\bU \e_k}{\bgamma_t(\p_k)}\,\p_k \le \bmu, \quad \forall t\in[T].
\end{eqalign}
}Both remain convex optimization problems. Setting $\b_t = w_t \1$ and  $\bmu = \mu\1$ recovers \ref{eq.cg.master}. Hence, we must have an improvement  $L_T^{\mathrm{AD}} \le L_T$. It is easy to see the separation problem for $(\bU_T, \bmu_T)$ becomes,
\begin{eqalign}{eq.cg.pricing.ad}
    \pi_{T+1}^{\mathrm{AD}} = \max_{j \in [n]}~ \max_{\bgamma \in \rH}~ \smallsum_{k=1}^K \inner{\p_{k,j} \cdot \bU_T \e_k}{\bgamma(\p_k)} - \mu_{T,j},
\end{eqalign}
We will later see that this extension is necessary in theory.

\section{Performance guarantees}
In CRM, we limit the number of classes in the universe $\rH$ to be finite: $\bar f < \infty$.
For simplicity, we specify the norm to be $\|\cdot\| \equiv \|\cdot\|_\infty$,
and we only verify the generalization risk for the homothetic androids as in \Cref{example.additive.homothetic}. For other androids (even non-homothetic ones), the generalization risk will be similar as long as their Rademacher complexity is bounded.
We make the following assumptions on the data and the hypothesis class.
\begin{assumption}[Boundedness]\label{asm.bound}
    There exists finite positive constants $D_\y \ge 0, D_\sigma \ge 0, D_\p \ge 0$.  For any $k \in [K]$, $\|\log(\p_k)\|_2 \le D_\p$.
    For any $\bgamma \in \classces \cup \classleon$, parameter $\sigma$ satisfies $\sigma \in [-1, D_\sigma]$, parameter $\y$ satisfies $\|\y\|_2 \le D_\y$. For any $\bgamma \in \classlin$, parameter $\y$ satisfies $\|\y\|_2 \le D_\y$.
\end{assumption}
Note the range for $\sigma$ implies $ -\infty \le  r \le \tfrac{D_\sigma}{1+D_\sigma} < 1$ or $r = 1.$
Limiting $\y$ to a bounded set is not restrictive since $\bgamma$ is shift invariant in $\y$. However, the boundedness of $\log(\p_k)$ cannot be omitted since setting $\p_k$ to zero leads to unboundedness of demand functions.
\subsection{Generalization risk and convergence}\label{sec.method.generalization}
Now, we present the following theorem on the generalization risk.
The proof is relegated to \Cref{sec.proof.gen_bound}.
\begin{theorem}[Generalization Bound]\label{thm.gen_bound}
    Under \Cref{asm.bound}, with probability at least $1-\delta$, if $\h \in \conv(\rH)$, that is, constant wealth functions $w_t(\p) \equiv w_t$ is used in the wealth redistribution, then the generalization risk is given by
    \begin{equation} \label{eq.gen.bound}
        \ex_{\tau}[\ell(\h(\p),\g(\p))] \le L(\h) + 2\sqrt{2} \cdot \left(\tfrac{\poly(n, D_\sigma, D_\y, D_\p)}{\sqrt{K}} + \sqrt{\tfrac{2\log \bar f}{K}}\right) + 6 \sqrt{\tfrac{\ln(2/\delta)}{2K}}.
    \end{equation}
\end{theorem}
The term in the middle is valid as long as $\bar f$ is finite and the android is chosen from a class with Rademacher complexity that is bounded in terms of the parameters. Note that if linear wealth is used (cf. \Cref{sec.method.linearw}), $\h \not\in \conv(\rH)$, but the generalization is similar.
\begin{corollary}[Generalization Bound with linear wealth]\label{cor.gen.bound.linwealth}
    Under the same assumptions as in \Cref{thm.gen_bound}, if the wealth functions are linear, i.e., $w_t(\p) = \inner{\p}{\b_t}$, then the generalization risk is given by:
    \begin{equation}\label{eq.gen.bound.linwealth}
        \ex_{\tau}[\ell(\h(\p),\g(\p))] \le L(\h) + 2\sqrt{2}\, n \cdot \left(\tfrac{\poly(n, D_\sigma, D_\y, D_\p)}{\sqrt{K}} + \sqrt{\tfrac{2\log \bar f}{K}}\right) + 6 \sqrt{\tfrac{\ln(2/\delta)}{2K}}.
    \end{equation}
\end{corollary}
Because we use more parameters, the second term inflates with the dimension $n$; see \Cref{sec.proof.gen_bound.linwealth}.
To discuss the convergence, we have the following fact.
\begin{fact}
    \Cref{alg.cg} converges to some $\widehat L^* \ge 0$, i.e., $\widehat L^* = \underset{T \to \infty}{\lim} L_T$ exists and $\widehat L^* \le L_T$.
\end{fact}
The proof is trivial since it is apparent that $\{L_T\}_{T=1}^\infty$ is monotonic and bounded from below. Using this fact, the empirical risk at some iteration $T$ can be decomposed into three parts:
\begin{eqalign}{eq.cg.convergence-decomposition}
    L_T := \underbrace{L_T - \widehat L^*}_{(\square_1)} + \underbrace{\widehat L^* - L^*}_{(\square_2)} +  \underbrace{L^*}_{(\square_3)},
\end{eqalign}
Only $(\square_1)$ is governed by the convergence rate of \Cref{alg.cg}.
The second term $(\square_2)$ indicates the gap between the best and the one achievable via \Cref{alg.cg}, while the last term $(\square_3)$ is the misspecification error, i.e., the smallest empirical risk attainable by the universe $\rH$.
We first present the following theorem for the term $(\square_1)$, which gives an explicit dependence on the number of androids $T$.
\begin{theorem}[Iteration complexity]\label{thm.iteration_complexity}
    If for all $T$, $(\bU_T, \mu_T)$ is chosen as the Analytic Center, then, \Cref{alg.cg} converges to some $\widehat L^* \ge L^* \ge 0$. Moreover, the number of androids added to the surrogate market is at most $\cO(nK\ln(\tfrac{1}{\epsilon}))$. When stopped, it holds that $L_T - \widehat L^* \le \epsilon$.
\end{theorem}
Here, we instantiate CRM using Analytic Centers; the estimate here is standard, see \citet{goffinMultipleCutsAnalytic2000,nesterovLecturesConvexOptimization2018} and the references therein. Briefly speaking, whenever an android $T+1$ is added to the surrogate market, it holds that $\pi_{T+1} - \mu_T > 0$. This validates a separating hyperplane of the dual space and the volume of the dual set is shrinked linearly. This can only happen in at most $O(nK\ln(\tfrac{1}{\epsilon}))$ times before reaching an accuracy of $\epsilon$. There is much room for further improvement, e.g., using stochastic approximations and variance reduction techniques.

\subsection{Hardness and solvability of the separation problem}\label{sec.method.hardness}
Usually, CPMs or Frank-Wolfe methods applied to ``simple'' sets where a separating hyperplane is easy to find (e.g., from the subdifferential), \ref{eq.cg.pricing} here is nonconvex. While \Cref{thm.iteration_complexity} itself does not require the solution quality of \ref{eq.cg.pricing}, the failure to solve it to global optimality can trigger false termination at some non-vanishing gap $(\square_2) > 0$.
The efficiency of solving \ref{eq.cg.pricing} is summarized in \Cref{tab.sep.complexity}; the relevant details are given in \Cref{sec.solve.sep}.
\begin{table}[h]
    \centering
    \caption{Efficiency in solving \ref{eq.cg.pricing} for a few exemplary classes.}
    \label{tab.sep.complexity}
    \footnotesize
    \begin{tabular}{r|c|c|c}
        \toprule
        Class        & $K = 1$         & $K \ge 2$                                              & Method for $K\ge 2$                        \\
        \cmidrule(lr){1-4}
        Cobb-Douglas &                 & Polynomial-time                                        & -                                          \\
        \cmidrule(lr){1-1}\cmidrule(lr){3-4}
        Linear       & Polynomial-time & Strongly NP-hard (\Cref{thm.nphard.sep.linear})        & MIP                                        \\
        \cmidrule(lr){1-1}\cmidrule(lr){3-4}
        CES          &                 & \multirow{2}{*}{NP-hard (\Cref{thm.nphard.sep.other})} & FPTAS (global, \Cref{thm.sep.fptas.other}) \\
        Leontief     &                 &                                                        & FPTAS (KKT, \Cref{remark.epsilon-kkt})     \\
        \bottomrule
    \end{tabular}
\end{table}

In general, \ref{eq.cg.pricing} is NP-hard whenever $K \ge 2$, even for simple additive homothetic classes (\Cref{example.additive.homothetic}, except for the Cobb-Douglas); this is likely to be true for more complex classes.
For the canonical case $\classlin$, we show that it is strongly NP-hard due to a reduction from the Maximum Acyclic Subgraph problem. The reduction here is in the same spirit of axioms of revealed preference \citep{varianNonparametricApproachDemand1982} that a rational android must admit no preference cycles. Mixed-integer programming (MIP) solvers can be used to find a global solution.
For classes like $\classces$, the separation problem is not strongly NP-hard, and we can design an FPTAS for a global solution (\Cref{alg.sep.fptas}).
It is also possible to use an interior-point method to find an approximate KKT solution in $\cO(\epsilon^{-1.5})$ time (cf. \Cref{remark.epsilon-kkt}).
While only of theoretical interest, if a global solution to \ref{eq.cg.pricing} can be found approximately, the gap $(\square_2)$ can be eliminated.
For practical purposes, the routine for solving SEP must take advantage of subsampling techniques because the complexity scales with $K$, even for classes beyond those considered here. Further development for solving SEP is beyond our current scope.

\subsection{Misspecification error}\label{sec.method.misspecification}
What remains in the risk estimates is the misspecification error $(\square_3)$.
It is not surprising that it cannot be eliminated if $\rH$ is not rich enough or the wealth functions used in the surrogate market do not match the real wealth functions. How should we choose the universe $\rH$ and the wealth functions $\rW$ when running CRM?
We begin by introducing the following assumption on the real market.
\begin{assumption}\label{asm.real.market}
    The excess demand (cf. \Cref{eq.def.z}) in the real market is homogeneous of degree zero: $\z(\lambda\p) = \z(\p)$ for all $\lambda > 0$ and $\p \in \cP$, and it satisfies the Walras's law: $\inner{\p}{\z(\p)} = 0$ for all $\p \in \cP$.
\end{assumption}
The above assumptions can help limit the discussion to $\cP = \Delta_n$ instead of a convex cone; consequently, the total wealth can also be normalized to 1:
\begin{eqalign}{}
    0 = \inner{\p}{\z(\p)} = \inner{\p}{\smallsum_{i\in[m]} \x_i(\p) - \1} = \smallsum_{i\in[m]}w_i(\p) - 1.
\end{eqalign}
However, this assumption holds in the Arrow-Debreu economy, but fails at constant and nonlinear wealth functions. For example, if all $\{w_i\}_{i\in[m]}$ are constant, it is clear that the demand at $\lambda \p, \lambda > 1$ will be smaller than $\d(\p)$ because everything becomes more expensive.
Fortunately, we can overcome this issue by constructing a lifted market that introduces money as the num\'eraire together with an additional agent, thereby satisfying the required conditions.
\begin{theorem}[Lifted market, informal]\label{thm.market.lift.informal}
    Let the wealth $\{w_i\}_{i\in[m]}$ in the real market be arbitrary continuous functions, with aggregate demand $\d$ on $\cP \subseteq \real_+^n$. There is a lifted market with $n+1$ goods and one extra agent that satisfies \Cref{asm.real.market} and preserves equilibria of the real market.
\end{theorem}
We leave a formal clarification to \Cref{sec.proof.lift}.
Based on this result, we show that it is sufficient to use homothetic $\rH$ and linear wealth functions in the surrogate market.
\begin{corollary}[Universality of homothetic classes]\label{thm.misspecification.linear}
    Let $\rH$ be the union of all homothetic classes and let the wealth functions of androids $\rW$ be linear, i.e., $w_t(\p) = \inner{\b_t}{\p}, \forall t \in [T]$. Then there exists a surrogate market such that the misspecification error is zero: $L^* = 0$.
\end{corollary}
It is not hard to see that simplifying $\rW$ to constant wealth is insufficient.
\begin{remark}
    The aggregate demand of homothetic androids with fixed budgets is the gradient of a convex potential function \citep{eisenbergConsensusSubjectiveProbabilities1959}, so the surrogate market cannot match any case that fails this integrability restriction (such as those from non-homothetic agents), no matter how many androids are introduced. This difficulty can be partly alleviated by using non-homothetic androids while not needed in theory.
\end{remark}

The universality in \Cref{thm.misspecification.linear} holds if $\rH$ consists of all homothetic classes. One might hope to narrow $\rH$ to the simple parametric classes like those in \Cref{example.additive.homothetic}.
We show that this is impossible even if general expenditure shares with bounded variation are involved; there exists a ``worst-case'' two-good, one-agent economy whose demand cannot be uniformly approximated by any surrogate market using such androids.
\begin{theorem}[Simple classes are not enough, informal]\label{thm.counterexample.informal}
    There exists a two-good, one-agent homothetic market with spending share $\g$ that is unapproachable by any expenditure share $\h$ constructed from androids with bounded variation and polynomial wealth functions, viz.
    \begin{eqalign}{eq.counterexample.informal.bound}
        \inf_{\h}~\sup_{\p\in\intp\Delta_2}~\|\h(\p)-\g(\p)\|_\infty \ge \frac{\tanh(1)}{2}\approx 0.3808.
    \end{eqalign}
    The bound is sharp and attainable by using only Cobb-Douglas consumers.
\end{theorem}
The verification is quite technical, and we defer it to \Cref{sec.proof.counterexample}. The bound here is a ``minimax''-type lower bound for the approximation errors.
We show that the same result holds even if we allow $\rW$ to be high-order, including constant and linear wealth functions as special cases. These results suggest no clear advantage of using more complex wealth functions. Adopting high-order $\rW$ cannot resolve the inherent obstacles, and would instead inflate the Rademacher complexity as we showed in \Cref{cor.gen.bound.linwealth}.
When approximating a demand function, richer homothetic $\rH$ instead of more complex $\rW$ should be considered, and it is sometimes unavoidable for a general market economy.

\begin{remark}[Sample and iteration complexity]\label{cor.perf.equilibrium}
    The above results show that CRM needs approximately $K = \Theta(\epsilon^{-2})$ samples and $T = \cO((\tfrac{1}{\epsilon})^2 \ln(\tfrac{1}{\epsilon}))$ androids, the quality of solution to \ref{eq.cg.pricing} and the richness of homothetic classes will determine the final prediction power of the surrogate market.
\end{remark}

\section{Numerical experiments}\label{sec.numerical.exp}
We conduct preliminary experiments for demonstration.
All experiments are conducted on a single machine with a 14-core Apple M4 Pro CPU and 48GB RAM using the Julia programming language.

\myparagraph{Methods.}
The wealth redistribution problems (\ref{eq.cg.semi-infinite-primal}, \ref{eq.cg.semi-infinite-dual}), and maximizing the weighted Nash social welfare \eqref{eq.def.welfare} can be formulated as convex optimization problems solvable by Cardinal Optimizer \citep{geCardinalOptimizerCopt2022}.
In the experiments, we \emph{only} use the androids in $\classces$ (including $\classcobbs$) to fit the surrogate market. The separation problem is solved by the nonlinear optimization package Optim.jl \citep{mogensenOptimMathematicalOptimization2018} for an approximate KKT point only.
We implement several methods that realize CRM, including
\begin{enumerate}
    \item A cutting plane method ($\cg$) and a Frank-Wolfe method ($\fw$) using constant wealth, cf. \Cref{sec.method.constantw}.  For simplicity, in $\cg$ we simply solve the wealth redistribution problem and obtain the dual solution as the extreme point. The implementation of $\fw$ follows the method in \citep{jaggiRevisitingFrankWolfe2013}.
    \item The counterparts with linear wealth (cf. \Cref{sec.method.linearw}): $\adcg$ and $\adfw$.
\end{enumerate}

\myparagraph{Real Market}
We generate a real market with $n$ goods and $m$ agents and produce a {training set} and a {test set} of revealed preference data. The total wealth is 1 and the price is normalized to be in $\Delta_n$.
The test set is only used for evaluation and is not used in \Cref{alg.cg}.
The dataset is denoted as $\bXi = \{(\p_k, \g_k)\}_{k=1}^K$.
\subsection{Prediction under different wealth functions}\label{sec.numerical.wealth}
We now compare the performance of CRM under different wealth functions in the real market.
Each of $m$ agents has a CES utility function defined as \Cref{example.additive.homothetic}: the parameter $r_i$ is uniformly selected from $[-3.5, 0.8]$, and a sparse coefficient vector $\c_i \in \real_+^n$ uniformly sampled between $[0, 30]^n$ with density $0.1$.  Hence, every agent's spending share lies in $\rH$.
We consider three different wealth functions in the real market:
\begin{enumerate}
    \item {Constant} / Zero-order: $w_i(\p) \equiv w_i$ with $\smallsum_{i\in[m]} w_i = 1$, i.e., the Fisher model.
    \item {First-order}: $w_i(\p) = \inner{\p}{\b_i}$ with $\b_i \in \real_+^n$ and $\smallsum_{i\in[m]} \b_i = \1$, i.e., the Arrow-Debreu model.
    \item {Second-order}: $w_i(\p) = \tfrac{\p^\top \bQ_i \p}{\smallsum_{j\in[m]} \p^\top \bQ_j \p}$ with symmetric positive semidefinite $\bQ_i \succeq 0$; thus we still have $\smallsum_{i\in[m]} w_i(\p) = 1$. The excess demand function does not have homogeneity properties.
\end{enumerate}

\Cref{tab.real.ces.wealth} reports the training risk, test risk, the number of androids $T$ with non-zero wealth, and the per-iteration time $t_{\mathrm{it}}$, for each of the three wealth functions.
\begin{table}[h]
    \centering
    \caption{Results on a CES real market ($n = 10$, $m = 30$, $K = 300$, mini-batch $50$, stopped if there is no improvement for $5$ consecutive iterations) under the three wealth functions. $t_{\mathrm{it}}$ is the mean per-iteration time (in milliseconds). Best train / test in each block in \textbf{\color{rosemary}bold}.}
    \label{tab.real.ces.wealth}
    \scriptsize
    \begin{tabular}{llrccr}
        \toprule
        \multirow{2}{*}{wealth}
         & \multirow{2}{*}{method}
         & \multirow{2}{*}{$T$}
         & \multicolumn{2}{c}{error}
         & \multirow{2}{*}{$t_{\mathrm{it}}$ (ms)} \\
        \cmidrule(lr){4-5}
         &
         &
         & train
         & test
         &                                         \\
        \midrule

        \multirow{4}{*}{constant}
         & $\cg$
         & $208$
         & $\emphr{2.863 \times 10^{-3}}$
         & $\emphr{5.575 \times 10^{-3}}$
         & $220.4$                                 \\
         & $\adcg$
         & $121$
         & $4.233 \times 10^{-3}$
         & $8.851 \times 10^{-3}$
         & $1549.1$                                \\
         & $\fw$
         & $311$
         & $1.125 \times 10^{-2}$
         & $3.031 \times 10^{-2}$
         & $32.7$                                  \\
         & $\adfw$
         & $2510$
         & $1.196 \times 10^{-2}$
         & $1.511 \times 10^{-2}$
         & $107.6$                                 \\
        \midrule

        \multirow{4}{*}{first-order}
         & $\cg$
         & $120$
         & $1.871 \times 10^{-2}$
         & $2.272 \times 10^{-2}$
         & $145.3$                                 \\
         & $\adcg$
         & $114$
         & $\emphr{2.949 \times 10^{-3}}$
         & $\emphr{6.297 \times 10^{-3}}$
         & $1694.2$                                \\
         & $\fw$
         & $559$
         & $2.181 \times 10^{-2}$
         & $6.164 \times 10^{-2}$
         & $26.6$                                  \\
         & $\adfw$
         & $2520$
         & $1.100 \times 10^{-2}$
         & $1.347 \times 10^{-2}$
         & $104.9$                                 \\
        \midrule

        \multirow{4}{*}{second-order}
         & $\cg$
         & $54$
         & $1.328 \times 10^{-1}$
         & $1.250 \times 10^{-1}$
         & $90.0$                                  \\
         & $\adcg$
         & $93$
         & $\emphr{3.111 \times 10^{-2}}$
         & $\emphr{4.457 \times 10^{-2}}$
         & $1601.8$                                \\
         & $\fw$
         & $54$
         & $8.118 \times 10^{-2}$
         & $1.943 \times 10^{-1}$
         & $17.5$                                  \\
         & $\adfw$
         & $1065$
         & $4.927 \times 10^{-2}$
         & $5.526 \times 10^{-2}$
         & $105.2$                                 \\
        \bottomrule
    \end{tabular}
\end{table}
We can see that all variants of the CRM can predict the aggregate demand function quite well. When the wealth functions in the real market are constant, there is no need to fit the data by redistributing the endowments. For more complicated wealth functions, the methods $\adcg$ and $\adfw$ help improve prediction accuracy, both improving their counterparts, $\cg$ and $\fw$. Not surprisingly, because the decision is now $\{\b_t\}_{t\in[T]}$ instead of $\{w_t\}_{t\in[T]}$, each iteration becomes more expensive. Overall, the Frank-Wolfe variants seem to be more suitable for exploring larger problems.
More importantly, the results of these experiments seem to confirm our theoretical properties, which match those of boosting methods \citep{bartlettBoostingMarginNew1998}. Even as the number of androids increases, the surrogate market becomes more complicated, the test risk will not deteriorate.

\subsection{Computing fair allocations}\label{sec.numerical.fair}
We apply CRM to compute a proportionally fair allocation (\ref{eq.def.prop.fair}) under fixed wealth functions. We use the following simple rule to allocate:
\begin{enumerate}
    \item After building the surrogate market, we compute the equilibrium price $\p^\ast$ (cf. \ref{eq.def.we}). This can be done efficiently by a price-adjustment process \citep{cheungTatonnementGrossSubstitutes2020,zhangSecondorderTatonnementDecentralized2025} if all the wealth functions are constants in the surrogate market.\footnote{While allowing linear wealth makes the problem PPAD-hard \citep{chenSpendingNotEasier2009,chenComplexityNonmonotoneMarkets2017}, it is possible to implement a vanilla Newton's method to solve \ref{eq.def.we} as a nonlinear least-squares problem, which is usually capable of obtaining a solution at a precision of $10^{-6}$ without complexity guarantees. We do not elaborate this option as using constant wealth suffices in our experiments.}
    \item Next, let each agent $i$ report its demand $\x_i(\p^\ast)$ at the posted price $\p^\ast$. Because it is possible that
          $\x_i(\p^\ast)$ does not clear the market, the final allocation is obtained by scaling the demand $\x_i(\p^\ast)$ to meet the supply:
          \begin{eqalign}{eq.def.weall}
              \widehat\x_i = \frac{\x_i(\p^\ast)}{\smallsum_{i\in[m]} \x_i(\p^\ast)}, \quad \widehat{\mathrm{NSW}} := \smallprod_{i\in[m]} u_i(\widehat\x_i)^{w_i}.
          \end{eqalign}
\end{enumerate}
By \eqref{eq.def.weall}, $\smallsum_{i\in[m]} \widehat\x_i = \1$, and so $\{\widehat\x_i\}_{i\in[m]}$ is a feasible allocation.
For comparison, we compute the optimal allocation $\{\x_i^{\mathrm{opt}}\}_{i\in[m]}$ by the following convex optimization problem:
\begin{eqalign}{eq.def.nsw.opt}
    \log\mathrm{NSW}^{\mathrm{opt}} = \max_{\x_i \in \real_+^n} ~ \smallsum_{i\in[m]} w_i \log u_i(\x_i),\quad \st~ \smallsum_{i\in[m]} \x_i \le \1.
\end{eqalign}
We must have $\mathrm{NSW}^{\mathrm{opt}} \ge \widehat{\mathrm{NSW}}$.
We report the relative gap between two logarithmic welfare values:
\begin{eqalign}{eq.def.delta.nsw}
    \Delta_{\mathrm{NSW}} := \frac{\log\mathrm{NSW}^{\mathrm{opt}} - \log\widehat{\mathrm{NSW}}}{\left|\log\mathrm{NSW}^{\mathrm{opt}}\right|}.
\end{eqalign}
Smaller values means that the allocation $\{\widehat\x_i\}_{i\in[m]}$ is close to the \ref{eq.def.prop.fair}.
We test two types of markets: $(1)$ CES market, generated in the same way as in \Cref{sec.numerical.wealth}; $(2)$ non-homothetic market using the following utility function,
\begin{eqalign}{eq.def.ges}
    u_i(\x) = \smallsum_{j\in[n]} c_{ij}\, x_j^{r_{ij}}, \quad \c_i \in \real_+^n, \quad r_{ij} \in (0,1).
\end{eqalign}
The utility can be seen as a generalization of the CES utility by allowing $r_{ij}$ to vary across goods $j$, making the agents non-homothetic. In CRM, we only use the CES androids and {constant wealth}.
\Cref{tab.real.ces.nsw} reports the results.
\begin{table}[h]
    \centering
    \caption{Fair allocation by $\cg$ on the constant-wealth markets ($K = 300$) as the problem size $n$ and the number of trades $m$ grow. All instances are run with time limit $500$ seconds. $T$: androids with non-zero wealth. $\Delta_{\mathrm{NSW}}$: cf. \eqref{eq.def.delta.nsw}.}
    \label{tab.real.ces.nsw}
    \scriptsize
    \begin{tabular}{llrrrrrrrr}
        \toprule
        \multirow{2}{*}{$m$}
         & \multirow{2}{*}{}
         & \multicolumn{4}{c}{CES}
         & \multicolumn{4}{c}{$\eqref{eq.def.ges}$} \\
        \cmidrule(lr){3-6}
        \cmidrule(lr){7-10}
         &
         & $n=10$
         & $20$
         & $50$
         & $100$
         & $n=10$
         & $20$
         & $50$
         & $100$                                    \\
        \midrule

        \multirow{2}{*}{$30$}
         & $T$
         & $242$
         & $213$
         & $166$
         & $163$
         & $486$
         & $318$
         & $173$
         & $172$                                    \\
         & $\Delta_{\mathrm{NSW}}$
         & $0.037\%$
         & $0.067\%$
         & $0.408\%$
         & $0.476\%$
         & $0.094\%$
         & $0.153\%$
         & $0.426\%$
         & $0.334\%$                                \\
        \midrule

        \multirow{2}{*}{$1000$}
         & $T$
         & $490$
         & $390$
         & $240$
         & $198$
         & $409$
         & $347$
         & $269$
         & $152$                                    \\
         & $\Delta_{\mathrm{NSW}}$
         & $0.001\%$
         & $0.008\%$
         & $0.031\%$
         & $0.080\%$
         & $0.403\%$
         & $0.391\%$
         & $0.035\%$
         & $0.029\%$                                \\
        \midrule

        \multirow{2}{*}{$5000$}
         & $T$
         & $477$
         & $400$
         & $237$
         & $204$
         & $426$
         & $362$
         & $269$
         & $157$                                    \\
         & $\Delta_{\mathrm{NSW}}$
         & $0.000\%$
         & $0.015\%$
         & $0.036\%$
         & $0.154\%$
         & $0.162\%$
         & $0.209\%$
         & $0.066\%$
         & $0.032\%$                                \\
        \bottomrule
    \end{tabular}
\end{table}

We see that the allocation produced by CRM (cf. \Cref{eq.def.weall}) is nearly optimal for both classes of markets. While the surrogate market is constructed using only CES androids with constant wealth, the resulting allocation achieves a welfare level very close to the optimal proportionally fair allocation computed from the true utilities. Across all instances, the welfare gap $\Delta_{\mathrm{NSW}}$ remains below $0.5\%$.

Furthermore, note that for a fixed number of goods $n$, the number of active androids $T$ remains relatively stable as $m$ increases from $30$ to $5000$. The overhead of CRM appears to depend primarily on the dimension of the commodity space rather than on the population size.
From a computational perspective, this is attractive because both the construction of the surrogate market and the subsequent allocation are driven by aggregate demand rather than individual preferences. By contrast, computing the optimal \ref{eq.def.prop.fair} by \Cref{eq.def.nsw.opt} involves all agents explicitly, whose size scales directly with $m$. These results therefore suggest that CRM may provide a scalable approach to computing approximately fair allocations in large markets where $m$ is large or personal utilities are unavailable, expanding applicability of price-adjustment mechanisms.

\section{Conclusion}
We study the Constructive Rationalization Method (CRM), which assembles a surrogate market from simple androids to rationalize collective revealed preferences.
Because the surrogate is itself a market, it is directly amenable to downstream analytics. One can implement the equilibrium price of it to the real market for an allocation that is approximately fair. Empirically, we show that using simple androids, CRM can recover near-optimal Nash social welfare in both homothetic and non-homothetic markets, without requiring full information of the agents.
Several questions remain open. As we show that there is a limit to using simple Androids like the linear or CES classes, it is not clear whether we can find a simple parametric class sufficient to approximate a general market. On the computational side, it is also our future work to strengthen the current implementation, which requires better separation oracles and stochastic approximation schemes instead of using all samples.

\bibliographystyle{plainnat}

\clearpage
\appendix
\etocdepthtag.toc{appendices}

\begingroup
\etocsettagdepth{mainmatter}{none}
\etocsettagdepth{appendices}{subsection}
\etocsetnexttocdepth{subsection}
\etocsettocstyle{\section*{Appendix}}{}
{\scriptsize
    \setlength{\parskip}{5pt}\setlength{\baselineskip}{0.5\baselineskip}
    \etocsetstyle{section}{}{}
    {\noindent\etocnumber~\etocname\dotfill\etocpage\par}{}
    \etocsetstyle{subsection}{}{}
    {\noindent\hspace{1.2em}\etocnumber~\etocname\dotfill\etocpage\par}{}
    \etocsetstyle{subsubsection}{}{}
    {\noindent\hspace{2.4em}\etocnumber~\etocname\dotfill\etocpage\par}{}
    \tableofcontents
}
\endgroup

\section*{Roadmap of the appendix}
The appendix is organized as follows.
\Cref{sec.proof.rh} derives basic facts about the expenditure shares for common utility classes. Next, we present the derivation of the primal-dual pair of the wealth redistribution problem in \Cref{sec.proof.master_dual}.
For the convenience of the reader, the derivation of performance measures is kept in \Cref{sec.proof.gen_bound} in consecutive order, while keeping in mind that whether these depend on how the separation oracle (\ref{eq.cg.pricing}) is implemented. The details about \ref{eq.cg.pricing} are presented in \Cref{sec.solve.sep} as a standalone section. The proofs to show the minimax risk bound are kept in \Cref{sec.proof.counterexample}.
\myparagraph{Notations.}
We use the following notations. The notation $\real,\bbZ,\bbN$ denotes the set of real, integer, and natural numbers; $\real_+^n$ and $\preal^n$ denote the sets of $n$-dimensional vectors with all nonnegative elements, and positive elements, respectively. For a subset $\cS \in \real^n$, we use $\intp(\cS)$ to denote the (relative) interior of $\cS$.
For a natural number $n\in \bbN$, we let $[n] = \{1, ..., n\}$.
We use $[a; b]$ (resp., $[a, b]$) to denote vertical (resp., horizontal) concatenation of arrays or numbers. For a subset $B \subseteq [n]$, we use $\x_{B}$ (resp., $\bA_{BB}$) to denote the subvector (resp., submatrix) of $\x$ (resp., $\bA$) obtained by keeping only the elements (resp., rows and columns) indexed by $B$.
We sometimes omit the brackets when there is no confusion. The symbol $\Delta_n$ means the $n$-dimensional simplex: $\Delta_n = \{\x:\inner{\x}{\1}=1,\x\in\real_+^n\}$. We use $\|\cdot\|$ to denote a proper semi-norm, and $\|\cdot\|_*$ to denote the dual norm; by default, they indicate the induced $\ell_2$ norm.

\section{Derivation of demand functions and the expenditure shares}\label{sec.proof.rh}
\subsection{CES consumer}
An agent or android is a said to be a CES consumer if in its utility function, $r \in (-\infty, 1)$.
The corresponding demand function is a single-valued mapping with the following analytical form
\begin{eqalign}{}
    \x(\p) = w\frac{\diag(\p)^{-(1+\sigma)}\c^{1+\sigma}}{\inner{\c^{1+\sigma}}{\p^{-\sigma}}}, \quad \sigma = \frac{r}{1-r} \ge -1.
\end{eqalign}
Here the quantity $\sigma + 1 = \tfrac{1}{1-r}$ is referred to as the elasticity of substitution \citep{mas-colellMicroeconomicTheory1995}, but a slight abuse of notation, we refer $\sigma$ as the elasticity for short if no confusion arises. The expenditure share $\bgamma(\p)$ is summarized as follows.
\begin{fact}\label{fact.demand.ces}
    For any $\c \in \real_+^n$, $r \in (-\infty, 1)$, define the following reparameterization:
    \begin{eqalign}{}
        \y = (1+\sigma)\log(\c) \in \real^n, \quad \sigma = \frac{r}{1-r} \ge -1.
    \end{eqalign}
    Also define the following mapping $\bgamma: \cP \mapsto \Delta_n$:
    \begin{eqalign}{eq.def.gamma.ces}
        \bgamma(\p) = \frac{\c^{1+\sigma} \circ \p^{-\sigma}}{\inner{\c^{1+\sigma}}{\p^{-\sigma}}} = \softmax{\y - \sigma \log(\p)},
    \end{eqalign}
    where $\mathrm{softmax}: \real^n \mapsto \Delta_n$ is the softmax transformation: $\softmax{\cdot} = \tfrac{\exp(\cdot)}{\inner{\exp(\cdot)}{\1}}.$
\end{fact}
The verification of this fact is straightforward so we leave it to the reader.
Motivated from \Cref{fact.demand.ces}, we will use the parameterization $(\y, \sigma)$ in most cases. Scaling the parameters $\c$ (or equivalently $\y$) produces the same expenditure share $\bgamma(\p)$, viz.
\begin{eqalign}{eq.softmax.shift.invariance}
    \softmax{\y - \sigma \log(\p)} = \softmax{\y + a \1 - \sigma \log(\p)}, \quad \forall a \in \real.
\end{eqalign}
The above shift invariance of the softmax operator is well-known.
Now, we can define $\classces$ formally.
\begin{eqalign}{}
    \classces := \left\{ \bgamma: \cP \to \Delta_n \mid (\y, \sigma) \in \real^n \times (-1, \infty),~ \bgamma(\p) =
    \softmax{\y - \sigma \log(\p)} \right\}.
\end{eqalign}
The symbol $\classces$ recognizes the fact that $r$ only takes values in $(-\infty, 1)$.
For the CES case, the output sequence $\{\bgamma(\p_k)\}_{k\in[K]}$ is unique given a price sequence $\{\p_k\}_{k\in[K]}$, since any $\bgamma$ is a single-valued mapping from $\cP$ to $\Delta_n$. Reversely, if given a set of vectors $\{\bgamma_1, \ldots, \bgamma_K\}$ and each $\bgamma_k \in \intp(\Delta_n)$, does there exist such a $\bgamma \in \classces$ that $\bgamma_k = \bgamma(\p_k)$ holds uniformly?
The following fact says the anwser is yes if $K=1$.
\begin{fact}\label{fact.y.sigma.exists}
    For any \emph{single} pair of price and expenditure share $(\p, \bgamma)$, where $\p \in \cP, \bgamma \in \intp(\Delta_n)$, there exists $(\y, \sigma)\in \real^n \times (-1, \infty)$ such that $\bgamma = \softmax{\y - \sigma \log(\p)}.$
\end{fact}
\begin{myproof}
    Taking log-ratios against $j=1$:
    \begin{eqalign}{eq.log-ratio-inversion}
        \log(\tfrac{\gamma_j}{\gamma_1}) = (y_j - y_1) - \sigma(\log \tfrac{p_j}{p_1}), \qquad j=2,\dots,n.
    \end{eqalign}
    Because of shift invariance \eqref{eq.softmax.shift.invariance}, we can set $y_1=0$ without loss of generality.
    This gives $n-1$ equations in $n$ unknowns $(y_2,\dots,y_n,\sigma)$, leaving $\sigma$ as a free parameter. For any fixed $\sigma > -1$, we recover
    \begin{eqalign}{}
        y_j = \log(\tfrac{\gamma_j}{\gamma_1}) + \sigma (\log \tfrac{p_j}{p_1}), \qquad j=2,\dots,n.
    \end{eqalign}
    The proof is complete.
\end{myproof}
In the meantime, with $K > 1$ data points, a linear system analogous to \eqref{eq.log-ratio-inversion} is overdetermined. It is not possible to recover the parameters reliably whenever there are more than two different samples.
\subsection{Leontief consumer}
In the Leontief case, the utility function is $u(\x) = \min_{j \in [n]} \frac{x_j}{c_j}, \c \in \preal^n,$
and the corresponding \ref{eq.def.ump} admits the perfect-complements demand $\x(\p) = \frac{w \c}{\inner{\p}{\c}}.$
We present the following fact for $\bgamma$ in parallel to \Cref{fact.demand.ces}.
\begin{fact}\label{fact.demand.leontief}
    For any $\c \in \preal^n$, set the reparameterization $\y = \log(\c) \in \real^n$. Then the expenditure share of a Leontief consumer at price $\p$ is
    $\bgamma(\p) = \frac{\diag(\p) \c}{\inner{\p}{\c}} = \softmax{\y + \log(\p)}.$
\end{fact}
The right-hand side recovers the CES expression \eqref{eq.def.gamma.ces} at the boundary $\sigma = -1$, where the substitution effect vanishes.
Similar to the CES case, $\classleon$ can be written as:
\begin{eqalign}{}
    \classleon := \left\{ \bgamma : \cP \to \Delta_n ~\Big|~ \y \in \real^n, ~ \bgamma(\p) = \softmax{\y + \log(\p)} \right\}.
\end{eqalign}
Analogously, a single CES observation is always representable.
\begin{fact}\label{fact.y.exists.leontief}
    For any single pair $(\p, \bgamma)$ with $\p \in \preal^n$ and $\bgamma \in \intp(\Delta_n)$, there exists $\y \in \real^n$ such that $\bgamma = \softmax{\y + \log(\p)}.$
\end{fact}
The proof is similar to that for \Cref{fact.y.sigma.exists}.
\subsection{Cobb-Douglas consumer}
The Cobb-Douglas consumer is the homothetic case $r \to 0$, equivalently the unit-elasticity case $\sigma = 0$ of \Cref{fact.demand.ces}. Its utility is
$u(\x) = \prod_{j=1}^n x_j^{c_j}, \c \in \preal^n,$
and the corresponding \ref{eq.def.ump} admits the demand $\x(\p) = \frac{w}{\inner{\c}{\1}}\diag(\p)^{-1}\c.$ We have the following immediately,
\begin{eqalign}{}
    \classcobbs := \left\{ \bgamma: \cP \to \Delta_n \mid \y \in \real^n,~ \bgamma(\p) \equiv \bgamma =
    \softmax{\y} \right\}.
\end{eqalign}
\subsection{Linear consumer}
For the linear case, its utility is given by $u(\x) = \inner{\c}{\x}, \c \in \real_+^n,$
and the corresponding \ref{eq.def.ump} admits the bang-per-buck demand $\x(\p) = \frac{w}{p_{j^\ast}}\e_{j^\ast}$, where  $j^\ast \in \arg\max_{j \in [n]} \frac{c_j}{p_j}.$
We present the following fact for $\bgamma$ in parallel to \Cref{fact.demand.ces}.
\begin{fact}\label{fact.demand.linear}
    For any $\c \in \preal^n$, set the reparameterization $\y = \log(\c) \in \real^n$. Then the expenditure share of a linear consumer at price $\p$ is
    $\bgamma(\p) = \e_{j^\ast}$ where $j^\ast \in \arg\max_{j \in [n]} \{y_j - \log p_j\}.$
\end{fact}
Strictly speaking, there could be multiple $j^\ast$ for a given $\p$. We force the concentration of expenditure, which means to select one $j^\ast$ by a tie-breaking rule.
Similar to the CES case, $\classlin$ can be written as:
\begin{eqalign}{}
    \classlin := \left\{ \bgamma : \cP \to \Delta_n ~\Big|~ \y \in \real^n, ~ \bgamma(\p) = \e_{j^\ast(\y, \p)}, ~ j^\ast(\y, \p) \in \arg\max_{j \in [n]} \{y_j - \log p_j\} \right\},
\end{eqalign}
Analogously, a single vertex observation is always representable.
\begin{fact}\label{fact.y.exists.linear}
    For any single pair $(\p, \bgamma)$ with $\p \in \preal^n$ and $\bgamma = \e_{j^\ast}$ is a vertex of $\Delta_n$, there exists $\bgamma \in \classlin$ such that
    $\y \in \real^n, \bgamma = \e_{j^\ast(\y, \p)}.$
\end{fact}
\begin{myproof}
    The observed vertex pins the bang-per-buck winner $j^\ast$. Any $\y$ satisfying
    \begin{eqalign}{}
        y_{j^\ast} - \log p_{j^\ast} \ge y_j - \log p_j, \qquad j \neq j^\ast,
    \end{eqalign}
    realizes $\bgamma$. Using the shift invariance of $\arg\max$, we can fix $y_{j^\ast} = 0$. The remaining $n-1$ inequalities define an closed non-empty half-space; e.g., $y_j = \log(\tfrac{p_j}{p_{j^\ast}}) - 1$ for $j \neq j^\ast$ suffices.
\end{myproof}
For $K > 1$ samples drawn from a single linear consumer, the recovery system stacks $K$ sets of inequalities; the inequalities are an analogue of the Afriat-Varian inequalities requiring cyclic consistency of preferences.
Checking whether the dataset $\{\p_k, \bgamma_k\}_{k \in [K]}$ is compatible with parameter $\y$ reduces to a single feasibility system by fixing $y_n = 0$:
\begin{eqalign}{eq.plc.mixture.lin.membership}
    & \inner{\1}{\bgamma_k} = 1, \quad \forall k \in [K], \\
    & y_j - \log p_{k,j} \ge y_{j'} - \log p_{k,j'} - M(1 - \gamma_{k,j}), \quad \forall j, j' \in [n], ~ k \in [K], \\
    & y_n = 0.
\end{eqalign}
The big-$M$ constant is set to be $M \ge 2\max_{k, j}|\log p_{k,j}|$. When $\bgamma_k$ are known, recovering $\y$ is only a matter of linear programming. Later this representation can be used in finding a separation hyperplane to the convex hull of $\classlin$.

\subsection{Homotheticity and the expenditure share}\label{sec.proof.homotheticity.share}

\begin{lemma}\label{lem.share.indep.iff.homothetic}
    The expenditure share $\bgamma(\p, w)$ is independent of $w$ iff the preferences are homothetic.
\end{lemma}

\begin{myproof}
    Suppose $u$ is homothetic, then there exists a strictly increasing function $\phi$ and a homogeneous of degree one function $\tilde u$ such that $u = \phi \circ \tilde u$. Substituting $\y = \tfrac{1}{\lambda}\x$ in the UMP at $(\p, \lambda w)$ and using the homogeneity of $\tilde u$,
    \begin{eqalign}{}
        \x(\p, \lambda w) = \arg\max_{\inner{\p}{\x} \le \lambda w} u(\x) = \lambda \arg\max_{\inner{\p}{\y} \le w} \phi(\lambda \tilde u(\y)).
    \end{eqalign}
    Both $\phi(\lambda \tilde u(\cdot))$ and $u(\cdot)$ are strictly increasing transforms of $\tilde u(\cdot)$, so the inner argmax is $\x(\p, w)$. Setting $\lambda = \tfrac{1}{w}$ gives $\x(\p, w) = w \x(\p, 1)$, and
    \begin{eqalign}{}
        \bgamma(\p, w) = \diag(\p) \x(\p, 1),
    \end{eqalign}
    independent of $w$.
    Conversely, suppose $\bgamma(\p, w) \equiv \bgamma(\p)$, equivalently the Walrasian demand is homogeneous of degree one in wealth,
    \begin{eqalign}{}
        \x(\p, w) = w\, \x(\p, 1), \qquad \forall \p \in \cP, ~ w > 0,
    \end{eqalign}
    so every wealth-expansion path $\{\x(\p, w) : w > 0\}$ is a ray through the origin.
    For continuous, locally non-satiated, and convex preferences (so that $\x(\p, w)$ is well defined), this is precisely the standard characterization of homotheticity: the Walrasian demand is homogeneous of degree one in wealth at every price if and only if $\succeq$ is homothetic \citep[Chapter 3]{mas-colellMicroeconomicTheory1995}.
    Hence the preferences are homothetic.
\end{myproof}

\section{Derivation of the primal and dual problems}\label{sec.proof.master_dual}

Rewrite the semi-infinite primal \eqref{eq.cg.semi-infinite-primal}:
\begin{eqalign}{}
    L^\ast = \min_{\omega\in \mathfrak M(\rH)} ~& \tfrac{1}{K}\smallsum_{k=1}^K \|\s_k\|, \\
    \st~& \s_k + \smallint_\rH \bgamma(\p_k)\,\diff\omega(\bgamma) = \g_k, \forall k \\
    &\s_k \in \real^n, \forall k.
\end{eqalign}
Write $\u_k = \bU \e_k \in \real^n$ for each sample $k$ and $\mu \in \real$ for the simplex constraint $\smallint_\rH \diff\omega = 1$. The Lagrangian is
\begin{eqalign}{eq.master.lagrangian}
    \cL(\omega, \s; \u, \mu) = \tfrac{1}{K}\bigg[
        & \smallsum_k \big(\|\s_k\| - \inner{\u_k}{\s_k}\big)
        + \smallint_\rH \Big(\mu - \smallsum_k \inner{\u_k}{\bgamma(\p_k)}\Big)\diff\omega(\bgamma) + \smallsum_k \inner{\u_k}{\g_k} - \mu \bigg].
\end{eqalign}
Note $\inf_{\s_k}\big[\|\s_k\| - \inner{\u_k}{\s_k}\big] > -\infty$ iff $\|\u_k\|_\ast \le 1$, so we can write the dual problem
\begin{eqalign}{}
    L^\ast \ge \max_{\u, \mu}~ & \tfrac{1}{K}\Big(\smallsum_k \inner{\u_k}{\g_k} - \mu\Big) \\
    \st~                           & \smallsum_{k=1}^K \inner{\u_k}{\bgamma(\p_k)} \le \mu, \quad \forall \bgamma \in \rH,                \\
    & \|\u_k\|_\ast \le 1, \quad \forall k \in [K],
\end{eqalign}
which is exactly \eqref{eq.cg.semi-infinite-dual}. Because the primal problem is convex, \Cref{lem.strong_duality} below justifies the strong duality.
\begin{lemma}[Strong duality]\label{lem.strong_duality}
    If the preferences are homothetic, then the primal \eqref{eq.cg.semi-infinite-primal} and dual \eqref{eq.cg.semi-infinite-dual} have the same optimal value.
\end{lemma}
The result is quite standard in the literature. Here we provide a hint for verifying the Slater's condition.
\begin{myproof}
    Note that the point $(\bU, \mu) = (\zero, 1)$ is strictly feasible; and thus
    Slater's condition holds for the dual, strong duality follows.
\end{myproof}
 \section{Proof of the performance guarantees}\label{sec.proof.gen_bound}
\subsection{Preliminaries}
We show the generalization bound \eqref{eq.gen.bound} for any $\h \in \conv(\rH)$ using a mixture of simple classes. Before that, we need a few auxiliary results.
\begin{lemma}[Lipschitz continuity of the loss function] \label{lem.lipschitz}
    For any fixed $\t \in \real^n$, the loss $\ell(\cdot, \t)$ is $1$-Lipschitz in the first argument:
    \begin{eqalign}{eq.lipschitz.loss}
        |\ell(\x, \t) - \ell(\x', \t)| \le \|\x - \x'\|_\infty \le \|\x - \x'\|_2, \quad \forall \x, \x' \in \real^n.
    \end{eqalign}
\end{lemma}
The proof is straightforward.
The following lemma \citep{maurerVectorContractionInequality2016} extends standard Lipschitz contraction inequality \citep{ledouxProbabilityBanachSpaces1991}.
\begin{lemma}[Vector-contraction inequality]\label{lem.maurer_contraction}
    Let $\rH$ be a class of functions from an input space $\cW$ to $\real^n$.
    For any $k = 1, \ldots, K$, let $\Phi_k: \real^n \mapsto \real$ be $M_0$-Lipschitz continuous function.
    For any sample $\bomega_1, \ldots, \bomega_K \in \cW$ of size $K$, the following inequality holds:
    \begin{eqalign}{}
        \frac{1}{K}\ex_{\varepsilon}\left[\sup_{\bgamma \in \rH}\smallsum_{k=1}^K \varepsilon_k\Phi_k(\bgamma(\bomega_k))\right]
        \le \frac{\sqrt{2}M_0}{K}\ex_{\bxi}\left[\sup_{\bgamma \in \rH}\smallsum_{k=1}^K \inner{\bxi_k}{\bgamma(\bomega_k)}\right].
    \end{eqalign}
    where $\bm \varepsilon = [\varepsilon_k]$ and $\bxi_k \in [\xi_{k, n}], \forall k \in [K]$ are independent Rademacher variables over $\{\pm 1\}$.
\end{lemma}
Specializing $\bomega_k = \p_k$, and $\Phi_k(\cdot) = \ell(\cdot, \g_k)$, we can write the above result using short-hand notations \citep{maurerVectorContractionInequality2016} for the empirical Rademacher complexity of $\rH$ and of the loss class $\ell \circ \rH$ as
\begin{eqalign}{eq.empirical.rademacher}
    \hat{\cR}_{K}(\rH) := \frac{1}{K} \ex_{\bxi}\left[\sup_{\bgamma \in \rH} \smallsum_{k=1}^K \inner{\bxi_k}{\bgamma(\p_k)}\right], \quad
    \hat{\cR}_{K}(\ell \circ \rH) := \frac{1}{K} \ex_{\varepsilon}\left[\sup_{\bgamma \in \rH} \smallsum_{k=1}^K \varepsilon_k \ell(\bgamma(\p_k), \g_k)\right],
\end{eqalign}
where $\varepsilon_k \in \{\pm 1\}$ is the scalar Rademacher variable. \Cref{lem.maurer_contraction} can be restated, with a slight abuse of notations, as follows:
\begin{eqalign}{eq.maurer-contraction}
    \hat{\cR}_K(\ell \circ \rH) \le \sqrt{2} \cdot\hat{\cR}_K(\rH).
\end{eqalign}
Since for any $k$, $\Phi_k$ is $1$-Lipschitz due to \Cref{lem.lipschitz}.
We will use this inequality in the sequel. Before that, let us bound the empirical Rademacher complexity of $\rH$, which depends on the component classes.

\subsection{Rademacher bounds}
As we use a finite family of $\bar f$ component classes to construct $\rH$, by definition, $\rH = \bigcup_{f \in [\bar f]} \rH^f$, we have the following lemma.
\begin{lemma}\label{lem.rad.union}
    For any finite union $\rH = \bigcup_{f \in [\bar f]} \rH^f$, it holds that
    \begin{eqalign}{eq.rad.union.bound}
        \hat{\cR}_{K}(\rH)
        \le \max_{f \in [\bar f]} \hat{\cR}_{K}(\rH^f) + \sqrt{\frac{2\log \bar f}{K}}.
    \end{eqalign}
\end{lemma}
\begin{myproof}
    Define
    \begin{eqalign}{}
        Z_f(\bxi) := \sup_{\bgamma \in \rH^f}\frac{1}{K}\smallsum_{k=1}^K \inner{\bxi_k}{\bgamma(\p_k)}, \quad
        Z(\bxi) := \max_{f \in [\bar f]} Z_f(\bxi),
    \end{eqalign}
    so that $\hat{\cR}_{K}(\rH^f) = \ex_{\bxi}[Z_f]$ and $\hat{\cR}_{K}(\rH) = \ex_{\bxi}[Z]$.
    Fix any sample index $k \in [K]$, and let $\bxi'$ agree with $\bxi$ except the $k$-th block $\bxi_k$. Then
    \begin{eqalign}{}
        Z_f(\bxi') - Z_f(\bxi)
        &=\sup_{\bgamma \in \rH^f}\frac{1}{K}\smallsum_{k=1}^K \inner{\bxi_k'}{\bgamma(\p_k)} - \sup_{\bgamma \in \rH^f}\frac{1}{K}\smallsum_{k=1}^K \inner{\bxi_k}{\bgamma(\p_k)} \\
        &\underset{(\square_1)}{\le} \sup_{\bgamma \in \rH^f} \tfrac{1}{K} \inner{\bxi_k' - \bxi_k}{\bgamma(\p_k)} \\
        &\le \sup_{\bgamma \in \rH^f} \tfrac{1}{K} \|\bxi_k' - \bxi_k\|_\infty \cdot \|\bgamma(\p_k)\|_1 \underset{(\square_2)}{\le} \tfrac{2}{K},
    \end{eqalign}
    where $(\square_1)$ uses $\sup F' - \sup F \le \sup(F' - F)$, and $(\square_2)$ is due to $\|\bxi_k' - \bxi_k\|_\infty \le 2$ and $\|\bgamma(\p_k)\|_1 = 1$.
    Exchanging $\bxi$ and $\bxi'$ gives the reverse inequality $Z_f(\bxi) - Z_f(\bxi') \le \tfrac{2}{K}$. Thus, $Z_f$ has bounded differences with constants $c_k=\tfrac{2}{K}$.
    By McDiarmid's inequality (e.g., \citet[Theorem 2.9.1]{vershyninHighdimensionalProbabilityIntroduction2018}),
    \begin{eqalign}{}
        \vbP_{\bxi}[Z_f(\bxi) - \ex_{\bxi} Z_f \ge t] \le \exp\left(\frac{-2t^2}{\smallsum_{k=1}^K c_k^2}\right) = \exp\left(\frac{-t^2}{{2}/{K}}\right);
    \end{eqalign}
    that is, $Z_f-\ex_{\bxi}Z_f$ is sub-Gaussian. Hence,
    \begin{eqalign}{eq.sub-gaussian.mgf}
        \ex_{\bxi}\left[\exp\left(t (Z_f-\ex_{\bxi}Z_f)\right)\right]
        \le
        \exp\left(\frac{t^2}{2K}\right),
        \quad \forall t\in\real.
    \end{eqalign}
    Therefore, for any $t>0$,
    \begin{eqalign}{}
        \ex_{\bxi}\left[\max_{f\in[\bar f]}(Z_f-\ex_{\bxi}Z_f)\right]
        &\underset{(\square_3)}{\le}\frac{1}{t}\log\left(\ex_{\bxi}\left[\exp\left(t\max_{f\in[\bar f]}(Z_f-\ex_{\bxi}Z_f)\right)\right]\right) \\
        &\underset{(\square_4)}{\le}\frac{1}{t}
        \log\left(\ex_{\bxi}\left[\smallsum_{f\in[\bar f]}\exp\left(t(Z_f-\ex_{\bxi}Z_f)\right)\right]\right) \\
        &\underset{\eqref{eq.sub-gaussian.mgf}}{\le}\frac{1}{t}\log\left(\exp(t^2/2K) \bar f\right) \le \frac{\log \bar f}{t}+\frac{t}{2K}.
    \end{eqalign}
    For $(\square_3)$, we note that by taking $W = \max_{f\in[\bar f]}(Z_f-\ex_{\bxi}Z_f)$, $e^{t\ex_{\bxi}[W]} \le \ex_{\bxi}[e^{tW}]$ because of Jensen's inequality; taking logarithms and dividing by $t$ gives $\ex_{\bxi}[W] \le \tfrac{1}{t}\log\ex_{\bxi}[e^{tW}]$ as claimed.
    Equation $(\square_4)$ is because each term is nonnegative.
    Optimizing the right-hand side over $t>0$ by taking $t=\sqrt{2K\log\bar f}$ yields
    \begin{eqalign}{}
        \ex_{\bxi}\left[\max_{f\in[\bar f]}(Z_f-\ex_{\bxi}Z_f)\right]
        \le
        \sqrt{\frac{2\log\bar f}{K}}.
    \end{eqalign}
    Finally,
    \begin{eqalign}{}
        \hat{\cR}_{K}(\rH)
        =
        \ex_{\bxi}\left[\max_{f\in[\bar f]}Z_f\right]
        &\le
        \max_{f\in[\bar f]}\ex_{\bxi}\left[Z_f\right]
        +
        \ex_{\bxi}\left[\max_{f\in[\bar f]}(Z_f-\ex_{\bxi}Z_f)\right]\\
        &\le
        \max_{f\in[\bar f]}\hat{\cR}_{K}(\rH^f)
        +
        \sqrt{\frac{2\log\bar f}{K}}.
    \end{eqalign}
    This proves \eqref{eq.rad.union.bound}.
\end{myproof}
Since $\bar f$ is fixed (e.g., $\bar f = 3$ for $\rH = \classces \cup \classleon \cup \classlin$), the union penalty $\sqrt{2\log \bar f / K}$ is $O(K^{-1/2})$ --- the same rate as the per-class Rademacher bound established next, and absorbed into its $\poly/\sqrt{K}$ envelope. It therefore suffices to verify, for each component class, that
\begin{eqalign}{eq.rad.box}
    \boxed{\hat{\cR}_{K}(\rH^f) \le \frac{\poly(n, D_\y, D_\sigma, D_\p)}{\sqrt{K}}.}
\end{eqalign}

\subsubsection{CES, Cobb-Douglas, and Leontief classes}
\begin{lemma}[Lipschitz continuity of the softmax operator] \label{lem.lipschitz.softmax}
    The softmax function is $\tfrac{1}{2}$-Lipschitz continuous:
    \begin{eqalign}{eq.lipschitz.softmax}
        \|\softmax{\x} - \softmax{\x'}\|_2 \le \tfrac{1}{2}\|\x - \x'\|_2, \quad \forall \x, \x' \in \real^n.
    \end{eqalign}
\end{lemma}
\begin{myproof}
    To show the result above, it suffices to bound the operator norm of the Jacobian mapping. Let $\s=\softmax{\x}$. Then $\nabla \softmax{\x} = \diag(\s)-\s\s^\top$.
    For any $\v\in\real^n$,
    \begin{eqalign}{}
        \inner{\v}{(\diag(\s)-\s\s^\top)\v}
        = \smallsum_i s_i v_i^2 - \left(\smallsum_i s_i v_i\right)^2,
    \end{eqalign}
    which is the covariance over a probability vector $\s$, hence $\|\nabla \softmax{\x}\|_2\le \tfrac{1}{2}$. Thus,
    \begin{eqalign}{}
        \|\softmax{\x'}-\softmax{\x}\|_2
        = \left \|\int_{0}^1 \nabla\softmax{\x + t(\x'-\x)}{(\x'-\x)} \diff t \right\|\\
        \le \int_{0}^1 \left\|\nabla\softmax{\x + t(\x'-\x)}\right\|\|\x'-\x\| \diff t = \tfrac{1}{2}\|\x-\x'\|_2.
    \end{eqalign}
    This completes the proof.
\end{myproof}
We will need the following comparison lemma for Rademacher and Gaussian averages.
\begin{lemma}[Comparison lemma]\label{lem.rad_gauss}
    Let $\bxi \in \{\pm 1\}^n$ be a Rademacher vector with i.i.d.\ coordinates and $\bzeta \sim \bN(\zero, \bI_n)$ a standard Gaussian vector. For any set $\cV \subseteq \real^n$,
    \begin{eqalign}{eq.rad_gauss}
        \ex_{\bxi}\left[\sup_{\v\in\cV} \inner{\bxi}{\v}\right]
        \le \sqrt{\tfrac{\pi}{2}}\cdot\ex_{\bzeta}\left[\sup_{\v\in\cV} \inner{\bzeta}{\v}\right].
    \end{eqalign}
\end{lemma}
\begin{myproof}
    We use the fact that by a Rademacher variable $\varepsilon_j$, $\varepsilon_j |\zeta_j| \sim \bN(0,1)$, then,
    \begin{eqalign}{eq.ex.abs.gaussian}
        \ex|\zeta_j| = \frac{1}{\sqrt{2\pi}}\int_{-\infty}^{\infty} |\zeta| \exp\{-\zeta^2/2\}d\zeta = \frac{2}{\sqrt{2\pi}}\int_{0}^{\infty} \zeta \exp\{-\zeta^2/2\}d\zeta = \sqrt{\frac{2}{\pi}}.
    \end{eqalign}
    Thus, we have
    \begin{eqalign}{}
        \ex_{\bzeta}\left[\sup_{\v\in\cV} \inner{\bzeta}{\v}\right]
        &= \ex_{\bzeta}\left[\sup_{\v\in\cV}\smallsum_{j=1}^n \zeta_j v_j\right] \\
        &= \ex_{\bm\varepsilon}\left[\left.\ex_{\bzeta}\left[\sup_{\v\in\cV}\smallsum_{j=1}^n \varepsilon_j |\zeta_j|\cdot v_j \right]~\right| \bm\varepsilon\right] \\
        &\underset{\square_1}{\ge} \ex_{\bm\varepsilon}\left[\sup_{\v\in\cV}\smallsum_{j=1}^n \ex[|\zeta_j|]\varepsilon_j v_j\right] \\
        &\underset{\eqref{eq.ex.abs.gaussian}}{=} \sqrt{\tfrac{2}{\pi}}\ex_{\bm\varepsilon}\sup_{\v\in\cV}\smallsum_{j=1}^n \varepsilon_j v_j = \sqrt{\tfrac{2}{\pi}}\cdot\ex_{\bxi}\sup_{\v\in\cV}\inner{\bxi}{\v},
    \end{eqalign}
    where $\square_1$ due to Jensen's inequality applied to the inner expectation and the convex supremum. Rearranging the terms gives \eqref{eq.rad_gauss}. This completes the proof.
\end{myproof}
The above lemma is well-known in the literature; see, e.g., \citet{ledouxProbabilityBanachSpaces1991,bartlettRademacherGaussianComplexities2003}. Perhaps an explicit specification of the constant $\sqrt{\tfrac{\pi}{2}}$, which is dimension-free, is less known in the literature (see a course \citep{duchiStatistics231CS229T2017} for more details), so we include it here for the completeness of presentation.
\begin{lemma}[Rademacher bound for $\classces \cup \classcobbs \cup \classleon$] \label{lem.rad_bound}
    If \Cref{asm.bound} holds, then
    \begin{eqalign}{}
        \hat{\cR}_K(\classces)
        \le \sqrt{\frac{\pi n(D_\y^2 + nD_\sigma^2)(n + D_\p^2)}{8K}}.
    \end{eqalign}
    The bound also holds for $\classleon$ and $\classcobbs$.
\end{lemma}
\begin{myproof}
    Observe that the expenditure share can be written as $\bgamma(\p) = \softmax{\bTheta\q(\p)}$, where
    \begin{eqalign}{}
        \bTheta := [\diag(\y), \sigma\bI_n] \in \real^{n\times 2n}, \quad \q(\p) := [\1; -\log(\p)] \in \real^{2n}.
    \end{eqalign}
    Under \Cref{asm.bound}:
    \begin{eqalign}{}
        \|\bTheta\|_F = \sqrt{\|\y\|_2^2 + n\sigma^2} \le \sqrt{D_\y^2 + nD_\sigma^2}, \quad
        \|\q(\p)\|_2 = \sqrt{n + \|\log(\p)\|_2^2} \le \sqrt{n + D_\p^2}.
    \end{eqalign}
    By definition \eqref{eq.empirical.rademacher}, $\bxi_1, \ldots, \bxi_K$ are i.i.d. Rademacher vectors. Hence, the concatenation $[\bxi_1; \ldots; \bxi_K]$ are i.i.d. Rademacher variables in dimension $nK$.
    Consider $\bzeta_k \sim \bN(\zero, \bI_n)$ for all $k \in [K]$, and similarly $[\bzeta_1; \ldots; \bzeta_K]$ are i.i.d. Gaussian variables in dimension $nK$.
    For convenience, define
    \begin{eqalign}{}
        \v = [\softmax{\bTheta\q(\p_1)}; \ldots; \softmax{\bTheta\q(\p_K)}] \in \real^{nK},
    \end{eqalign}
    so by \Cref{lem.rad_gauss} applied to the class $\classces$ at the points $\p_1, \ldots, \p_K$,
    we have the following inequality:
    \begin{eqalign}{eq.radgauss.applied}
        K \cdot \hat{\cR}_K(\classces)
        =  \ex_{\bxi} \left[ \sup_{\bTheta} \smallsum_{k=1}^K \inner{\bxi_k}{\softmax{\bTheta\q(\p_k)}} \right] \\
        \le \sqrt{\tfrac{\pi}{2}}\cdot\ex_{\bzeta} \left[ \sup_{\bTheta} \smallsum_{k=1}^K \inner{\bzeta_k}{\softmax{\bTheta\q(\p_k)}} \right].
    \end{eqalign}
    To the fact that $\bzeta_1, ..., \bzeta_K$ are i.i.d. Gaussian variables, the following inequalities hold:
    \begin{subalign}{eq.ex.inner.properties}
        &\ex\left[\inner{\bzeta_k}{\bzeta_k}\right] = \trace(\ex[\bzeta_k\bzeta_k^\top]) = \trace(\bI_n) = n \\
        \forall k \neq k', \quad  &\ex\left[\inner{\bzeta_k}{\bzeta_{k'}}\right] = 0,  \\
        \label{eq.ex.inner.squared}  \forall \v \in \real^n, \quad   &\ex\left(\inner{\bzeta_k}{\v}^2\right)
        =\ex\left(\trace\left(\bzeta_k \bzeta_k^\top\v\v^\top\right)\right) =\trace\left(\ex\left(\bzeta_k \bzeta_k^\top\right)\v\v^\top\right) = \|\v\|_2^2.
    \end{subalign}
    Define the Gaussian processes indexed by $\bTheta$,
    \begin{eqalign}{}
        X_{\bTheta} := \smallsum_{k=1}^K \inner{\bzeta_k}{\softmax{\bTheta\q(\p_k)}}, \quad
        Y_{\bTheta} := \tfrac{1}{2}\smallsum_{k=1}^K \inner{\bzeta_k}{\bTheta\q(\p_k)}.
    \end{eqalign}
    It is easy to see $\ex[X_{\bTheta}] = \ex[Y_{\bTheta}] = 0.$
    For any $\bTheta, \bTheta'$, let
    \begin{eqalign}{}
        \v_k := \softmax{\bTheta\q(\p_k)} - \softmax{\bTheta'\q(\p_k)}.
    \end{eqalign}
    So we have,
    \begin{eqalign}{}
        \ex[(X_{\bTheta} - X_{\bTheta'})^2]
        &= \ex\left[\smallsum_{k=1}^K \inner{\bzeta_k}{\softmax{\bTheta\q(\p_k)}}
            - \smallsum_{k=1}^K \inner{\bzeta_k}{\softmax{\bTheta'\q(\p_k)}}\right]^2 \\
        &= \ex\left[\smallsum_{k=1}^K \inner{\bzeta_k}{\softmax{\bTheta\q(\p_k)}-\softmax{\bTheta'\q(\p_k)}}\right]^2 \\
        &= \ex\left[\smallsum_{k=1}^K \inner{\bzeta_k}{\v_k}\right]^2 = \ex\left[\smallsum_{k,k'=1}^K \inner{\bzeta_k}{\v_k}\inner{\bzeta_{k'}}{\v_{k'}}\right] = \ex\left[\smallsum_{k,k'=1}^K \trace\left(\bzeta_k\bzeta_{k'}^\top\v_{k'}\v_k^\top\right)\right] \\
        &\underset{\eqref{eq.ex.inner.properties}}{=} \smallsum_{k=1}^K \ex\inner{\bzeta_k}{\v_k}^2 \underset{\eqref{eq.ex.inner.squared}}{=} \smallsum_{k=1}^K \|\v_k\|_2^2
        \underset{\eqref{eq.lipschitz.softmax}}{\le} \tfrac{1}{4}\smallsum_{k=1}^K \|(\bTheta - \bTheta')\q(\p_k)\|_2^2
        = \ex\left[(Y_{\bTheta} - Y_{\bTheta'})^2\right].
    \end{eqalign}
    By Sudakov-Fernique inequality; see, for example, \citet[Theorem 3.15]{ledouxProbabilityBanachSpaces1991}, we have
    \begin{eqalign}{}
        \ex\left[\sup_{\bTheta} X_{\bTheta}\right] \le \ex\left[\sup_{\bTheta} Y_{\bTheta}\right].
    \end{eqalign}
    In view of \eqref{eq.radgauss.applied}, we have,
    \begin{eqalign}{}
        \hat{\cR}_K(\classces)
        \le \sqrt\frac{\pi}{8}\frac{1}{K} \ex_{\bzeta} \left[ \sup_{\bTheta} \smallsum_{k=1}^K \inner{\bzeta_k}{\bTheta\q(\p_k)} \right].
    \end{eqalign}
    The rest of the proof is to bound the RHS from Gaussian variables.
    Indeed, take $\bM := \smallsum_{k=1}^K \q(\p_k)\bzeta_k^\top$,
    \begin{eqalign}{eq.intermediate1}
        \sup_{\bTheta} \smallsum_{k=1}^K \inner{\bzeta_k}{\bTheta\q(\p_k)}
        = \sup_{\bTheta} \trace(\bM\bTheta)
        \le \|\bTheta\|_F \|\bM\|_F
        \le \sqrt{D_\y^2 + nD_\sigma^2}\|\bM\|_F.
    \end{eqalign}
    Note that,
    \begin{eqalign}{}
        \ex \left[\|\bM\|_F^2\right]
        &= \ex \left[\trace(\bM^\top\bM)\right]
        = \smallsum_{k,k'=1}^K \inner{\q(\p_k)}{\q(\p_{k'})}\ex\left[\inner{\bzeta_k}{\bzeta_{k'}}\right] \\
        &\underset{\eqref{eq.ex.inner.properties}}{=} n \smallsum_{k=1}^K \|\q(\p_k)\|_2^2
        \le nK(n + D_\p^2),
    \end{eqalign}
    Combining, we have, $\hat{\cR}_K(\classces) \le \sqrt{\frac{\pi n(D_\y^2 + nD_\sigma^2)(n + D_\p^2)}{8K}}.$
    This completes the proof.
\end{myproof}
Apparently, the Rademacher bound for $\classces$ is polynomial in $n, D_\y, D_\sigma, D_\p$. The same holds for $\classleon$ because the parameterization is also valid for $\classleon$ except that $\sigma$ is fixed to be $-1$.

\subsubsection{Linear class}
The linear class $\classlin$ is different because we are able to show it is a finite class. The Rademacher bound is given as follows.
\begin{lemma}[Rademacher bound for $\classlin$]
    \label{cor.rad_bound_lin}
    Under \Cref{asm.bound},
    \begin{eqalign}{eq.rad.lin}
        \hat{\cR}_K(\classlin)
        \le \sqrt{\frac{2n\log\left(\frac{e(n-1)K}{2}\right)}{K}}.
    \end{eqalign}
\end{lemma}

\begin{myproof}
    By definition \eqref{eq.empirical.rademacher},
    \begin{align}\label{eq.lin.rad.start}
        K\cdot\hat{\cR}_K(\classlin)
        = \ex_{\bxi}\left[
            \sup_{\|\y\|_2\le D_\y}
            \smallsum_{k=1}^K \inner{\bxi_k}{\e_{j^\ast(\y,\p_k)}}
            \right].
    \end{align}
    For fixed samples \(\p_1,\ldots,\p_K\), we introduce the following mapping:
    \begin{eqalign}{}
        \bJ(\y)
        :=
        \left[
            j_1,
            \ldots,
            j_K
            \right]
        \in \cJ \subseteq [n]^K,\quad \text{where }   j_k := j^\ast(\y,\p_k)
        =
        \arg\max_{j\in[n]}
        \big\{ y_j - \log p_{k,j} \big\} ,
    \end{eqalign}
    matching the parameterization of $\classlin$. Assume a deterministic tie-breaking rule for the argmax, then the mapping $\bJ : \real^n \to \cJ$ is single-valued and piecewise constant; its value changes only when, for some sample $k$, two coordinates $j \ne j'$ become tied:
    \begin{eqalign}{}
        y_j - \log p_{k,j} = y_{j'} - \log p_{k,j'}.
    \end{eqalign}
    That is, the values depends on the homogenous hyperplane in $\y$
    \begin{eqalign}{eq.lin.hyperplanes}
        H_{jj'k}
        :=
        \Bigl\{
        \y \in \real^n :
        y_j - y_{j'}
        =
        \log p_{k,j} - \log p_{k,j'}
        \Bigr\},
        \qquad
        j < j', k \in [K].
    \end{eqalign}
    In other words, by the affine hyperplanes \eqref{eq.lin.hyperplanes}, $\real^n$ is partitioned into finitely many regions, in each of which $\bJ$ is constant. This means the image $\cJ$ is finite.
    By \citet{coverGeometricalStatisticalProperties1965} (see, also \citet[6.1.1]{matousekLecturesDiscreteGeometry2002}), we can bound the number of possible regions generated by these hyperplanes, since there are at most $\binom{n}{2} K$ of them in all:
    \begin{eqalign}{eq.region.bound}
        |\cJ|
        \le
        \smallsum_{i=0}^{n}\binom{\binom{n}{2} K}{i}
        \le
        \left(\frac{e\binom{n}{2} K}{n}\right)^{n}
        =
        \left(\frac{e\binom{n}{2}K}{n}\right)^{n}
        =
        \left(\frac{e(n-1)K}{2}\right)^{n},
    \end{eqalign}
    where the second inequality is due to
    \citet[Lemma~A.5]{shalev-shwartzUnderstandingMachineLearning2014}.
    Writing the Rademacher vectors into
    $\bxi:=[\bxi_1;\cdots;\bxi_K]\in\{\pm1\}^{nK}$, we can associate
    each labeling $\bJ\in\cJ$ with the vector
    \begin{eqalign}{}
        \e(\bJ) := [\e_{j_1};\cdots;\e_{j_K}]\in\real^{nK},
        \qquad \|\e(\bJ)\|_2 = \sqrt{K}.
    \end{eqalign}
    Hence, we have,
    \begin{eqalign}{eq.sup.max}
        \ex_{\bxi}\left[
            \sup_{\|\y\|_2\le D_\y}
            \smallsum_{k=1}^K \inner{\bxi_k}{\e_{j^\ast(\y,\p_k)}}
            \right]
        \underset{(\square_1)}{\le}
        \ex_{\bxi}\left[
            \max_{\bJ\in\cJ}~\inner{\bxi}{\e(\bJ)}
            \right]
        \underset{(\square_2)}{\le}
        \sqrt{K}\cdot\sqrt{2\log|\cJ|}.
    \end{eqalign}
    Here, part $\square_1$ is because, $\inner{\bxi_k}{\e_{j^\ast(\y,\p_k)}}$ depends on $\y$ only through $\bJ(\y)$.
    Part $\square_2$ is due to the Massart lemma
    \citep[Lemma~5.2]{massartApplicationsConcentrationInequalities2000}.
    Substituting \eqref{eq.region.bound} into \eqref{eq.sup.max},
    \begin{eqalign}{}
        \hat{\cR}_K(\classlin)
        \le
        \frac{\sqrt{K}\cdot\sqrt{2\log|\cJ|}}{K}
        =\sqrt{\frac{2\log|\cJ|}{K}}
        \le
        \sqrt{\frac{2n\log\left(\frac{e(n-1)K}{2}\right)}{K}}.
    \end{eqalign}
\end{myproof}
Clearly, it is bounded by a polynomial in $n$.

\subsection{Proof of \Cref{thm.gen_bound}}
We are now ready to show the main result, based on the fact that each component class has a Rademacher bound in polynomial size.
\begin{myproof}
    By Theorem 3.3 in \cite{mohriFoundationsMachineLearning2018}, for any $\delta > 0$, with probability at least $1-\delta$, the generalization bound on the loss function $\ell$ holds:
    \begin{equation} \label{eq.rad_base}
        \ex_{\tau}[\ell(\h(\p), \diag(\p)\d)] \le L_{\bXi}(\h) + 2\hat{\cR}_{K}(\ell \circ \conv(\rH)) + 3\sup |\ell(\h(\p), \diag(\p)\d)| \sqrt{\frac{\ln(2/\delta)}{2K}},
    \end{equation}
    where the supremum is taken over $\h \in \conv(\rH)$ and $(\p,\d) \in \cP \times \real_+^n$. It follows that from the contraction lemma (cf. \Cref{lem.maurer_contraction}),
    \begin{align}
        \hat{\cR}_{K}(\ell \circ \conv(\rH)) \underset{\eqref{eq.maurer-contraction}}{\le} \sqrt{2}\hat{\cR}_{K}(\conv(\rH)) = \sqrt{2}\hat{\cR}_{K}(\rH),
    \end{align}
    Recall that \eqref{eq.maurer-contraction} is by setting $\Phi_k(\cdot) = \ell(\cdot, \g_k)$, which is $1$-Lipschitz in the first argument (cf. \Cref{lem.lipschitz}).
    The second equation for the convex hull is well-known; see, e.g., \citet[Lemma 7.4]{mohriFoundationsMachineLearning2018}.
    Moreover,
    \begin{eqalign}{}
        \sup |\ell| = \sup \|\h(\p) - \diag(\p)\d\|_\infty \le \|\h(\p) - \diag(\p)\d\|_1 \le \|\diag(\p)\d\|_1 + \|\h(\p)\|_1 \le 1 + \|\w\|_1\|\bgamma(\p)\|_1 \le 2.
    \end{eqalign}
    Combining,
    \begin{eqalign}{eq.final_bound}
        \ex_{\tau}[\ell(\h(\p), \diag(\p)\d)] &\le L_{\bXi}(\h) + 2\sqrt{2} \cdot \hat{\cR}_K(\rH) + 6 \sqrt{\frac{\ln(2/\delta)}{2K}}.
    \end{eqalign}
    This completes the proof.
\end{myproof}

\subsection{Generalization risk under linear wealth functions}\label{sec.proof.gen_bound.linwealth}
\Cref{thm.gen_bound} concerns the exogenous wealth, where we can write $\h \in \conv(\rH)$ with fixed weights $\w \in \Delta_T$. For linear wealth where each android $t$ the wealth $w_t(\p) = \inner{\p}{\b_t}$. The predictor \eqref{eq.cg.predictor.ad} reads
\begin{eqalign}{eq.lin.predictor}
    \h(\p) &= \smallsum_{t\in[T]} \inner{\p}{\b_t}\,\bgamma_t(\p), \qquad \b_t \ge \zero, \quad \smallsum_{t\in[T]} \b_t = \1\\
    &= \smallsum_{j\in[n]} p_{j}\, \hat\bgamma_j(\p), \quad\text{where}\quad \hat\bgamma_j(\p) := \smallsum_{t\in[T]} b_{t,j}\,\bgamma_t(\p).
\end{eqalign}
The second equation simply changes the order of summation.
Since $\smallsum_t \b_t = \1$, we have for each $j$, $\hat\bgamma_j \in \conv(\rH)$.
Let $\rH^{\mathrm{AD}}$ denote the class of all such predictors.
The next lemma shows that the empirical Rademacher complexity of $\rH^{\mathrm{AD}}$ remains controlled by that of the base class $\rH$, up to a multiplicative constant that does not depend on the number of androids $T$.
\begin{lemma}[Rademacher bound under linear wealth]\label{lem.rad.linwealth}
    Under \Cref{asm.bound}, it holds that
    \begin{eqalign}{eq.rad.linwealth}
        \hat{\cR}_K(\rH^{\mathrm{AD}}) \le n\,\hat{\cR}_K(\rH).
    \end{eqalign}
\end{lemma}
\begin{myproof}
    By the definition of the empirical Rademacher complexity and \eqref{eq.lin.predictor},
    \begin{eqalign}{}
        K \cdot \hat{\cR}_K(\rH^{\mathrm{AD}})
        = \ex_{\bxi}\left[\sup_{\h\in\rH^{\mathrm{AD}}} \smallsum_{k=1}^K \inner{\bxi_k}{\h(\p_k)}\right]
        \underset{(\square_1)}{\le} \smallsum_{j\in[n]} \ex_{\bxi}\left[\sup_{\hat\bgamma_j\in\conv(\rH)} \smallsum_{k=1}^K p_{k,j}\inner{\bxi_k}{\hat\bgamma_j(\p_k)}\right],
    \end{eqalign}
    where $(\square_1)$ uses $\sup\smallsum_j F_j \le \smallsum_j \sup F_j$ and lets each $\hat\bgamma_j$ range freely over $\conv(\rH)$.
    Fix a good $j$ and define, for $\a = [a_1, \ldots, a_K] \in \real^K$,
    \begin{eqalign}{}
        g(\a) := \ex_{\bxi}\left[\sup_{\bgamma\in\conv(\rH)} \smallsum_{k=1}^K a_k \inner{\bxi_k}{\bgamma(\p_k)}\right].
    \end{eqalign}
    The map $g$ is convex, being an expectation of a supremum of linear functions of $\a$; moreover, since each $\bxi_k$ is symmetric, flipping the sign of $a_k$ is equivalent to flipping $\bxi_k$ and leaves $g$ unchanged, so $g$ is even in each $a_k$. Since $\p_k \in \Delta_n$, the coefficients $a_k = p_{k,j} \in [0, 1]$, so $g$ attains its maximum over the box $[-1, 1]^K$ at a vertex, where $|a_k| = 1$; by symmetry every vertex shares the value $\ex_{\bxi}[\sup_{\bgamma}\smallsum_k \inner{\bxi_k}{\bgamma(\p_k)}]$. Hence
    \begin{eqalign}{}
        \ex_{\bxi}\left[\sup_{\hat\bgamma_j\in\conv(\rH)} \smallsum_{k=1}^K p_{k,j}\inner{\bxi_k}{\hat\bgamma_j(\p_k)}\right]
        \le K\hat{\cR}_K(\conv(\rH))
        = K\hat{\cR}_K(\rH),
    \end{eqalign}
    where the last equality is the convex-hull identity \citep[Lemma 7.4]{mohriFoundationsMachineLearning2018}. Summing over $j \in [n]$ gives \eqref{eq.rad.linwealth}.
\end{myproof}
With \Cref{lem.rad.linwealth} in place, the generalization bound follows naturally.
\begin{equation}
    \ex_{\tau}[\ell(\h(\p),\g(\p))] \le L(\h) + 2\sqrt{2}\, n \cdot \left(\tfrac{\poly(n, D_\sigma, D_\y, D_\p)}{\sqrt{K}} + \sqrt{\tfrac{2\log \bar f}{K}}\right) + 6 \sqrt{\tfrac{\ln(2/\delta)}{2K}}.
\end{equation}
The rest of the proof is trivial so we leave to the interested readers.

\subsection{Proof of \Cref{thm.market.lift.informal}}\label{sec.proof.lift}
Here we prove \Cref{thm.market.lift.informal}. We describe the construction of the lifted market first, then, a formal version is provided.
The price space $\cP \subseteq \real_+^n$ of the original market is a closed convex cone, and a price $\q \in \cP$ may be taken at any scale. Its continuous aggregate demand $\d: \cP \mapsto \real_+^n$ need not be homogeneous of degree zero, since the wealth functions $\{w_i\}$ are arbitrary; that is, $\d(\lambda\q)$ and $\d(\q)$ may differ. We use the num\'eraire trick: take money as the $(n{+}1)$-th commodity and define the lifted market as follows.

\myparagraph{Money supply.} By local non-satiation each agent exhausts its budget, so the value of aggregate demand equals the total wealth, $\inner{\q}{\d(\q)} = \smallsum_i w_i(\q) = W(\q)$. We fix a money supply that dominates the wealth in excess of the value of supply,
\begin{eqalign}{eq.lifted.money.supply}
    M \ge \max\left\{0, ~\sup_{\q \in \cP}\left(W(\q) - \inner{\q}{\1}\right)\right\}\ge 0.
\end{eqalign}
Note $M < \infty$ only if $\q$ is bounded. Strictly speaking, $M$ is dependent on $\cP$ and we should let $\q \in \cP \cap \{\q: \|\q\| \le R\}$ for some sufficiently large $R < \infty$. We omit this for brevity.

\myparagraph{Lifted prices and demand.} For $\q \in \cP$ define the normalized lifted price and the lifted price set
\begin{eqalign}{}
    \bar\p(\q) := \frac{[\q; 1]}{1 + \inner{\q}{\1}} \in \bar\cP, \qquad
    \bar\cP := \left\{\bar\p(\q) : \q \in \cP\right\} \subseteq \Delta_{n+1}.
\end{eqalign}
Writing $\bar\p = [\p; \pi]$, the money price is $\pi = \tfrac{1}{1 + \inner{\q}{\1}} > 0$ and $\q = \tfrac{\p}{\pi}$ recovers the original price. We define the lifted demand:
\begin{eqalign}{eq.def.lifted}
    \bar\d(\bar\p) &:= [\d(\q); ~d_{n+1}(\q)], \\
    d_{n+1}(\q) &:= M + \inner{\q}{\1} - W(\q).
\end{eqalign}
By \eqref{eq.lifted.money.supply}, $d_{n+1}(\q) \ge 0$, so $\bar\d(\bar\p) \in \real_+^{n+1}$.

\myparagraph{Lifted agents.} To ensure $\bar\d$ can be rationalized by rational agents, keep each original agent $i \in [m]$ with the same utility and a rescaled budget, and add a single \emph{rentier}, an agent indexed $0$ who only cares about money:
\begin{eqalign}{}
    \bar u_i(\x, m) &:= u_i(\x), \quad \bar w_i(\bar\p) := \pi\, w_i(\q), \\
    \bar u_0(\x, m) &:= m, \quad \bar w_0(\bar\p) := \pi\, d_{n+1}(\q).
\end{eqalign}
Each $\bar u_i$ inherits local non-satiation and concavity from $u_i$, and $\bar u_0(\x, m) = m$ is linear, hence locally non-satiated and concave.
It is easy to see that any agent $i \in [m]$ has no interest in money:
\begin{eqalign}{}
    \bar\x_i &= \underset{\bar\x \in \real_+^{n+1}}{\arg\max} ~\bar u_i(\bar\x, m) ~~\st~~ \inner{\bar\p}{\bar\x} \le \bar w_i(\bar\p) \\
    &= \underset{\bar\x \in \real_+^{n+1}}{\arg\max} ~u_i(\x) ~~\st~~ \inner{\p}{\x} + \pi m \le \pi w_i(\q) = [\x_i(\q); 0];
\end{eqalign}
For the rentier,
\begin{eqalign}{}
    \bar\x_0 &= \underset{\bar\x \in \real_+^{n+1}}{\arg\max} ~\bar u_0(\bar\x, m) ~~\st~~ \inner{\bar\p}{\bar\x} \le \bar w_0(\bar\p) \\
    &= \underset{\bar\x \in \real_+^{n+1}}{\arg\max} ~m ~~\st~~ \inner{\p}{\x} + \pi m \le \pi d_{n+1}(\q) = [\zero; d_{n+1}(\q)].
\end{eqalign}
Therefore, the aggregate demand of the lifted agents is $\bar\d$.
\begin{theorem}\label{thm.market.lift}
    Let the real market have arbitrary continuous wealth functions $\{w_i\}_{i\in[m]}$, with aggregate demand $\d$ on $\cP \subseteq \real_+^n$. The lifted market \eqref{eq.def.lifted} satisfies
    \begin{enumerate}
        \item The price $\bar \p \in \Delta_{n+1}$ and demand $\bar\d(\bar\p) \in \real_+^{n+1}$.
              Besides, let $\bar\z: \Delta_{n+1} \mapsto \real^{n+1}$ be the excess demand function:
              \begin{eqalign}{}
                  \bar\z(\bar\p) = \bar\d(\bar\p) - [\1; M].
              \end{eqalign}
              Then, it is homogeneous of degree zero and satisfies the Walras's law.
        \item If $[\p^*; \pi^*] \in \intp(\bar\cP)$ is a Walrasian equilibrium of the lifted market, then $\q^* = \tfrac{\p^*}{\pi^*}$ is a Walrasian equilibrium of the original market. The reverse direction is also true.
    \end{enumerate}
\end{theorem}
\begin{myproof}
    Nonnegativity of $\bar\d$ was shown above. For homogeneity, the map $\bar\p = [\p; \pi] \mapsto \q = \p/\pi$ is invariant under $\bar\p \mapsto \lambda\bar\p$ for $\lambda > 0$, so $\bar\d(\lambda\bar\p) = \bar\d(\bar\p)$ and
    \begin{eqalign}{eq.lifted.homog}
        \bar\z(\lambda\bar\p) = \bar\d(\lambda\bar\p) - [\1; M] = \bar\z(\bar\p).
    \end{eqalign}
    For Walras's law, write $\bar\p = [\p; \pi]$ with $\p = \pi\q$, so that
    \begin{eqalign}{eq.lifted.walras}
        \tfrac{1}{\pi}\inner{\bar\p}{\bar\z(\bar\p)}
        &= \inner{[\q; 1]}{[\d(\q); d_{n+1}(\q)] - [\1; M]} \\
        &= \inner{[\q; 1]}{[\d(\q); d_{n+1}(\q)]} - \inner{[\q; 1]}{[\1; M]} \\
        &= \inner{\q}{\d(\q)} + d_{n+1}(\q) - \inner{\q}{\1} - M \\
        &\underset{\eqref{eq.def.lifted}}{=} \inner{\q}{\d(\q)} + \left(M + \inner{\q}{\1} - \inner{\q}{\d(\q)}\right) - \inner{\q}{\1} - M = 0,
    \end{eqalign}
    and $\pi > 0$ gives $\inner{\bar\p}{\bar\z(\bar\p)} = 0$. This proves part $(1)$.

    For part $(2)$, the first $n$ coordinates of $\bar\d(\bar\p^*) = [\1; M]$ read $\d(\q^*) = \1$; and when $\d(\q^*) = \1$, the last coordinate satisfies $d_{n+1}(\q^*) = M + \inner{\q^*}{\1} - \inner{\q^*}{\1} = M$ automatically, so the money market clears as well. The two clearing conditions are therefore equivalent. Since $\pi^* > 0$, the ray $\q^* = \tfrac{\p^*}{\pi^*}$ is well defined and lies in $\cP$. Clearly, if $\q^* \in \cP$ is an equilibrium, then $\bar\p^* = [\tfrac{\q^*}{1+\inner{\q^*}{\1}}; \tfrac{1}{1+\inner{\q^*}{\1}}] \in \intp(\bar\cP)$ is an equilibrium of the lifted market.
\end{myproof}

\subsection{Proof of \Cref{thm.misspecification.linear}}\label{sec.proof.misspecification}
We need the following result.
\begin{theorem}[\cite{mantelHomotheticPreferencesCommunity1976}]\label{thm.mantel}
    Suppose that $\z: \Delta_n \mapsto \cX$ is twice differentiable and satisfies the Walras's law: $\inner{\p}{\z(\p)} = 0$, and let $\cK \subset \intp(\Delta_n)$ be a compact set. Then there exists an exchange economy with $n$ consumers, in which each consumer's preferences are convex, homothetic, and monotonic, such that the sum of consumer's excess demand function coincides with $\z$ on $\cK$.
\end{theorem}
The above theorem is extension of the Sonnenschein-Mantel-Debreu theorem \citep{shaferChapter14Market1982} which says for any continuous function $\z$ satisfying the Walras's law and homogeneity of degree zero, there is an exchange economy whose aggregate excess demand coincides with $\z$ on a compact set. Mantel's result strengthens the result by showing that homothetic agents are sufficient. This directly implies the existence of a surrogate market with no misspecification error.
\begin{myproof}
    Let $\z = \d - \1$ be the market excess demand of the real market, which is homogeneous of degree zero and satisfies the Walras's law $\inner{\p}{\z(\p)} = 0$. By \Cref{thm.mantel}, there exists an exchange economy of $n$ homothetic consumers whose excess demand coincides with $\z$ on $\cK$. Hence, there exists a surrogate market using $\rH$ and linear wealth functions such that
    \begin{eqalign}{}
        \diag(\p)(\d(\p) - \1) = \diag(\p)\z(\p) = \h(\p) - \p,
    \end{eqalign}
    therefore, the misspecification error is zero: $L^* = 0$.
\end{myproof}
In Mantel's construction, only excess demand matches $\z$ but the supply is scaled up $\delta$ times, that is, the total supply becomes $\delta \1$, for some $\delta \ge 0$.
For clarity, in the proof we assume that $\delta=1$.
For general $\delta \neq 1$, we can scale the total supply in \eqref{eq.cg.predictor.ad}, by letting $\smallsum_{t\in[n]} \b_t = \delta \1$,
and setting $\g_k \leftarrow \g_k + (\delta - 1) \p_k$.
Alternatively, we can implement it in \Cref{alg.cg} as a variable to be optimized. In that case, the wealth redistribution problem remains to be convex.
Either of the two approaches produces the same result.

\section{Methods of solving the separation oracle}\label{sec.solve.sep}

For the convenience of our presentation, we drop the subscript $T$ in this section. Let us recall the separation oracle \eqref{eq.cg.pricing} below needed for finding new android at each iteration:
\begin{eqalign}{eq.cg.sep.equiv}
    \textrm{\ref{eq.cg.pricing}}: \pi^* &= \max_{\bgamma \in \rH} \sum_{k=1}^K \inner{\bU \e_k}{\bgamma(\p_k)}.
\end{eqalign}
It can be written using the parameterization via $(\y, \sigma)$; see \Cref{sec.proof.rh}. The following fact about \ref{eq.cg.pricing} will be repeatedly used in this section.
\begin{fact}\label{fact.sep.invariant}
    For any $a \in \real$, \ref{eq.cg.pricing} is invariant under affine addition, viz.,
    \begin{eqalign}{}
        \pi^* &= \max_{\bgamma \in \rH} \sum_{k=1}^K \inner{\bU \e_k}{\bgamma(\p_k)} =\max_{\bgamma \in \rH} \sum_{k=1}^K \inner{\bU \e_k + a \1}{\bgamma(\p_k)} - a\cdot K.
    \end{eqalign}
\end{fact}
Hence, we can always assume that $\bU \e_k \ge \zero$ for all $k \in [K]$.
In the following, we show that \ref{eq.cg.pricing} splits into two categories: \begin{enumerate}
    \item The linear case $r=1, \sigma = \infty$, cf. \Cref{sec.sep.linear}. This reduces to a mixed-integer programming problem, and is strongly NP-hard due to a reduction from Maximum Acyclic Subgraph (MAS) problem.
    \item The remaining cases, including the CES case $\sigma \in (-1, \infty), r \in (-\infty, 1)$ (including the Cobb-Douglas class $\sigma = 0$) and the Leontief case $\sigma = -1, r = -\infty$. We show the reduction to a Linear Fractional Programming (LFP) problem in \Cref{sec.sep.remaining} and a FPTAS for global optimal solution in \Cref{sec.pfs.remaining.fptas}.
\end{enumerate}
For the first category, we will solve it by standard mixed-integer linear programming solvers.
Because the practical difficulty of solving \ref{eq.cg.pricing} globally in the latter category, we show an approximation to KKT point is easy to find (cf. \Cref{sec.practical.epsilon-kkt}). In theory, this also allows a FPTAS (cf. \Cref{remark.epsilon-kkt}), and off-the-shelf interior point solvers can be used.

\subsection{Linear case}\label{sec.sep.linear}
When the agents have linear utilities (that is, it belongs to $\classlin$ and $\sigma = \infty$), the spending share collapses to a vertex of $\Delta_n$ (cf. \Cref{fact.demand.linear}), viz.,
\begin{eqalign}{}
    \bgamma(\p_k) = \e_{j_k^\ast}, \quad j_k^\ast \in \arg\max_j \tfrac{c_j}{p_{k,j}} = \arg\max_j (y_j - \log p_{k,j}), \forall k \in [K].
\end{eqalign}
By \eqref{eq.plc.mixture.lin.membership}, \ref{eq.cg.pricing} becomes a mixed-integer program with the big-$M$ constant: $M \ge 2 \max_{k,j} |\log p_{k,j}|$:
\begin{equation}\label{eq.cg.sep.linear}
    \tag*{SEP-$\classlin$}
    {\small
        \begin{aligned}
            \pi^* = \max_{\bgamma_1, \ldots, \bgamma_K}~
                  & \sum_{k=1}^K \inner{\bU \e_k}{\bgamma_k}                                                                       \\
            \st~~ & \bgamma_k \in \{0,1\}^n, ~ \inner{\1}{\bgamma_k} = 1, \quad \forall k \in [K],                                 \\
                  & y_j - \log p_{k,j} \ge y_{j'} - \log p_{k,j'} - M(1 - \gamma_{k,j}), \quad \forall j, j' \in [n], ~ k \in [K], \\
                  & y_n = 0.
        \end{aligned}
    }
\end{equation}
The problem has $Kn$ binaries and $n$ continuous variables.
If for every sample, $\gamma_{k,j_k^\ast} = 1$ is fixed for some $j_k^\ast$. Then the system becomes,
\begin{eqalign}{eq.sep.linear.diffconstr}
    y_{j_k^\ast} - y_{j'} \ge \log p_{k,j_k^\ast} - \log p_{k,j'}, \quad \forall j' \ne j_k^\ast, ~ k \in [K].
\end{eqalign}
A vector $\y$ satisfying all of \eqref{eq.sep.linear.diffconstr} exists if and only if the directed graph that places an edge $j' , j_k^\ast$ of weight $\log p_{k,j_k^\ast} - \log p_{k,j'}$ for each inequality has no positive-weight cycle.
We will later see the connection between \ref{eq.cg.sep.linear} and the Maximum Acyclic Subgraph (MAS) problem defined as follows.
\begin{definition}[MAS]\label{def.mas}
    Given a directed graph $G = (V,E)$, the MAS is to find the largest acyclic edge subset $S \subseteq E$.
\end{definition}
MAS is equivalent to minimum feedback edge set and is strongly NP-hard \citep{karpReducibilityCombinatorialProblems1972}.
\begin{theorem}\label{thm.nphard.sep.linear}
    \ref{eq.cg.pricing} for linear utility \eqref{eq.cg.sep.linear} is polynomial-time solvable for $K = 1$ and \emph{strongly} NP-hard for $K \ge 2$.
\end{theorem}
\begin{myproof}
    \myparagraph{Case 1: $K = 1$.}
    A linear android spends its whole budget on a single good, so $\bgamma(\p_1) = \e_{j^\ast}$ for some $j^\ast$, and the objective equals $u_{1,j^\ast}$. The optimum is therefore $j^\ast \in \arg\max_j u_{1,j}$, found in $O(n)$ time, and \Cref{fact.y.exists.linear} guarantees a parameter $\y$ that realizes this single choice.

    \myparagraph{Case 2: $K \ge 2$.}
    We establish strong NP-hardness by a reduction from the maximum acyclic subgraph (MAS) problem to \ref{eq.cg.sep.linear}.
    Consider a graph $G = (V, E)$. Index the edges as $E = \{e_1, \ldots, e_{|E|}\}$ and create one sample per edge, so $K = |E|$ and sample $k$ corresponds to $e_k$; write $e_k = (a_k , b_k)$, where $a_k \in V, b_k \in V$. The goods are one per vertex of $V$, plus one extra \emph{skip good} $d_k$ used only by sample $k$. Fix integers $Y > K + 1$ and $B > 2Y + 1$. For each sample $k$, and the edge $e_k$, set the data $(\p_k, \u_k)$ as
    \begin{eqalign}{eq.data.reduction}
        &\log p_{k,a_k} = 0, \quad \log p_{k,b_k} = -1, \quad \log p_{k,d_k} = 0, \quad \log p_{k,\ell} = B ~~ (\ell \ne a_k, b_k, d_k),\\
        &u_{k,a_k} = 1, \quad u_{k,\ell} = 0 ~~ (\ell \ne a_k).
    \end{eqalign}
    A feasible solution (with respect to \ref{eq.cg.sep.linear}) with data \eqref{eq.data.reduction} chooses a vertex $j_k^\ast \in V$ for each sample $k$.  If $j_k^\ast = a_k$, reward $u_{k,a_k} = 1$ is obtained.

    Now, let us show that $\pi^* \le \mathrm{MAS}(G)$. Take any feasible $(\bgamma, \y)$ and let $S = \{k : \gamma_{k,a_k} = 1\}$. It is apparent that the objective value equals to $|S|$. For each $k \in S$, let $j' = b_k$ gives
    \begin{eqalign}{}
        y_{a_k} - y_{b_k} \ge 1.
    \end{eqalign}
    Summing the above inequalities over any directed cycle $C \subseteq S$, the left side telescopes to $0$ while the right side is $|C| \ge 1$, a contradiction. Hence $S \subseteq E$ must be cycle free, so $|S| \le \mathrm{MAS}(G)$.

    Conversely, let us show that $\pi^* \ge \mathrm{MAS}(G)$. Take any acyclic $S \subseteq E$, and build the solution pair $(\bgamma, \y)$ as follows: set $\gamma_{k,a_k} = 1$ for $k \in S$ and $\gamma_{k,d_k} = 1$ for $k \in E \setminus S$. Set $y_v$ to the number of edges on the longest directed path of $(V,S)$ starting at $v$, so $0 \le y_v \le K$, since there are at most $K$ edges in $G$.
    Set $y_{d_k} = -1$ if $k \in S$, $y_{d_k} = Y$ otherwise. So the following inequalities hold by our construction \eqref{eq.data.reduction}:
    \begin{enumerate}
        \item For $k \in S$, $a_k$ is chosen:
              \begin{eqalign}{}
                  y_{a_k} - y_{b_k} &\ge 1                        &&= \log p_{k,a_k} - \log p_{k,b_k}, \\
                  y_{a_k} - y_{d_k} &= y_{a_k} + 1 \ge 0      &&= \log p_{k,a_k} - \log p_{k,d_k}, \\
                  y_{a_k} - y_\ell  &\ge -K > -B                  &&= \log p_{k,a_k} - \log p_{k,\ell}, \quad \ell \in V \setminus \{a_k, b_k\}.
              \end{eqalign}
              The first inequality is because the edge $(a_k, b_k)$ extends any path out of $b_k$.
        \item For $k \in E \setminus S$, $d_k$ is chosen:
              \begin{eqalign}{}
                  y_{d_k} - y_{a_k} &= Y - y_{a_k} \ge Y - K > 0  &&= \log p_{k,d_k} - \log p_{k,a_k}, \\
                  y_{d_k} - y_{b_k} &= Y - y_{b_k} \ge Y - K > 1  &&= \log p_{k,d_k} - \log p_{k,b_k}, \\
                  y_{d_k} - y_\ell  &= Y - y_\ell  \ge Y - K > -B &&= \log p_{k,d_k} - \log p_{k,\ell}, \quad \ell \in V \setminus \{a_k, b_k\}.
              \end{eqalign}
    \end{enumerate}
    Thus, the objective value equals to $|S|$ and $(\bgamma, \y)$ is feasible. By taking $S$ to be a maximum acyclic subgraph, we have $\pi^* \ge \mathrm{MAS}(G)$.
    Together, we have proved that $\pi^* = \mathrm{MAS}(G)$. The reduction is exact in objective value.

    The construction runs in time polynomial in $|V| + |E|$, and all numbers are polynomially bounded in $|V| + |E|$: $Y, B = O(|E|)$, the log-prices lie in $\{-1, 0, B\}$, the dual entries in $\{0, 1\}$, and $D_\y = Y$ and $D_\p = B$. By the strong NP-hardness of MAS, \ref{eq.cg.sep.linear} is strongly NP-hard for $K \ge 2$.
\end{myproof}
In this paper, we can tackle \ref{eq.cg.sep.linear} by a mixed-integer programming solver.
\subsection{CES, Cobb-Douglas, and Leontief cases: I, Hardness results}\label{sec.sep.remaining}
We focus on rest of the cases, for which $\sigma \in [-1, D_\sigma]$. We solve
\ref{eq.cg.pricing} in a slightly more restrictive domain than declared in \Cref{asm.bound}:
\begin{eqalign}{}
    \|\y\|_\infty \le \|\y\|_2 \le D_\y, \quad |\sigma| \le D_\sigma.
\end{eqalign}
The Leontief and Cobb-Douglas classes are the special cases.  The \ref{eq.cg.pricing} can be compactly written as follows:
\begin{equation}
    \pi^* = \max_{\y \in \real^n} \pi(\y, \sigma) = \smallsum_{k=1}^K \inner{\bU \e_k}{\softmax{\y - \sigma \log \p_k}}.
\end{equation}
Now, we first consider the case when $\sigma$ is fixed:
\begin{eqalign}{}
    \pi^*(\sigma) = \max_{\|\y\|_\infty \le D_\y} \pi(\y, \sigma),
\end{eqalign}
which separates into:
\begin{enumerate}
    \item If $\sigma = 0$, then we consider the Cobb-Douglas class, and it becomes,
          \begin{eqalign}{}
              \pi^* = \max_{\|\y\|_\infty \le D_\y} \inner{\smallsum_{k=1}^K \bU \e_k}{\softmax{\y}};
          \end{eqalign}
          there is no difference between $K=1$ and $K\ge 1$ and is always globally solvable in polynomial time.
    \item Otherwise, it is equivalent to solving the following Linear Fractional Programming (LFP) problem with nonnegative coefficients:
          \begin{equation}\label{eq.pricing.linearfrac}
              \tag{LFP} \vartheta^* = \max_{\q \in \cQ}~ \vartheta(\q) = \sum_{k=1}^K \frac{\inner{\q}{\btheta_k}}{\inner{\q}{\f_k}},
          \end{equation}
          where $\btheta_k \in \real_+^n$, $\f_k \in \real_+^n$ and $\cQ$ is a compact set in $\real^n$ specified in the following result.
          \begin{lemma}[\ref{eq.cg.pricing} with fixed $\sigma$ is equivalent to \ref{eq.pricing.linearfrac}]\label{lem.pricing.linearfrac}
              For any fixed $\sigma \in [-1, D_\sigma]$ and dual iterate $\bU \in \real^{n\times K}$,
              define the following quantities:
              \begin{eqalign}{}
                  \q = \exp(\y) \in \preal^n, \quad \f_k = \p_k^{-\sigma} \in \real_+^n, \quad \btheta_k = \diag(\f_k)\bU\e_k \in \real_+^n,
              \end{eqalign}
              and $\cQ = \left[e^{-D_\y}, e^{D_\y}\right]^n \subseteq \preal^n$. Then, it holds that, $\pi^*(\sigma) = \vartheta^*.$
          \end{lemma}
          \begin{myproof}
              Substituting $\bgamma(\p_k) = \softmax{\y - \sigma \log \p_k}$ and $\q = \exp(\y)$, each summand satisfies
              \begin{eqalign}{}
                  \inner{\bU\e_k}{\bgamma(\p_k)}
                  &= \frac{\inner{\bU\e_k}{\exp(\y - \sigma\log\p_k)}}{\inner{\1}{\exp(\y - \sigma\log\p_k)}}
                  = \frac{\inner{\bU\e_k}{\diag(\p_k^{-\sigma})\exp(\y)}}{\inner{\1}{\diag(\p_k^{-\sigma})\exp(\y)}} \\
                  &= \frac{\inner{\q}{\diag(\f_k)\bU\e_k}}{\inner{\q}{\f_k}}
                  = \frac{\inner{\q}{\btheta_k}}{\inner{\q}{\f_k}},
              \end{eqalign}
              giving the desired reformulation. In view of \Cref{fact.sep.invariant}, $\f_k \in \real_+^n$, and thus $\btheta_k \in \real_+^n$.
          \end{myproof}
\end{enumerate}
As for \ref{eq.pricing.linearfrac}, here is what we have known about its computational complexity:
\begin{enumerate}
    \item If $K=1$, \ref{eq.pricing.linearfrac} can be solved in polynomial time via the homogeneous transformation \citep{charnesProgrammingLinearFractional1962}. Note that, this happens to be the cases where it is possible to directly recovered the parameters; see, e.g., \Cref{fact.y.sigma.exists}.
    \item If $K\ge 2$, \ref{eq.pricing.linearfrac} is NP-hard \citep{matsuiNPhardnessLinearMultiplicative1996,freundSolvingSumofRatiosProblem2001} due to the reduction from set partitioning problem. It is not strongly NP-hard, and there is an FPTAS provided in \citet{depetriniApproximationLinearFractionalmultiplicative2011} if the problem data can be encoded by integers.
\end{enumerate}
Based on this, for any fixed $\sigma \neq 0$ it is NP-hard to solve \ref{eq.cg.pricing} globally.
\begin{theorem}\label{thm.nphard.sep.other}
    Fix $\sigma \in [-1, D_\sigma]$. Solving \ref{eq.cg.pricing} is polynomial-time solvable for $K = 1$. For $K \ge 2$, it is polynomial-time solvable at $\sigma = 0$ (the Cobb-Douglas class) and NP-hard for every $\sigma \in [-1, D_\sigma] \setminus \{0\}$.
\end{theorem}
\begin{myproof}
    The case $K=1$ is the same as the proof of \Cref{thm.nphard.sep.linear}, due to \Cref{fact.y.sigma.exists,fact.y.exists.leontief}.
    For $K \ge 2$ and $\sigma = 0$, the expenditure share $\softmax{\y}$ is independent of the price, so \ref{eq.cg.pricing} is the Cobb-Douglas problem treated above, which is globally solvable in polynomial time.
    \noindent For $K \ge 2$ and $\sigma \neq 0$, we establish NP-hardness by reducing \ref{eq.pricing.linearfrac} to \ref{eq.cg.pricing}, i.e., by inverting the map of \Cref{lem.pricing.linearfrac}. Given an instance of \ref{eq.pricing.linearfrac} with coefficients $\btheta_k, \f_k \in \real_+^n$ and feasible box $\cQ$, set
    \begin{eqalign}{}
        \p_k = \f_k^{-1/\sigma}, \qquad \bU\e_k = c\, \diag(\f_k)^{-1}\btheta_k, \qquad \forall k \in [K],
    \end{eqalign}
    where $c > 0$ is small enough that $\|\bU\e_k\|_* \le 1$, and $D_\y, D_\p$ are large enough that $\cQ \subseteq [e^{-D_\y}, e^{D_\y}]^n$ and $\|\log\p_k\|_2 \le D_\p$.  Hence, for any LFP instance with polynomially bounded data, the reduction is in polynomial time.
\end{myproof}

\subsection{CES, Cobb-Douglas, and Leontief cases: II, An FPTAS for global maximum}\label{sec.pfs.remaining.fptas}
Based on the ``AppLFMP'' algorithm in \citet{depetriniApproximationLinearFractionalmultiplicative2011}, we can construct a FPTAS to the joint problem in $(\y, \sigma)$ as shown in \Cref{alg.sep.fptas}.
\begin{myalgo}[alg.sep.fptas]{Grid-Search FPTAS for \ref{eq.cg.pricing}}
    \small
    \KwIn{Dual iterate $\bU$; tolerance $\epsilon \in (0,1)$.}
    Set $h \gets \tfrac{\epsilon}{K D_\p}$ and the set of grid points:
    \begin{eqalign}{}
        \Sigma_h \gets \{-1, -1+h, -1+2h, \ldots\} \cap [-1, D_\sigma].
    \end{eqalign}

    \ForEach{$\sigma' \in \Sigma_h$}{
        Solve \ref{eq.pricing.linearfrac} at parameter $\sigma'$ via ``AppLFMP'' at relative tolerance $\delta \gets \tfrac{\epsilon}{2K}$, obtaining $\tilde\q(\sigma') \in \cQ$\;
        Set $\tilde\y(\sigma') \gets \log \tilde\q(\sigma')$\;
    }
    Set:
    \begin{eqalign}{}
        \hat\sigma \gets \arg\max_{\sigma' \in \Sigma_h} \pi(\tilde\y(\sigma'), \sigma'), \quad
        \hat\y \gets \tilde\y(\hat\sigma).
    \end{eqalign}

    \KwOut{$(\hat\y, \hat\sigma)$.}
\end{myalgo}
The idea is simple. By constructing a proper one-dimensional grid of $\sigma$, we solve the pricing problem for a fixed $\sigma$ at each each of the grid point to the relative tolerance $\tfrac{\epsilon}{2K}$. Then, we aggregate the results to get the final solution by picking the best grid point.
We show that \Cref{alg.sep.fptas} is an FPTAS.
\begin{theorem}\label{thm.sep.fptas.other}
    For any $\epsilon \in (0,1)$ and any fixed $K$, \Cref{alg.sep.fptas} returns an approximate solution $(\hat\y, \hat\sigma)$ in $\tilde\cO(\epsilon^{-(K+1)})$ time\footnote{We use $\tilde\cO(\cdot)$ to hide polynomial terms in fixed parameters.} such that $\pi(\hat\y, \hat\sigma) \ge \pi^* - \epsilon.$
\end{theorem}
The first step is to show the complexity for a fixed $\sigma$.
\subsubsection{An FPTAS for \ref{eq.cg.pricing} at fixed $\sigma$.}
Note that \citet{depetriniApproximationLinearFractionalmultiplicative2011}
considers the linear fractional minimization problem (we are solving a maximization problem). They provided a FPTAS, called ``{\bf AppLFMP}'', that returns a ``relatively'' approximate optimal solution if the problem data is integral.
To use AppLFMP here, we need to justify that the methods finds an approximate optimal solution measured in \emph{absolute error} for \emph{real-valued data}.
We first present the result to guarantee absolute error if the problem data is integral, matching the integral data used in \citet{depetriniApproximationLinearFractionalmultiplicative2011}.
\begin{lemma}\label{lem.fptas.add}
    For any $\sigma \in [-1, D_\sigma]$ and any $\eta \in (0, 1)$, the FPTAS of \citet{depetriniApproximationLinearFractionalmultiplicative2011} applied to \ref{eq.pricing.linearfrac} at relative tolerance $\delta = \tfrac{\eta}{K}$ returns $\tilde\y \in [-D_\y, D_\y]^n$ with
    \begin{eqalign}{eq.fptas.add}
        \pi(\tilde\y, \sigma) \ge \pi^*(\sigma) - \eta,
    \end{eqalign}
    in $\tilde\cO(\eta^{-K})$ arithmetic operations.
\end{lemma}
\begin{myproof}
    First, we justify that the optimal value of \ref{eq.pricing.linearfrac} is bounded from above.
    Since $\bgamma_k \in \Delta_n$ and $\|\bU \e_k\|_2 \le 1$, each summand of $\pi(\cdot, \sigma)$ is bounded by $1$, so the optimal value of \ref{eq.pricing.linearfrac} is bounded by the sample size.
    \begin{eqalign}{eq.fptas.upper}
        \vartheta^* = \pi^*(\sigma) \le K.
    \end{eqalign}
    That is, solving \ref{eq.pricing.linearfrac} is equivalent to solving
    \begin{eqalign}{eq.linfrac.min}
        \min_{\q \in \cQ} \; K - \vartheta(\q).
    \end{eqalign}
    \citet{depetriniApproximationLinearFractionalmultiplicative2011} considers a $\delta$-relatively optimal solution $\tilde{\q}$ for the minimization problem, which to be precise, refers to $\tilde\q$ satisfying the following inequality when applying to \eqref{eq.linfrac.min}:
    \begin{eqalign}{}
        (K - \vartheta(\tilde{\q})) \leq (1+\delta) \cdot (K - \vartheta^*),
    \end{eqalign}
    which is equivalent to say,
    \begin{eqalign}{eq.delta.opt}
        \vartheta(\tilde{\q}) + \delta K \ge (1+\delta) \cdot \vartheta^*.
    \end{eqalign}
    It is shown that AppLFMP outputs a solution $\tilde{\q}$ satisfying \eqref{eq.delta.opt} in $\tilde\cO(\delta^{-K})$ arithmetic operations.
    By setting $\delta = \tfrac{\eta}{K}$, AppLFMP returns some $\tilde\q \in \cQ$ satisfying the following inequality:
    \begin{eqalign}{}
        \vartheta(\tilde\q) \ge (1 + \delta) \vartheta^* - \delta K \ge \vartheta^* - \delta K = \vartheta^* - \eta,
    \end{eqalign}
    where the second inequality uses $\delta \vartheta^* \ge 0$. By \Cref{lem.pricing.linearfrac}, we recover the parameter via $\tilde\y = \log\tilde\q$, which gives the stated bound. The runtime is $\tilde\cO(\delta^{-K}) = \tilde\cO(\eta^{-K})$.
\end{myproof}

Next, we show how to apply ``AppLFMP'' to real-valued coefficients $(\f_k, \btheta_k)$.
The idea is to find a rounded instance $(\hat \f_k, \hat \btheta_k)$ with bounded encoding length and apply the method to it.
Namely, we choose some $\nu \in (0, 1), \eta_1 \in (0, 1)$,
\begin{eqalign}{eq.fptas.nu}
    \nu := \frac{\eta_1}{8K} \exp\left(-2 D_\y - D_\sigma D_\p\right) \le \eta_1;
\end{eqalign}
and based on that, set coefficients and objective as follows:
\begin{eqalign}{eq.fptas.round.def}
    &\hat\f_k :=\left\lfloor \tfrac{1}{\nu}\f_k + \tfrac{1}{2}\1 \right\rfloor \in \bbN^n, \qquad \hat\btheta_k := \left\lfloor \tfrac{1}{\nu}\btheta_k + \tfrac{1}{2}\1 \right\rfloor \in \bbN^n, \qquad \forall k \in [K] \\
    &\hat\vartheta(\q) := \sum_{k=1}^K \frac{\inner{\q}{\hat\btheta_k}}{\inner{\q}{\hat\f_k}}.
\end{eqalign}
It is legitimate to do so since the ratio computed from $(\hat\f_k, \hat\btheta_k)$ produces the same value as the one from $(\nu\hat\f_k, \nu\hat\btheta_k)$. Using the choice \eqref{eq.fptas.nu}, we show that $(\nu\hat\f_k, \nu\hat\btheta_k)$ is close to the original data $(\f_k, \btheta_k)$, and the error of a rounding instance can be bounded as follows.
\begin{lemma}\label{lem.fptas.rounding}
    Under \Cref{asm.bound}, consider the rounded instance \eqref{eq.fptas.round.def}, then it satisfies, for all $\q \in \cQ$ and $k \in [K]$,
    \begin{eqalign}{eq.fptas.round.bound}
        \|\nu \hat\f_k - \f_k\|_\infty \le \tfrac{\nu}{2}, \quad \|\nu \hat\btheta_k - \btheta_k\|_\infty \le \tfrac{\nu}{2}, \quad |\vartheta(\q) - \hat\vartheta(\q)| \le \tfrac{\eta_1}{2}.
    \end{eqalign}
    Each entry of $\hat\f_k, \hat\btheta_k$ has encoding length in at most $\cO(\log(\tfrac{K}{\eta_1}) + D_\y + D_\sigma D_\p)$.
\end{lemma}
\begin{myproof}
    Under \Cref{asm.bound}, and $\bU\e_k \in [0,1]^n$, we have
    \begin{eqalign}{}
        \|\log\p_k\|_\infty \le \|\log\p_k\|_2 \le D_\p, \quad |\sigma| \le D_\sigma, \quad \cQ = [\exp(-D_\y), \exp(D_\y)]^n.
    \end{eqalign}
    Thus, it holds that
    \begin{eqalign}{eq.fptas.fbounds}
        \exp(-D_\sigma D_\p)\1 \le \f_k \le \exp(D_\sigma D_\p)\1, \qquad \zero \le \btheta_k \le \f_k.
    \end{eqalign}
    By definition \eqref{eq.fptas.round.def}, $\hat f_{k,j} = \left\lfloor \tfrac{f_{k,j}}{\nu} + \tfrac{1}{2}\right\rfloor$, so
    \begin{eqalign}{eq.fptas.round.entry}
        \left| \hat f_{k,j} - \tfrac{f_{k,j}}{\nu} \right| \le \tfrac{1}{2} \quad \Longrightarrow \quad |\nu \hat f_{k,j} - f_{k,j}| \le \tfrac{\nu}{2}, \qquad \forall j \in [n].
    \end{eqalign}
    Taking the max over $j$ gives $\|\nu \hat\f_k - \f_k\|_\infty \le \tfrac{\nu}{2}$. $\|\nu \hat\btheta_k - \btheta_k\|_\infty \le \tfrac{\nu}{2}$ follows from a similar argument.
    Since $\q \in \cQ$, take
    \begin{eqalign}{}
        A_k := \inner{\q}{\btheta_k}, \quad B_k := \inner{\q}{\f_k}, \quad \hat A_k := \inner{\q}{\hat\btheta_k}, \quad \hat B_k := \inner{\q}{\hat\f_k}.
    \end{eqalign}
    Then by \eqref{eq.fptas.fbounds} and \eqref{eq.fptas.round.entry},
    \begin{subalign}{eq.round.b_k}
        \label{eq.round.b_k.1}    B_k \ge n \exp(-D_\y - D_\sigma D_\p), \quad A_k \le B_k, \\
        \label{eq.round.b_k.2}    |A_k - \nu\hat A_k| \le \|\q\|_1 \tfrac{\nu}{2} \le \tfrac{n}{2} \exp(D_\y) \nu \\
        \label{eq.round.b_k.3}   |B_k - \nu\hat B_k| \le \|\q\|_1 \tfrac{\nu}{2} \le \tfrac{n}{2} \exp(D_\y) \nu.
    \end{subalign}
    By the choice \eqref{eq.fptas.nu}, we note,
    \begin{eqalign}{}
        \tfrac{n}{2} \exp(D_\y) \nu
        &=\tfrac{n}{2} \exp(D_\y) \tfrac{\eta_1}{8K} \exp\left(-2 D_\y - D_\sigma D_\p\right) \le \tfrac{B_k}{2}.
    \end{eqalign}
    Hence,
    \begin{eqalign}{eq.round.b_k.4}
        \nu\hat B_k
        &\underset{\eqref{eq.round.b_k.3}}{\ge} B_k - \tfrac{n}{2} \exp(D_\y) \nu\\
        &\ge \tfrac{B_k}{2} \ge \tfrac{n}{2} \exp(-D_\y - D_\sigma D_\p).
    \end{eqalign}
    Therefore
    \begin{eqalign}{eq.fptas.ratio.diff}
        \left| \frac{A_k}{B_k} - \frac{\nu\hat A_k}{\nu\hat B_k} \right|
        &= \left| \frac{A_k \cdot \nu \hat B_k - B_k \cdot \nu \hat A_k}{\nu\hat B_k \cdot B_k} \right| = \left| \frac{A_k (\nu\hat B_k - B_k) - B_k (\nu \hat A_k - A_k)}{\nu\hat B_k \cdot B_k} \right| \\
        & \le \frac{A_k}{\nu\hat B_k \cdot B_k} |B_k - \nu\hat B_k| + \frac{1}{\nu\hat B_k} |A_k - \nu\hat A_k| \underset{\eqref{eq.round.b_k}}\le \left(\frac{A_k}{B_k} + 1\right)\cdot \frac{1}{\nu\hat B_k} \cdot \frac{n}{2} \exp(D_\y) \nu \\
        &\le \frac{n}{\nu\hat B_k} \cdot \exp(D_\y) \nu  \underset{\eqref{eq.round.b_k.4}}{\le} \frac{n \exp(D_\y) \nu}{\tfrac{n}{2} \exp(-D_\y - D_\sigma D_\p)} \cdot \le 2 \exp\left(2 D_\y + D_\sigma D_\p\right) \nu = \frac{\eta_1}{4K}.
    \end{eqalign}
    Summing over $k \in [K]$ yields \eqref{eq.fptas.round.bound}. For the encoding length, each entry of $\hat\f_k, \hat\btheta_k$ is a non-negative integer bounded by $\nu^{-1}\exp(D_\sigma D_\p)$, so its binary encoding takes
    \begin{eqalign}{eq.fptas.encoding}
        \text{(bit-length)}
        & = \cO\left(\log_2(\nu^{-1}\exp(D_\sigma D_\p))\right) = \cO\left(\log_2 \nu^{-1} + D_\sigma D_\p\right)                            \\
        & \underset{\eqref{eq.fptas.nu}}{=}
        \cO\left(\log_2\left(\frac{8K}{\eta_1} \exp\left(2 D_\y + D_\sigma D_\p\right)\right) + D_\sigma D_\p\right) = \cO\left(\log(\tfrac{K}{\eta_1}) + D_\y + D_\sigma D_\p\right).
    \end{eqalign}
    This completes the proof.
\end{myproof}
The complexity for fixed $\sigma$ is apparent from above results.
\begin{corollary}[``AppLFMP'' on real-valued data]\label{cor.fptas.real}
    Under \Cref{asm.bound}, applying ``AppLFMP'' to the rounded instance \eqref{eq.fptas.round.def} with $\eta_1 = \tfrac{2\eta}{3}$. Set relative tolerance $\delta = \tfrac{\eta}{3K}$, then it returns $\tilde\y \in [-D_\y, D_\y]^n$ in $\tilde\cO(\eta^{-K})$ time such that
    \begin{eqalign}{eq.fptas.real.add}
        \pi(\tilde\y, \sigma) \ge \pi^*(\sigma) - \eta.
    \end{eqalign}
\end{corollary}
\begin{myproof}
    By \Cref{lem.fptas.rounding},
    \begin{eqalign}{}
        |\vartheta(\q) - \hat\vartheta(\q)| \le \tfrac{\eta_1}{2} = \tfrac{\eta}{3}.
    \end{eqalign}
    By \Cref{lem.fptas.add} applied to the rounded instance, ``AppLFMP'' returns $\tilde\q \in \cQ$ such that
    \begin{eqalign}{}
        \hat\vartheta(\tilde\q) \ge \hat\vartheta^* - \tfrac{\eta}{3}.
    \end{eqalign}
    Combining,
    \begin{eqalign}{}
        \vartheta(\tilde\q)
        & \underset{\text{\eqref{eq.fptas.round.bound}}}{\ge} \hat\vartheta(\tilde\q) - \tfrac{\eta}{3}
        \underset{\text{\Cref{lem.fptas.add}}}{\ge} \hat\vartheta^* -  \tfrac{2\eta}{3} \underset{\text{\eqref{eq.fptas.round.bound}}}{\ge} \vartheta^* - \eta.
    \end{eqalign}
    Setting $\tilde\y := \log\tilde\q$ and using $\pi(\y, \sigma) = \vartheta(\exp(\y))$ from \Cref{lem.pricing.linearfrac} gives \eqref{eq.fptas.real.add}. The runtime follow from \Cref{lem.fptas.rounding}.
\end{myproof}
\subsubsection{Proof of \Cref{thm.sep.fptas.other}.}
Finally, we are ready to show the worst-case complexity of \Cref{alg.sep.fptas}.
\begin{myproof}
    Recall that $\bgamma_k(\y, \sigma) := \softmax{\y - \sigma \log \p_k}$, and
    \begin{eqalign}{}
        \pi(\y, \sigma) := \sum_{k=1}^K \inner{\bU \e_k}{\bgamma_k(\y, \sigma)}, \quad \pi^*(\sigma) := \max_{\y} \pi(\y, \sigma).
    \end{eqalign}
    By \Cref{lem.lipschitz}, the softmax is $\tfrac{1}{2}$-Lipschitz continuous, so for any fixed $\y$ and any $\sigma_1, \sigma_2$,
    \begin{eqalign}{eq.fptas.gamma.lip}
        \|\bgamma_k(\y, \sigma_1) - \bgamma_k(\y, \sigma_2)\|_2 \le \tfrac{1}{2} \|\log \p_k\|_2 |\sigma_1 - \sigma_2| \le \tfrac{D_\p}{2} |\sigma_1 - \sigma_2|,
    \end{eqalign}
    Combined with the dual-norm constraint $\|\bU \e_k\|_* \le 1$,
    \begin{eqalign}{eq.fptas.lip}
        |\pi(\y, \sigma_1) - \pi(\y, \sigma_2)|
        &\le \sum_{k=1}^K \|\bU \e_k\|_2 \|\bgamma_k(\y, \sigma_1) - \bgamma_k(\y, \sigma_2)\|_2 \\
        &\le \tfrac{K D_\p}{2} |\sigma_1 - \sigma_2|.
    \end{eqalign}
    Since the bound is uniform in $\y$, picking $\y_i^* \in \arg\max_\y \pi(\y, \sigma_i)$ for $i = 1, 2$ yields
    \begin{eqalign}{}
        \pi^*(\sigma_1) - \pi^*(\sigma_2)
        &= \pi(\y_1^*, \sigma_1) - \pi(\y_2^*, \sigma_2) \le \pi(\y_1^*, \sigma_1) - \pi(\y_1^*, \sigma_2) \le \tfrac{K D_\p}{2} |\sigma_1 - \sigma_2|, \\
        \pi^*(\sigma_2) - \pi^*(\sigma_1)
        &= \pi(\y_2^*, \sigma_2) - \pi(\y_1^*, \sigma_1) \le \pi(\y_2^*, \sigma_2) - \pi(\y_2^*, \sigma_1) \le \tfrac{K D_\p}{2} |\sigma_1 - \sigma_2|,
    \end{eqalign}
    and thus
    \begin{eqalign}{eq.fptas.pi.lip}
        |\pi^*(\sigma_1) - \pi^*(\sigma_2)| \le \tfrac{K D_\p}{2} |\sigma_1 - \sigma_2|.
    \end{eqalign}
    By definition, in \Cref{alg.sep.fptas} with $h = \tfrac{\epsilon}{K D_\p}$, the size of grid is
    \begin{eqalign}{}
        |\Sigma_h| = \cO(\tfrac{D_\sigma + 1}{h}) = \cO(\tfrac{K D_\p D_\sigma}{\epsilon}).
    \end{eqalign}
    For every $\sigma \in [-1, D_\sigma]$, there exists $\sigma' \in \Sigma_h$ with $|\sigma - \sigma'| \le h$, hence by \eqref{eq.fptas.pi.lip}
    \begin{eqalign}{}
        |\pi^*(\sigma) - \pi^*(\sigma')| \le \tfrac{K D_\p}{2} h = \tfrac{\epsilon}{2}.
    \end{eqalign}
    Let $\tilde\y(\sigma')$ be the output of ``AppLFMP'' in \Cref{alg.sep.fptas}. By \Cref{cor.fptas.real}, in $\tilde\cO(\epsilon^{-K})$ operations,
    \begin{eqalign}{eq.fptas.inner}
        \pi(\tilde\y(\sigma'), \sigma') \ge \pi^*(\sigma') - \tfrac{\epsilon}{2}.
    \end{eqalign}
    Let $(\y^*, \sigma^*)$ be a joint maximizer. Then, there exists $\sigma'_* \in \Sigma_h$ with $|\sigma^* - \sigma'_*| \le h$; this $\sigma'_*$ need not equal $\hat\sigma$. So we have,
    \begin{eqalign}{}
        \pi(\hat\y, \hat\sigma)
        & \ge \pi(\tilde\y(\sigma'_*), \sigma'_*)
        \underset{\eqref{eq.fptas.inner}}{\ge} \pi^*(\sigma'_*) - \tfrac{\epsilon}{2}
        \underset{\eqref{eq.fptas.pi.lip}}{\ge} \pi^*(\sigma^*) - \tfrac{K D_\p}{2} h - \tfrac{\epsilon}{2}    \\
        & = \pi(\y^*, \sigma^*) - \epsilon.
    \end{eqalign}
    The total cost is $|\Sigma_h| \cdot \tilde\cO(\epsilon^{-K}) = \tilde\cO(\epsilon^{-(K+1)})$. This completes the proof.
\end{myproof}

\subsection{CES, Cobb-Douglas, and Leontief cases: III, KKT points}\label{sec.practical.epsilon-kkt}
While the global solution via \Cref{alg.sep.fptas} is not very practical, it worth noting that an $\epsilon$-approximate KKT point with complexity guarantees and can be found in using many existing interior-point codes. Except for the linear case, we can solve \ref{eq.cg.pricing} jointly in $(\y, \sigma)$,
\begin{eqalign}{}
    \max_{\y \in \real^n, \sigma \in \real}
    \pi(\y, \sigma)~&= \sum_{k=1}^K \inner{\bU \e_k}{\softmax{\y - \sigma \log(\p_k)}}, \\
    \st~& \y \in [-D_\y, D_\y]^n, \sigma \in [-1, D_\sigma],
\end{eqalign}
which has a box constraint on $\y$ and $\sigma$.
An $\epsilon$-approximate KKT point is defined as follows,
\begin{equation}\label{eq.fptas.kkt}
    ~\begin{array}{ll}
        \left|\partial_{y_j} \pi(\y, \sigma)\right| \le \epsilon                & \text{if } y_j \in (-D_\y, D_\y),     \\
        \max\left\{\partial_{y_j} \pi(\y, \sigma),\, 0\right\} \le \epsilon     & \text{if } y_j = -D_\y,               \\
        \max\left\{-\partial_{y_j} \pi(\y, \sigma),\, 0\right\} \le \epsilon    & \text{if } y_j = D_\y                 \\
        \left|\partial_{\sigma} \pi(\y, \sigma)\right| \le \epsilon             & \text{if } \sigma \in (-1, D_\sigma), \\
        \max\left\{\partial_{\sigma} \pi(\y, \sigma),\, 0\right\} \le \epsilon  & \text{if } \sigma = -1,               \\
        \max\left\{-\partial_{\sigma} \pi(\y, \sigma),\, 0\right\} \le \epsilon & \text{if } \sigma = D_\sigma.
    \end{array}
\end{equation}
\begin{remark}[Hardness of $\epsilon$-approximate KKT points]\label{remark.epsilon-kkt}
    It is known that an $\epsilon$-approximate KKT point \eqref{eq.fptas.kkt} can be found in $\cO(\epsilon^{-1.5})$ time \citep{bianComplexityAnalysisInterior2015}.
\end{remark}

\section{Proof of \Cref{thm.counterexample.informal}}\label{sec.proof.counterexample}
Here prove the lower bound by an example with $n=2$ goods. Because of that, we can focus on the expenditure share of the first good (i.e., $g_1(\p)$). We consider androids that are \emph{simple} if its share $\gamma_{1}(\p)$ has bounded total variation, and write $\TV(\bgamma_{1})\le G < \infty$.

\subsection{Two-good economy and total variation}\label{sec.counterexample.prelim}
For a homothetic agent with with two goods, we can write the quantities through the log-ratio of prices.  Consider the quantity ratio $r_x$ and the expenditure-share ratio $r_\gamma$ as function of the log-ratio of prices,
\begin{eqalign}{eq.def.twogood.coords}
    \tau(\p) := \log\frac{p_1}{p_2}, \quad r_x(\tau) := \frac{x_1}{x_2}, \quad
    r_\gamma(\tau) &:= \frac{\gamma_1(\tau)}{\gamma_2(\tau)} = \exp(\tau) r_x(\tau).
\end{eqalign}
Write $\ell(z) := \tfrac{\exp(z)}{1+\exp(z)}$ for the logistic map, so that $\gamma_1 = \ell(\log r_\gamma)$. We can define the elasticity of substitution is defined locally by
\begin{eqalign}{}
    \frac{\diff\log r_\gamma(\tau)}{\diff\tau} = \frac{\gamma_1'(\tau)}{\gamma_1(1-\gamma_1)} = -\sigma(\tau), \qquad 1+\sigma(\tau)\ge0,
\end{eqalign}
where the logistic factor uses $\gamma_1 = \ell(\log r_\gamma)$.

For a function $f:\cI\mapsto\real$ on an interval $\cI\subseteq\real$, let $\TV_\cI(f)$ be the total variation of $f$ on $\cI$, defined as the supremum of the sum of the absolute differences of the function over all possible partitions of the interval. Namely, for a finite subset $S_m = \{\tau_0, \cdots, \tau_m\} \subseteq \cI$ with $\tau_0 < \dots < \tau_m \in \cI$,
\begin{eqalign}{}
    \TV_\cI(f) = \sup_{S_m} \sum_{j=1}^m |f(\tau_j)-f(\tau_{j-1})|.
\end{eqalign}
We write $\TV(f) := \TV_\real(f)$ for short. The following properties are standard; see \citet[\S3.5]{follandRealAnalysisModern1999}.
\begin{lemma}[Total variation]\label{lem.tv}
    The following properties hold for the total variation of a function $f:\cI\mapsto\real$ on an interval $\cI\subseteq\real$:
    \begin{enumerate}
        \item If $f$ is absolutely continuous, then $\TV_\cI(f) = \int_\cI |f'(\tau)|\diff\tau$.
        \item If $f$ is monotone, then $\TV_\cI(f) = \sup_\cI f - \inf_\cI f$; if further $f$ is valued in $[0,1]$, then $\TV_\cI(f) \le 1$.
        \item $\TV(af) = |a|\TV(f)$, $\TV\left(\sum_k \lambda_k f_k\right) \le \sum_k \lambda_k \TV(f_k)$ for $\lambda_k \ge 0$.
        \item If $f_m \to f$ pointwise, then $\TV(f) \le \liminf_m \TV(f_m)$.
    \end{enumerate}
\end{lemma}
\begin{remark}
    A CES agent has bounded TV and $G=1$: its log-odds is affine in $\tau$, so its share is monotone, hence $\TV\le1$ by \Cref{lem.tv}(2). A linear agent also has finite variation: in two goods, linear utility $u(\x)=a_1x_1+a_2x_2$ with $a_1,a_2>0$ yields
    \begin{eqalign}{}
        \gamma_1(\tau)
        =
        \begin{cases}
            1, & \tau \le \log\tfrac{a_1}{a_2}, \\
            0, & \tau > \log\tfrac{a_1}{a_2},
        \end{cases}
    \end{eqalign}
    with the tie value arbitrary, so $\TV(\gamma_1)=1$.
\end{remark}

\subsection{The role of wealth functions}\label{sec.counterexample.wealth}
Now consider the wealth function $w_t(\p)$ of the android $t$.
We focus on polynomial wealth functions: each android's wealth $w_t$ is a nonnegative combination of the degree-$q$ Bernstein basis in the price of the first good $p_1 = \ell(\tau)$ (a subclass of the nonnegative degree-$q$ polynomials). That is, $w_t \in \rB^q$, defined as follows.
\begin{equation}
    \tag*{$(\rB^q)$}
    \begin{aligned}[b]
        \hat w_k(\tau) & := B_k^q(p_1(\tau)), \qquad B_k^q(s) := \binom{q}{k}s^k(1-s)^{q-k}, \quad k=0,\ldots,q, \\
        w_t(\p)        & := \sum_{k=0}^q c_{t,k} \hat w_k(\tau), \qquad c_{t,k}\ge0.
    \end{aligned}
\end{equation}
We keep $\sum_t^T c_{t,k}=1$ for every $k$, so the expenditure share regroups from a sum over the $T$ androids into a sum over the $q+1$ weights,
\begin{eqalign}{eq.q.order.mixture}
    h_1 = \sum_{t=1}^T w_t(\p)\gamma_{t1} = \sum_{k=0}^q \hat w_k(\tau)\hat g_{k1}, \qquad \hat g_{k1} := \sum_{t=1}^T c_{t,k}\gamma_{t1}.
\end{eqalign}
For each $k$, $\hat g_{k1}$ is a convex combination of the base shares $\gamma_{t1}$ with the constant weights $c_{t,k}$; by \Cref{lem.tv}(3), we immediately have
\begin{eqalign}{}
    \TV(\hat g_{k1})\le G.
\end{eqalign}
We need the following facts about the Bernstein basis.
\begin{lemma}[Bernstein basis]\label{lem.bernstein}
    The following holds for the degree-$q$ Bernstein basis $B_k^q(s)$ on $[0,1]$:
    \begin{enumerate}
        \item $\sum_{k=0}^q B_k^q(s) = 1$.
        \item $(B_k^q)'(s) = q\left(B_{k-1}^{q-1}(s) - B_k^{q-1}(s)\right)$, with out-of-range terms zero.
        \item $\int_0^1 \sum_{k=0}^q |(B_k^q)'(s)|\diff s \le 2q$.
    \end{enumerate}
\end{lemma}
\begin{myproof}
    Statements (1) and (2) are standard. For (3), we note
    \begin{eqalign}{}
        \sum_{k=0}^q |(B_k^q)'(s)| \le q\sum_{k=0}^q\left(B_{k-1}^{q-1}(s)+B_k^{q-1}(s)\right) = 2q,
    \end{eqalign}
    and integrating over $[0,1]$ gives the bound.
\end{myproof}

\begin{lemma}[$q$-th order additive wealth shares]\label{prop.q.order.weights}
    Let $n=2$. Any $h_1$ following \eqref{eq.q.order.mixture} satisfies
    \begin{eqalign}{}
        \TV(h_1)\le q + (q+1)G,
    \end{eqalign}
    independently of the number of androids.
\end{lemma}
\begin{myproof}
    We argue directly from the partition definition of total variation, so the base shares $\hat g_{k1}$ need only be of bounded variation (the weights $\hat w_k$ are smooth). Fix a partition $S_m = \{\tau_0<\cdots<\tau_m\}$ and write $\Delta_j f := f(\tau_j)-f(\tau_{j-1})$. Since $\sum_k\hat w_k\equiv1$ by \Cref{lem.bernstein}(1), we have $\sum_k\Delta_j\hat w_k = 0$. Summation by parts on \eqref{eq.q.order.mixture} gives
    \begin{eqalign}{}
        \Delta_j h_1
        &= \sum_{k=0}^q \left[\hat w_k(\tau_j)\,\Delta_j\hat g_{k1} + \hat g_{k1}(\tau_{j-1})\,\Delta_j\hat w_k\right] \\
        &= \sum_{k=0}^q \hat w_k(\tau_j)\,\Delta_j\hat g_{k1} + \sum_{k=0}^q\left(\hat g_{k1}(\tau_{j-1})-\tfrac12\right)\Delta_j\hat w_k,
    \end{eqalign}
    the second line subtracting $\tfrac12\sum_k\Delta_j\hat w_k = 0$. Since $0\le\hat w_k\le1$ and $|\hat g_{k1}-\tfrac12|\le\tfrac12$,
    \begin{eqalign}{}
        |\Delta_j h_1| \le \sum_{k=0}^q |\Delta_j\hat g_{k1}| + \tfrac12\sum_{k=0}^q |\Delta_j\hat w_k|.
    \end{eqalign}
    Summing over $j$ and taking the supremum over partitions,
    \begin{eqalign}{}
        \TV(h_1) \le \sum_{k=0}^q \TV(\hat g_{k1}) + \tfrac12\sum_{k=0}^q \TV(\hat w_k).
    \end{eqalign}
    Each $\hat g_{k1}$ satisfies $\TV(\hat g_{k1})\le G$, so the first sum is at most $(q+1)G$. The weights $\hat w_k = B_k^q(p_1(\tau))$ are absolutely continuous, and since $p_1=\ell(\tau)$ is monotone onto $(0,1)$, the change of variables $s=p_1$ with \Cref{lem.bernstein}(3) gives
    \begin{eqalign}{}
        \sum_{k=0}^q\TV(\hat w_k) = \int_\real\sum_{k=0}^q |\hat w_k'(\tau)|\diff\tau = \int_0^1\sum_{k=0}^q |(B_k^q)'(s)|\diff s \le 2q.
    \end{eqalign}
    Therefore $\TV(h_1)\le (q+1)G + q$.
\end{myproof}
The Fisher and Arrow-Debreu markets are only special cases.
\begin{itemize}
    \item \emph{$q=0$ (Fisher).} The only basis function is $B_0^0\equiv1$, so $\hat w_0\equiv1$ and the wealth
          \begin{eqalign}{}
              w_t = c_{t,0}
          \end{eqalign}
          is a price-independent constant (fixed budget). Then $h_1 = \sum_t c_{t,0}\gamma_{t1}$, giving $\TV(h_1)\le G$.
    \item \emph{$q=1$ (Arrow-Debreu).} The basis is $B_0^1(s)=1-s$ and $B_1^1(s)=s$, so with $s=p_1$ and $p_1+p_2=1$ on $\Delta_2$ the weights are $\hat w_0 = p_2$, $\hat w_1 = p_1$, and the wealth
          \begin{eqalign}{}
              w_t = c_{t,0}p_2 + c_{t,1}p_1 = \inner{\p}{\b_t}, \qquad \b_t = (c_{t,1}, c_{t,0}),
          \end{eqalign}
          is the value of the endowment $\b_t$. Then $h_1 = \sum_t \inner{\p}{\b_t}\gamma_{t1}$, giving the Arrow-Debreu bound $\TV(h_1)\le 2G+1$.
\end{itemize}

\subsection{A homothetic agent with unbounded variation}\label{sec.counterexample.example}
\begin{figure}[h]
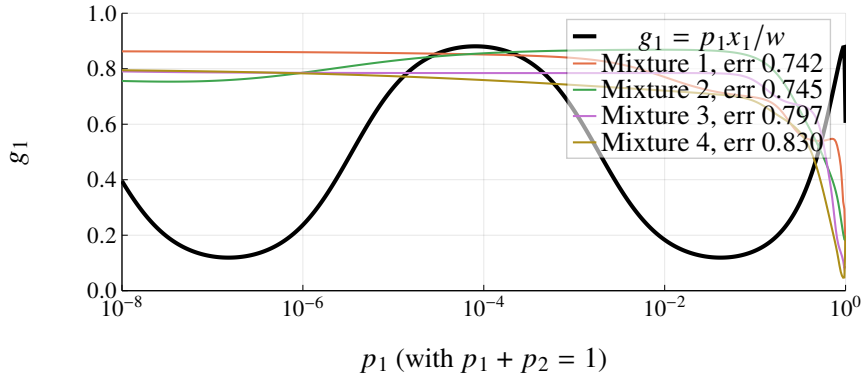

    \centering
    \resizebox{0.65\textwidth}{!}{


 }
    \caption{Expenditure share of \Cref{ex.oscillating.target}. Four different random mixtures of CES agents are presented for comparison.}\label{fig.counterexample.share}
\end{figure}

In the following one-agent economy, the market share $\g$ is the agent's own share $\bgamma$. We present its share versus the price of the first good $p_1$ in \Cref{fig.counterexample.share}.
\begin{example}[A homothetic ``oscillating'' agent]\label{ex.oscillating.target}
    Let $r_x(\tau) := \exp\left(-\tau + 2\sin(\tfrac\tau2)\right).$
    Then it can be rationalized by the utility
    \begin{eqalign}{eq.counterexample.utility}
        u(x_1,x_2)
        :=
        x_2\phi\left(\frac{x_1}{x_2}\right), \quad \text{where} \quad
        \log\phi(\rho)
        :=
        \int_1^\rho
        \frac{\diff s}{s + \exp(-r_x^{-1}(s))} < \infty.
    \end{eqalign}
    Furthermore, it holds that
    \begin{enumerate}
        \item $u$ is homogeneous of degree one, increasing, and concave.
        \item The expenditure share has log-odds
              \begin{eqalign}{eq.counterexample.logodds}
                  \log\frac{g_1(\tau)}{g_2(\tau)} = 2\sin(\tfrac\tau2),
              \end{eqalign}
              hence $\TV(g_1) = \infty$.
    \end{enumerate}
\end{example}
\begin{myproof}
    We first verify that the integral is well-defined. The exponent $\log r_x(\tau) = -\tau + 2\sin(\tfrac\tau2)$ has derivative $\cos(\tfrac\tau2)-1\le0$, so $r_x$ is a strictly decreasing continuous bijection from $\real$ onto $(0,\infty)$, and $r_x^{-1}$ is defined and continuous on $(0,\infty)$. The integrand is continuous on $(0,\infty)$ and satisfies $0 < [s+\exp(-r_x^{-1}(s))]^{-1} \le \tfrac1s$. Hence for any $\rho>0$,
    \begin{eqalign}{}
        |\log\phi(\rho)|
        = \left|\int_1^\rho \frac{\diff s}{s + \exp(-r_x^{-1}(s))}\right|
        \le \left|\int_1^\rho \frac{\diff s}{s}\right|
        = |\log\rho| < \infty,
    \end{eqalign}
    so the integral in \eqref{eq.counterexample.utility} is finite and $\log\phi(\rho)$ is well-defined.

    \noindent For part $(1)$, homogeneity is immediate from \eqref{eq.counterexample.utility}. Let
    \begin{eqalign}{}
        \rho := \frac{x_1}{x_2}, \quad D(\rho) := \rho + \exp(-r_x^{-1}(\rho)) > \rho,
    \end{eqalign}
    so that \eqref{eq.counterexample.utility} reads $\frac{\phi'(\rho)}{\phi(\rho)} = \frac{1}{D(\rho)}$. With $\diff_{x_1}\rho = 1/x_2$ and $\diff_{x_2}\rho = -\rho/x_2$, differentiating $u = x_2\phi(\rho)$ gives
    \begin{eqalign}{}
        u'_1 &= x_2\phi'(\rho)\diff_{x_1}\rho = \phi'(\rho), \\
        u'_2 &= \phi(\rho) + x_2\phi'(\rho)\diff_{x_2}\rho = \phi(\rho)-\rho\phi'(\rho) = \phi(\rho)\frac{D(\rho)-\rho}{D(\rho)},
    \end{eqalign}
    both positive (since $\phi>0$ and $D(\rho)>\rho$), so $u$ is increasing, with marginal rate of substitution
    \begin{eqalign}{eq.counterexample.mrs.value}
        \frac{u'_1}{u'_2} = \frac{\phi'(\rho)}{\phi(\rho)-\rho\phi'(\rho)} = \exp(r_x^{-1}(\rho)).
    \end{eqalign}
    For concavity we show $\phi$ is concave; then $u(x_1,x_2)=x_2\phi(x_1/x_2)$, its perspective, is concave on $\{x_2>0\}$. Since $r_x^{-1}$ is decreasing, $\exp(-r_x^{-1}(\rho))$ is increasing, so $D(\rho)-\rho = \exp(-r_x^{-1}(\rho))$ is increasing and $\diff D \ge \diff\rho$ as Stieltjes measures. From $\phi' = \phi/D$ we have $\log\phi' = \log\phi - \log D$, so for $0<a<b$, using $(\log\phi)' = 1/D$ and $\diff\log D = \diff D/D$,
    \begin{eqalign}{}
        \log\frac{\phi'(b)}{\phi'(a)}
        = \int_a^b \frac{\diff\rho}{D(\rho)} - \int_{(a,b]}\frac{\diff D(\rho)}{D(\rho)}
        \le 0,
    \end{eqalign}
    since $\diff D \ge \diff\rho$ and $D>0$. Hence $\phi'$ is nonincreasing, $\phi$ is concave, and so is $u$. Part $(1)$ is proved.

    \noindent Since $u$ is increasing, the budget binds, and the KKT condition for utility maximization reads
    \begin{eqalign}{}
        - \nabla u + \lambda \p = 0 \qlq \frac{u'_1}{u'_2} = \frac{p_1}{p_2} = \exp(\tau).
    \end{eqalign}
    By \eqref{eq.counterexample.mrs.value} this is $\exp(r_x^{-1}(\rho)) = \exp(\tau)$, so $\rho = r_x(\tau)$ and the optimizer is
    \begin{eqalign}{}
        x_1(\p,w) &=\frac{w r_x(\tau)}{p_1 r_x(\tau)+p_2}, \quad
        x_2(\p,w) = \frac{w}{p_1 r_x(\tau)+p_2}.
    \end{eqalign}
    Then
    \begin{eqalign}{}
        \frac{g_1(\tau)}{g_2(\tau)}
        =
        \frac{p_1x_1}{p_2x_2}
        =
        \exp(\tau)r_x(\tau)
        =
        \exp\left(2\sin(\tfrac\tau2)\right),
    \end{eqalign}
    which gives the log-odds \eqref{eq.counterexample.logodds}. Inverting the logistic map,
    \begin{eqalign}{}
        g_1(\tau)
        =
        \ell\left(2\sin(\tfrac\tau2)\right).
    \end{eqalign}
    This share oscillates between $\ell(-2)$ and $\ell(2)$ every $4\pi$ units of log-price. On the partition $S_{2N}$ with $\tau_j = \pi + 2\pi j$, $j = 0, \ldots, 2N$, the identity $2\sin(\tfrac{\tau_j}{2}) = 2(-1)^j$ gives $g_1(\tau_j) = \ell(2(-1)^j)$, so consecutive values alternate between $\ell(2)$ and $\ell(-2)$ and
    \begin{eqalign}{}
        \sum_{j=1}^{2N} |g_1(\tau_j)-g_1(\tau_{j-1})| = 2N\left(\ell(2)-\ell(-2)\right),
    \end{eqalign}
    which diverges as $N\to\infty$. Hence $\TV(g_1)=\infty$.
\end{myproof}
Economically, the elasticity in \Cref{ex.oscillating.target} is $\sigma(\tau) = -\cos(\tfrac\tau2)$, oscillating over $[-1,1]$. As $p_1$ rises the share $g_1$ alternately rises and falls, so the first good is neither a global complement nor a global substitute.

\subsection{A sharp lower bound}\label{sec.counterexample.lower}
\begin{lemma}\label{lem.grid.lower}
    Let $h:\real\mapsto[0,1]$ with $\TV(h)\le C$. Then for every partition $S_m = \{\tau_0<\cdots<\tau_m\}$ and fixed $f:\real\mapsto[0,1]$,
    \begin{eqalign}{eq.grid.lower}
        \sup_{\tau\in\real}|f(\tau)-h(\tau)|
        \ge
        \frac{1}{2m}
        \max\left\{
        \sum_{j=1}^m |f(\tau_j)-f(\tau_{j-1})| - C,
        ~0
        \right\}.
    \end{eqalign}
\end{lemma}
\begin{myproof}
    Bounding each pointwise gap $|f(\tau_i)-h(\tau_i)|$ by $\sup_{\tau\in\real}|f(\tau)-h(\tau)|$, for each adjacent pair,
    \begin{eqalign}{}
        |f(\tau_j)-f(\tau_{j-1})|
        &\le |h(\tau_j)-h(\tau_{j-1})|
        + |f(\tau_j)-h(\tau_j)|
        + |f(\tau_{j-1})-h(\tau_{j-1})| \\
        &\le |h(\tau_j)-h(\tau_{j-1})| + 2\sup_{\tau\in\real}|f(\tau)-h(\tau)|.
    \end{eqalign}
    Summing over $S_m$ gives
    \begin{eqalign}{}
        \sum_{j=1}^m |f(\tau_j)-f(\tau_{j-1})|
        &\le \sum_{j=1}^m |h(\tau_j)-h(\tau_{j-1})| + 2m\sup_{\tau\in\real}|f(\tau)-h(\tau)| \\
        &\le \TV(h) + 2m\sup_{\tau\in\real}|f(\tau)-h(\tau)| \\
        &\le C + 2m\sup_{\tau\in\real}|f(\tau)-h(\tau)|.
    \end{eqalign}
    Rearranging gives \eqref{eq.grid.lower}; if the bracketed term is negative, the bound holds trivially since $\sup_{\tau\in\real}|f(\tau)-h(\tau)|\ge0$.
\end{myproof}

\begin{corollary}[Sharp lower bound for fixed finite order]\label{cor.counterexample.lower}
    Fix $q<\infty$ and $1\le T\le\infty$, and let $g_1$ be the target of \Cref{ex.oscillating.target}. The best uniform approximation over admissible pairs of wealth functions and base shares is
    \begin{eqalign}{}
        \inf_{\substack{w_t\in\rB^q,\ \sum_t w_t\equiv1 \\ \gamma_{t1}:\real\mapsto[0,1],\ \TV(\gamma_{t1})\le G}}
        ~\sup_{\tau\in\real}\left|g_1(\tau)-\sum_{t=1}^T w_t(\p)\gamma_{t1}(\tau)\right|
        =
        \frac{\ell(2)-\ell(-2)}{2}
        =
        \frac{\tanh(1)}{2}
        \approx 0.3808,
    \end{eqalign}
    The bound is attained if every android $t$ is uses a Cobb-Douglas share.
\end{corollary}
\begin{myproof}
    Take the partition $S_m$ with $\tau_j := \pi + 2\pi j,~ j=0,\ldots,m.$
    Then $\sin(\tau_j/2)=(-1)^j$, so for any $j$,
    \begin{eqalign}{}
        |g_1(\tau_j)-g_1(\tau_{j-1})|
        = \ell(2)-\ell(-2).
    \end{eqalign}
    Since $w_t \in \rB^q$ for every $t$, \Cref{prop.q.order.weights} gives $\TV(h_1)\le q+(q+1)G$. Applying \Cref{lem.grid.lower} to $S_m$ with $C=q+(q+1)G$ gives
    \begin{eqalign}{}
        \sup_{\tau\in\real}|g_1(\tau)-h_1(\tau)|
        \ge
        \frac{m(\ell(2)-\ell(-2))-(q+(q+1)G)}{2m}.
    \end{eqalign}
    Taking $m\to\infty$ yields
    \begin{eqalign}{}
        \sup_{\tau\in\real}|g_1(\tau)-h_1(\tau)|
        \ge
        \frac{\ell(2)-\ell(-2)}{2}.
    \end{eqalign}
    Finally,
    \begin{eqalign}{}
        \ell(2)-\ell(-2)
        =
        \frac{\exp(2)-1}{\exp(2)+1}
        =
        \tanh(1),
    \end{eqalign}
    and $\ell(-2)=1-\ell(2)$, so the constant share $h_1\equiv\tfrac12$ has uniform error $(\ell(2)-\ell(-2))/2$. This constant is attainable by taking every android Cobb-Douglas such that $\gamma_{t1}\equiv\tfrac12$, and thus,
    \begin{eqalign}{}
        \gamma_{t1}(\tau) &\equiv \tfrac12, \quad
        h_1(\tau)
        &=
        \sum_{t} w_t(\p)\gamma_{t1}(\tau)
        =
        \tfrac12\sum_{t} w_t(\p)
        =
        \tfrac12.
    \end{eqalign}
    Since $2\sin(\tfrac\tau2)$ ranges over $[-2,2]$ and $\ell$ is increasing with $\ell(-2)=1-\ell(2)$, the share $g_1(\tau)=\ell(2\sin(\tfrac\tau2))$ is farthest from $\tfrac12$ at the extremes, so
    \begin{eqalign}{}
        \sup_{\tau\in\real}\left|g_1(\tau)-h_1(\tau)\right|
        = \sup_{\tau\in\real}\left|\ell\left(2\sin(\tfrac\tau2)\right)-\tfrac12\right|
        = \ell(2)-\tfrac12
        = \frac{\ell(2)-\ell(-2)}{2}
        = \frac{\tanh(1)}{2}.
    \end{eqalign}
    This matches the lower bound established above.
\end{myproof}

\end{document}